\documentclass[11pt,a4paper]{article}
\pdfoutput=1
\usepackage{jheppub} 
\usepackage{url}
\usepackage{multirow}
\usepackage{array,booktabs,colortbl,multirow}
\usepackage{colortbl,xcolor,colordvi,color}
\usepackage{rotating}
\usepackage{placeins}

\newcommand{\cmrule}{\midrule[0.25mm]}
\newcommand{\colcell}{\cellcolor[rgb]{0.85,0.85,0.85}}
\newcommand{\mt}[1]{\mathrm{#1}}
\newcolumntype{C}[1]{>{\centering\let\newline\\\arraybackslash\hspace{0pt}}m{#1}}

\title{Electroweak precision observables and Higgs-boson signal strengths in
the Standard Model and beyond: present and future}

\author[a]{J. de Blas,}
\author[b]{M. Ciuchini,}
\author[a]{E. Franco,}
\author[c]{S. Mishima,}
\author[d]{M. Pierini,}
\author[e,f]{L. Reina,}
\author[a]{and L. Silvestrini}

\affiliation[a]{INFN, Sezione di Roma,\\ Piazzale A. Moro 2, I-00185 Rome, Italy}
\affiliation[b]{INFN, Sezione di Roma Tre,\\ Via della Vasca Navale 84, I-00146 Roma, Italy}
\affiliation[c]{Theory Center, Institute of Particle and Nuclear Studies (IPNS), \\
High Energy Accelerator Research Organization (KEK), 
1-1 Oho, Tsukuba, 305-0801, Japan}
\affiliation[d]{CERN,\\ Geneva, Switzerland}
\affiliation[e]{Physics Department, Florida State University,\\ Tallahassee, FL 32306-4350, USA}
\affiliation[f]{Kavli Institute for Theoretical Physics, University of
  California,\\
Santa Barbara, CA 93106-4030, USA}

\emailAdd{jorge.deblasmateo@roma1.infn.it}
\emailAdd{marco.ciuchini@roma3.infn.it}
\emailAdd{enrico.franco@roma1.infn.it}
\emailAdd{satoshi.mishima@kek.jp}
\emailAdd{maurizio.pierini@cern.ch}
\emailAdd{reina@hep.fsu.edu}
\emailAdd{luca.silvestrini@roma1.infn.it}

\abstract{We present results from a state-of-the-art fit of
  electroweak precision observables and Higgs-boson signal-strength
  measurements performed using 7 and 8 TeV data from the Large Hadron
  Collider.  Based on the {\tt HEPfit} package, our study updates the
  traditional fit of electroweak precision observables and extends it
  to include Higgs-boson measurements. As a result we obtain
  constraints on new physics corrections to both electroweak
  observables and Higgs-boson couplings. We present the projected
  accuracy of the fit taking into account the expected sensitivities
  at future colliders.}

\notoc
\begin{document}
\preprint{KEK-TH-1919}
\maketitle
\flushbottom

\section{Introduction}
\label{sec:introduction}

Looking for indirect evidence of physics beyond the Standard
Model (SM) has become a strong component of the Large Hadron Collider
(LHC) physics program. Run I of the LHC revealed the existence of a
Higgs boson ($H$) with characteristics very similar to the Higgs boson
of the SM. Identifying the $H$ particle with the SM Higgs boson fully
determines the SM Lagrangian, so that all electroweak precision
observables (EWPO) and all Higgs-boson couplings can be predicted
within the SM. Thus, EWPO and Higgs-boson observables play a key role in
constraining extensions of the SM and in searching for new physics (NP).

In this paper we present a global fit of both EWPO and Higgs-boson
signal strengths, based on results obtained at LEP, SLC, the Tevatron,
and during Run I of the LHC, at both 7 and 8 TeV center-of-mass
energies. The fit is carried out using the {\tt HEPfit}
package~\cite{hepfit,hepfitsite}, a general tool to combine direct and
indirect constraints on the SM and its extensions. In
particular, we use {\tt HEPfit} to perform a statistical analysis of
EWPO and Higgs-boson signal-strength
measurements in the SM and beyond.
Most importantly, we obtain
constraints on possible deviations of the Higgs-boson couplings to
both gauge bosons and fermions from the SM prediction.  Finally, we
investigate the impact of the high-luminosity upgrade of the LHC and 
of future $e^+ e^-$ colliders 
on the precision of the fit in the SM and beyond.  Our analysis
updates the study of ref.~\cite{Ciuchini:2013pca} and extends it to
include recent Higgs-boson physics results. Results from the initial
stages of this project were presented
in~\cite{deBlas:2014ula,Ciuchini:2014dea,Reina:2015yuh,hepfit:LP15} and have by now
been updated to reflect all the most recent developments in
theoretical calculations and experimental measurements.  A
model-independent study of NP effects on both EWPO and Higgs-boson
couplings based on an effective-field-theory approach will be
presented in a forthcoming paper~\cite{hepfit:2015daje}.

Recent updates of global fits to EWPO in the SM and beyond, as well as
constraints on Higgs-boson couplings, have been presented in
refs.~\cite{Baak:2014ora,Agashe:2014kda}. In spite of the different
statistical methods and of the different inputs, we obtain compatible
results for the EWPO fit. We however consider more NP
parameterizations, implement constraints from Higgs-boson signal
strengths, and extend the analysis of future accuracies to more
scenarios.

The paper is organized as follows.
In section~\ref{sec:hepfit} we briefly describe the {\tt
  HEPfit} package.  In sections~\ref{sec:ew-precision-fit-SM} and
\ref{sec:ew-precision-fit-BSM} we summarize results for the
electroweak (EW) precision fits of the SM and its extensions, while we
illustrate in section~\ref{sec:Higgs-coupl-constraints} the
constraints we obtain for non-standard Higgs-boson couplings. The
impact of future colliders on our analysis is discussed in
section~\ref{sec:future}.  In section~\ref{sec:conclusions} we present
our conclusions.

\section{The {\tt HEPfit} package}
\label{sec:hepfit}
The {\tt HEPfit} package\footnote{Formerly known as {\tt SUSYfit}, the
  package has now grown to include more physical observables and
  multiple models.} is a general tool to combine direct and indirect
constraints on the Standard Model and its extensions, available under
the GNU General Public License (GPL) ~\cite{hepfitsite}. The {\tt
  HEPfit} code can be extended to include new observables and NP
models  which can be added to the main core as
external modules.  Exploiting the Markov-Chain Monte-Carlo
implementation provided in the Bayesian Analysis Toolkit
(BAT)~\cite{Caldwell:2008fw}, {\tt HEPfit} can be used as a standalone
program to perform Bayesian statistical analyses.  Alternatively, it
can be used in library mode to compute observables in any implemented
model, allowing for phenomenological analyses in any statistical
framework. The interested reader can find more details on {\tt HEPfit}
in refs.~\cite{hepfit,hepfitsite}.  The first application of the {\tt
  HEPfit} code has been to 
update the EW precision fit presented in ref.~\cite{Ciuchini:2014dea},
a detailed explanation of which can be found in
\cite{Ciuchini:2013pca} and references therein.

In this paper we use {\tt HEPfit} to perform a Bayesian statistical
analysis of EWPO and Higgs-boson observables in the SM and beyond.
The code for the EWPO and Higgs observables has been written from
scratch. The EWPO results have been successfully validated against {\tt
  ZFITTER}~\cite{Akhundov:2013ons}.

\section{Electroweak precision fit in the Standard Model}
\label{sec:ew-precision-fit-SM}

\begin{table}[t]
{\footnotesize
\setlength\tabcolsep{5pt}
\begin{center}
\begin{tabular}{lcccccc}
\toprule
& Ref. & Measurement & Posterior & Prediction &\!1D Pull\! &\!nD Pull\! \\
\cmrule
$\alpha_s(M_Z)$ &\cite{Agashe:2014kda} & $ 0.1179 \pm 0.0012 $ & $ 0.1180 \pm 0.0011 $  & $ 0.1185 \pm 0.0028 $ & -0.2  & \\ 
$\Delta\alpha_{\rm had}^{(5)}(M_Z)$ & \cite{Burkhardt:2011ur} & $ 0.02750 \pm 0.00033 $ & $ 0.02747 \pm 0.00025 $  & $ 0.02743 \pm 0.00038 $ & 0.04  & \\ 
\rowcolor[gray]{.85}$M_Z$ [GeV] & \cite{ALEPH:2005ab} & $ 91.1875 \pm 0.0021 $ & $ 91.1879 \pm 0.0020 $  & $ 91.199 \pm 0.011 $ & -1.0 & \\ 
$m_t$ [GeV] & \cite{ATLAS:2014wva} & $ 173.34 \pm 0.76 $ & $ 173.61 \pm 0.73 $  & $ 176.6 \pm 2.5 $ & -1.3  & \\ 
$m_H$ [GeV] & \cite{Aad:2015zhl} & $ 125.09 \pm 0.24 $ & $ 125.09 \pm 0.24 $  & $ 102.8 \pm 26.3 $ & 0.8  & \\
\cmrule
$M_W$ [GeV] & \cite{Group:2012gb} & $ 80.385 \pm 0.015 $ & $ 80.3644 \pm 0.0061 $  & $ 80.3604 \pm 0.0066 $ & 1.5  & \\ 
\cmrule 
$\Gamma_{W}$ [GeV] & \cite{ALEPH:2010aa} & $ 2.085 \pm 0.042 $ & $2.08872 \pm 0.00064 $  & $ 2.08873 \pm 0.00064 $ & -0.2  & \\ 
\cmrule 
$\sin^2\theta_{\rm eff}^{\rm lept}(Q_{\rm FB}^{\rm had})$\!\!\!\!\! & \cite{ALEPH:2005ab} & $ 0.2324 \pm 0.0012 $ & $ 0.231464 \pm 0.000087 $  & $ 0.231435 \pm 0.000090 $ & 0.8  & \\
\cmrule 
$P_{\tau}^{\rm pol}\!=\!\mathcal{A}_\ell$ & \cite{ALEPH:2005ab} & $ 0.1465 \pm 0.0033 $ & $ 0.14748 \pm 0.00068 $  & $ 0.14752 \pm 0.00069 $ & -0.4  & \\ 
\cmrule
\rowcolor[gray]{.85}$\Gamma_{Z}$ [GeV] & \cite{ALEPH:2005ab} & $ 2.4952 \pm 0.0023 $ & $ 2.49420 \pm 0.00063 $  & $ 2.49405 \pm 0.00068 $ & 0.5 & \\ 
\rowcolor[gray]{.85}$\sigma_{h}^{0}$ [nb] & \cite{ALEPH:2005ab} & $ 41.540 \pm 0.037 $ & $ 41.4903 \pm 0.0058 $  & $ 41.4912 \pm 0.0062 $ & 1.3 & 0.7\\
\rowcolor[gray]{.85}$R^{0}_{\ell}$ & \cite{ALEPH:2005ab} & $ 20.767 \pm 0.025 $ & $ 20.7485 \pm 0.0070 $  & $ 20.7472 \pm 0.0076 $ & 0.8 & \\ 
\rowcolor[gray]{.85}$A_{\rm FB}^{0, \ell}$ & \cite{ALEPH:2005ab} & $ 0.0171 \pm 0.0010 $ & $ 0.01631 \pm 0.00015 $  & $ 0.01628 \pm 0.00015 $ & 0.8 & \\ 
\cmrule
\rowcolor[gray]{.75}$\mathcal{A}_{\ell}$ (SLD) & \cite{ALEPH:2005ab} & $ 0.1513 \pm 0.0021 $ & $ 0.14748 \pm 0.00068 $  & $ 0.14765 \pm 0.00076 $ & 1.7 & \\ 
\rowcolor[gray]{.75}$\mathcal{A}_c$ & \cite{ALEPH:2005ab} & $ 0.670 \pm 0.027 $ & $ 0.66810 \pm 0.00030 $  & $ 0.66817 \pm 0.00033 $ & 0.02 & \\ 
\rowcolor[gray]{.75}$\mathcal{A}_b$ & \cite{ALEPH:2005ab} & $ 0.923 \pm 0.020 $ & $ 0.934650 \pm 0.000058 $  & $ 0.934663 \pm 0.000064 $ & -0.6 & \\ 
\rowcolor[gray]{.75}$A_{\rm FB}^{0, c}$ & \cite{ALEPH:2005ab} & $ 0.0707 \pm 0.0035 $ & $ 0.07390 \pm 0.00037 $  & $ 0.07399 \pm 0.00042 $ & -0.9 & 1.5\\ 
\rowcolor[gray]{.75}$A_{\rm FB}^{0, b}$ & \cite{ALEPH:2005ab} & $ 0.0992 \pm 0.0016 $ & $ 0.10338 \pm 0.00048 $  & $ 0.10350 \pm 0.00054 $ & -2.6 & \\ 
\rowcolor[gray]{.75}$R^{0}_{c}$ & \cite{ALEPH:2005ab} & $ 0.1721 \pm 0.0030 $ & $ 0.172228 \pm 0.000023 $  & $ 0.172229 \pm 0.000023 $ & -0.05 & \\ 
\rowcolor[gray]{.75}$R^{0}_{b}$ & \cite{ALEPH:2005ab} & $ 0.21629 \pm 0.00066 $ & $ 0.215790 \pm 0.000028 $  & $ 0.215788 \pm 0.000028 $ & 0.7 & \\ 
\cmrule 
$\sin^2\theta_{\rm eff}^{ee}$ & \cite{Aaltonen:2016nuy} & $ 0.23248 \pm 0.00052 $ & \multirow{6}{*}{$0.231464 \pm 0.000087$}  & \multirow{6}{*}{$0.231435 \pm 0.000090$} &  2.1 \\
$\sin^2\theta_{\rm eff}^{\mu\mu}$ & \cite{Aaltonen:2014loa} & $ 0.2315 \pm 0.0010 $ &  & & 0.07  \\ 
$\sin^2\theta_{\rm eff}^{ee}$ & \cite{Abazov:2014jti} & $ 0.23146 \pm 0.00047 $ &  & & 0.1  \\ 
$\sin^2\theta_{\rm eff}^{ee,\mu\mu}$ & \cite{Aad:2015uau} & $ 0.2308 \pm 0.0012 $ &  & & -0.5  \\
$\sin^2\theta_{\rm eff}^{\mu\mu}$ & \cite{Chatrchyan:2011ya} & $ 0.2287 \pm 0.0032 $ &  & & -0.8  \\
$\sin^2\theta_{\rm eff}^{\mu\mu}$ & \cite{Aaij:2015lka} & $ 0.2314 \pm 0.0011 $ &  & & -0.1  \\ 
\bottomrule
\end{tabular}
\end{center}
}
\caption{
  Experimental measurement, result, prediction, and pull for the five input
  parameters ($\alpha_s(M_Z)$, $\Delta \alpha^{(5)}_{\mathrm{had}}(M_Z)$, $M_Z$,
  $m_t$, $m_H$), and for the set of EWPO considered in the SM EW fit. The values in
  the column \emph{Prediction} are determined without using the
  corresponding experimental information (see text). Pulls are calculated
  both as individual pulls (\emph{1D Pull}) and as global pulls
  (\emph{nD Pull}) for sets of correlated observables (see text), and are given in units of standard
  deviation. Groups of correlated observables
  are identified by shades of grey.} 
\label{tab:SMfit}
\end{table}

In this section we update the fit of EWPO presented in
refs.~\cite{Ciuchini:2013pca,Ciuchini:2014dea}, where all relevant
formul{\ae} and a detailed overview of the literature can be found.  With
respect to ref.~\cite{Ciuchini:2013pca}, we include the full two-loop
fermionic EW corrections to $Z$ partial decay widths computed in
ref.~\cite{Freitas:2014hra}, and the four-loop approximate QCD
corrections to the $W$ mass computed in
ref.~\cite{Schroder:2005db,Chetyrkin:2006bj,Boughezal:2006xk} (we use
the updated semi-analytical formula given in
ref.~\cite{Awramik:2003rn}).

Among the input parameters, $G_\mu$ and
$\alpha$ are fixed ($G_\mu=1.1663787\times10^{-5}$~GeV$^{-2}$, and
$\alpha=1/137.035999139$~\cite{Agashe:2014kda}), while $\alpha_s(M_Z)$,
$\Delta\alpha^{(5)}_{\mathrm{had}}(M_Z)$, $M_Z$, $m_t$, and $m_H$ are
taken as floating.  
We use flat priors for all the SM input parameters, and include the
information of their experimental measurements in the likelihood. We
assume all experimental distributions are Gaussian.
Parameters and results for the various EWPO included in the fit are
summarized in table~\ref{tab:SMfit}, where we also give the references
from which the measurements have been taken.  

With respect to refs.~\cite{Ciuchini:2013pca,Ciuchini:2014dea}, we
have updated $m_H$~\cite{Aad:2015zhl} and we use the top-quark mass as
given by the most up-to-date world average~\cite{ATLAS:2014wva}. The
values of $M_Z$~\cite{ALEPH:2005ab} and
$\Delta\alpha_\mathrm{had}^{(5)}(M_Z)$~\cite{Burkhardt:2011ur} are
unchanged.  Concerning $\alpha_s(M_Z)$, we notice that the most recent
PDG average~\cite{Agashe:2014kda}
($\alpha_s(M_Z)=0.1179\pm 0.0012$), excluding the result of the EW fit is compatible
  but sizeably different from the previous one
($\alpha_s(M_Z)=0.1185\pm 0.0006$), showing both a
  lower central value and a larger uncertainty.  This is mainly due to
  the fact that now the uncertainty of the combined lattice result for
  $\alpha_s(M_Z)$ is calculated using the same unweighted average procedure
  as done for the determination of $\alpha_s(M_Z)$ in other sub-fields
  (e.g. hadronic $\tau$ decays, DIS, etc.). A $\chi^2$ averaging
  procedure is then applied to combine the values of
  $\alpha_s(M_Z)$ from the different sub-fields.  Previously, the
  PDG world average for $\alpha_s(M_Z)$ had been obtained using a
  $\chi^2$ averaging procedure also to obtain the value of $\alpha_s(M_Z)$
  from lattice QCD alone~\cite{Agashe:2014kda}.  The new procedure
turns out to be more conservative and increases the
uncertainty on the lattice determination of $\alpha_s(M_Z)$ (which was
previously dominating the average), leading to a larger
  final uncertainty of the new world average, and to a reduced fixing
  power towards the central average value that is now shifted towards
  lower values induced by measurements in other sub-fields, like the
  newly added CMS measurement of the $t\bar{t}$ cross section at
  $\sqrt{s}=7$~TeV~\cite{Chatrchyan:2013haa}.  Oddly, the new error
of the PDG lattice average is comparable to the uncertainty of
$\alpha_s(M_Z)$ by FLAG~\cite{Aoki:2016frl}, although the FLAG error
is dominated by an estimate of the uncertainty associated with the
truncation of the perturbative series.

In the following, we will use the new PDG average (obtained excluding the EW fit
determination) $\alpha_s(M_Z)=0.1179\pm0.0012$
as a reference value. However, in view of the impact on the EW fit of the increased
error, we also present the results for the SM fit with the previous PDG average
$\alpha_s(M_Z)=0.1185\pm0.0005$ to allow the reader to appreciate the effect
of the new average.
Finally, we have included in the fit the
latest determinations of the effective leptonic angle,
$\sin^2\theta_{\mathrm{eff}}^{\mathrm{lept}}$, obtained at the
Tevatron and at Run I of the LHC.

For each observable, we give the experimental information
used as input (\emph{Measurement}), together with the output of the
combined fit (\emph{Posterior}), and the \emph{Prediction} of the same
quantity. The latter is obtained from the posterior predictive
distribution derived from a combined analysis of all the other
quantities. The compatibility of the constraints is then tested
computing the \emph{Pull} for each observable as the difference
between the corresponding prediction and measurement in units of the
combined standard deviation (\emph{1D Pull}). Care must be taken in
defining the pull for experimentally correlated observables.
In this case, we
remove from the fit one set of correlated observables at a time and
compute the prediction for the set of observables together with their
correlation matrix. Adding the experimental covariance matrix to the
one obtained from the fit, we compute the log likelihood and the
corresponding $p$-value, which we then convert into a global pull for
the correlated set of observables assuming Gaussian distributions
(\emph{nD Pull}).

\begin{figure}
  \centering
  \includegraphics[width=.35\textwidth]{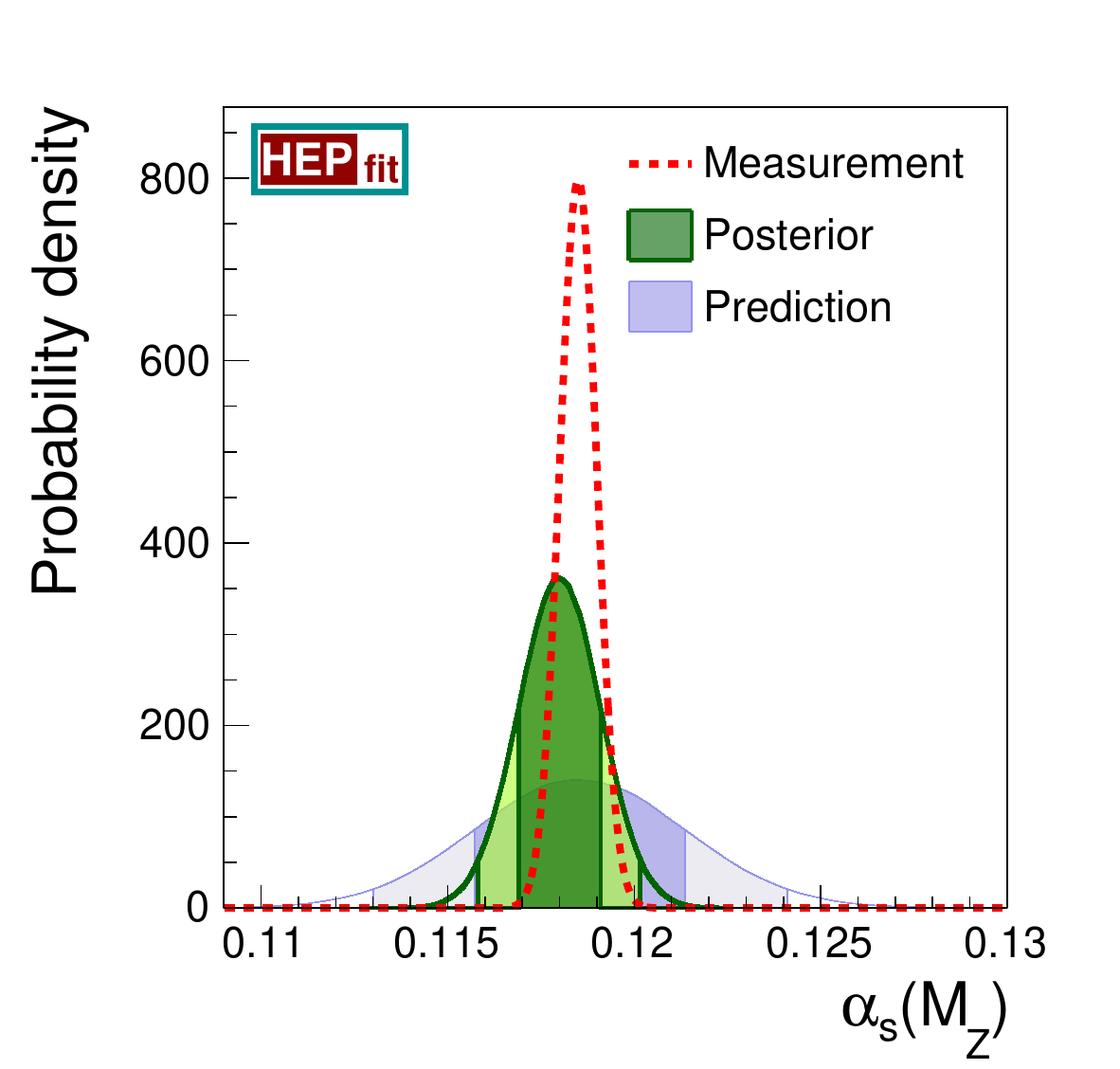}
  \hspace{-7mm}
  \includegraphics[width=.35\textwidth]{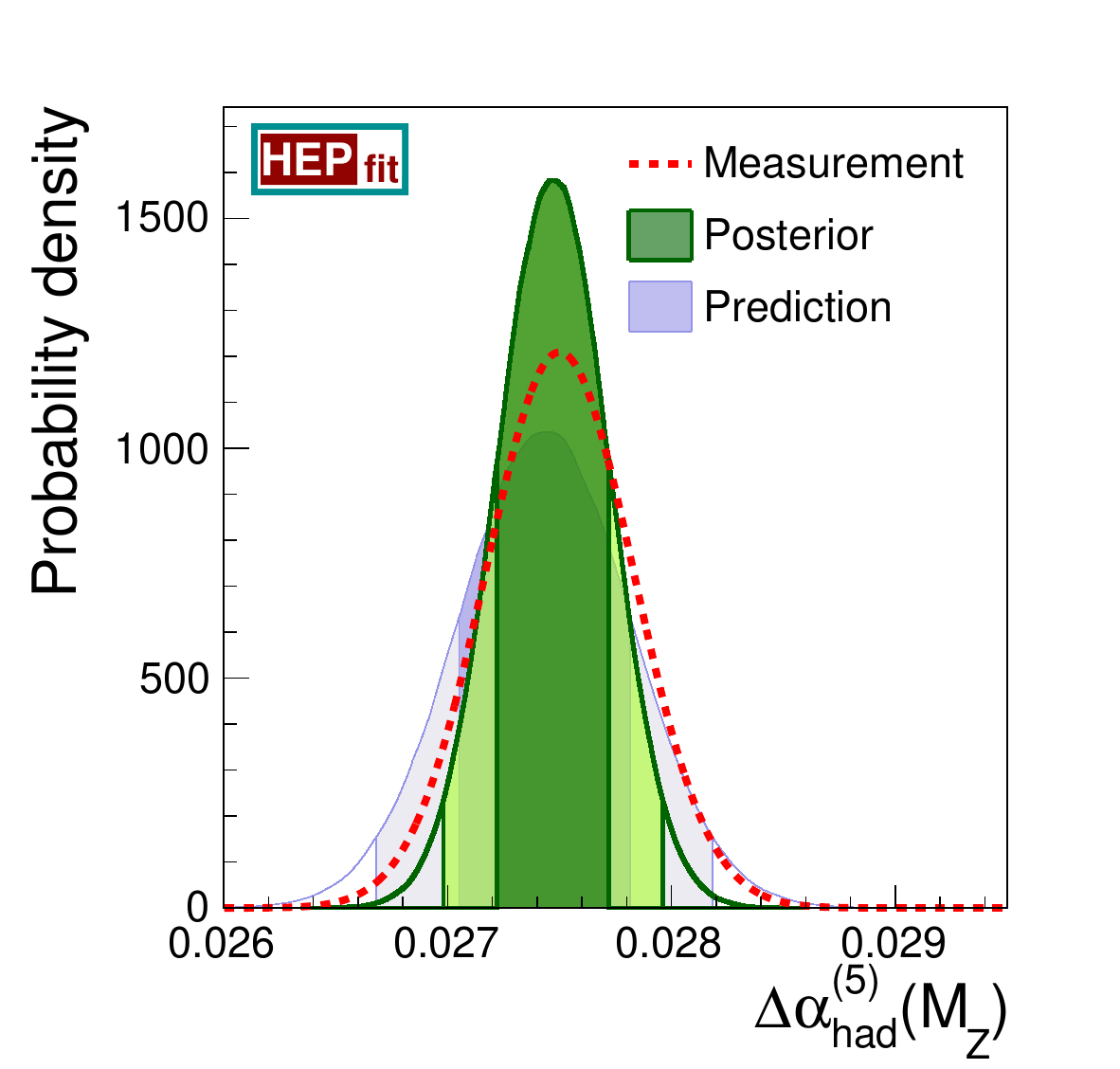}
  \hspace{-7mm}
  \includegraphics[width=.35\textwidth]{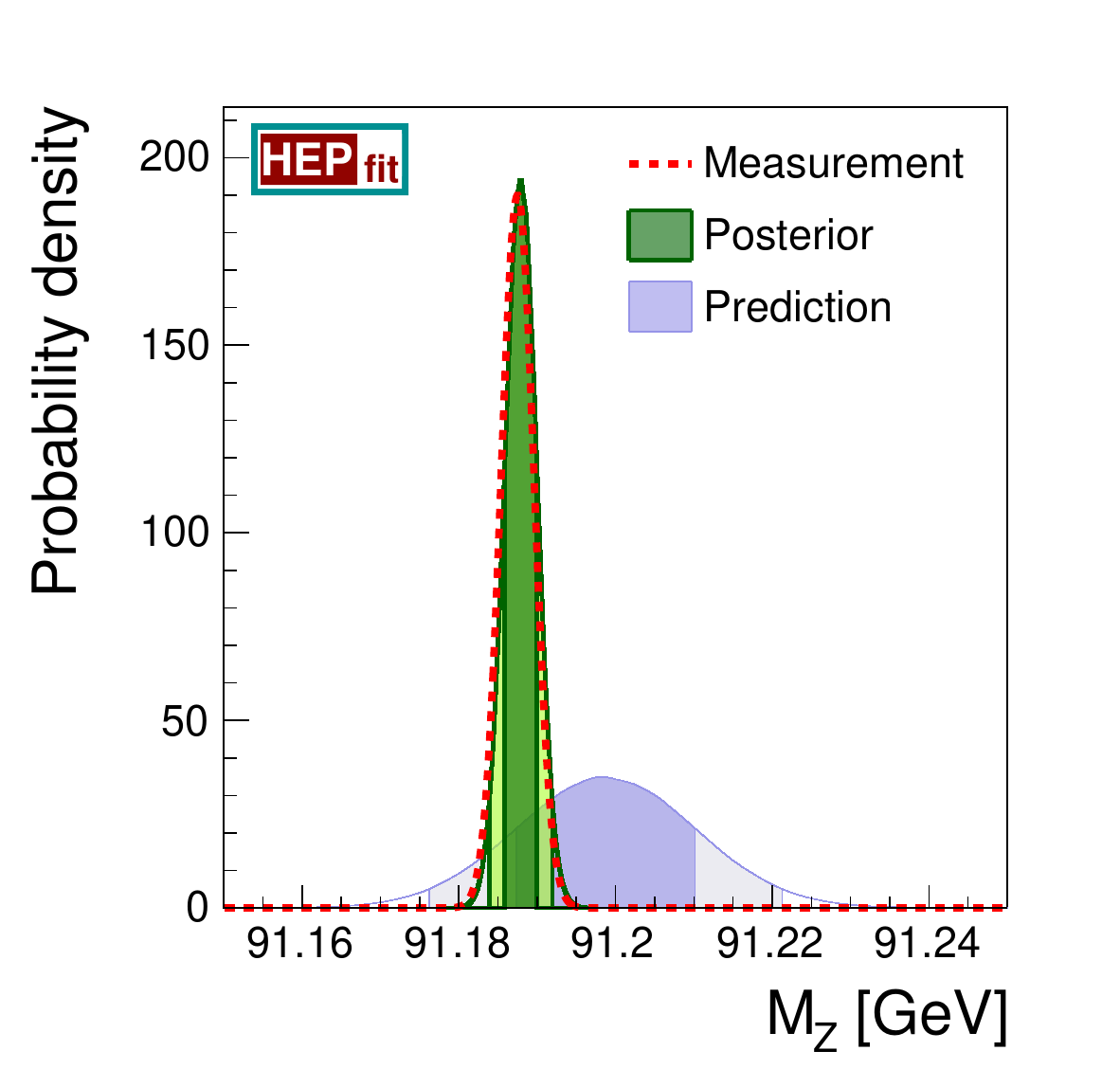}
  \\
  \includegraphics[width=.35\textwidth]{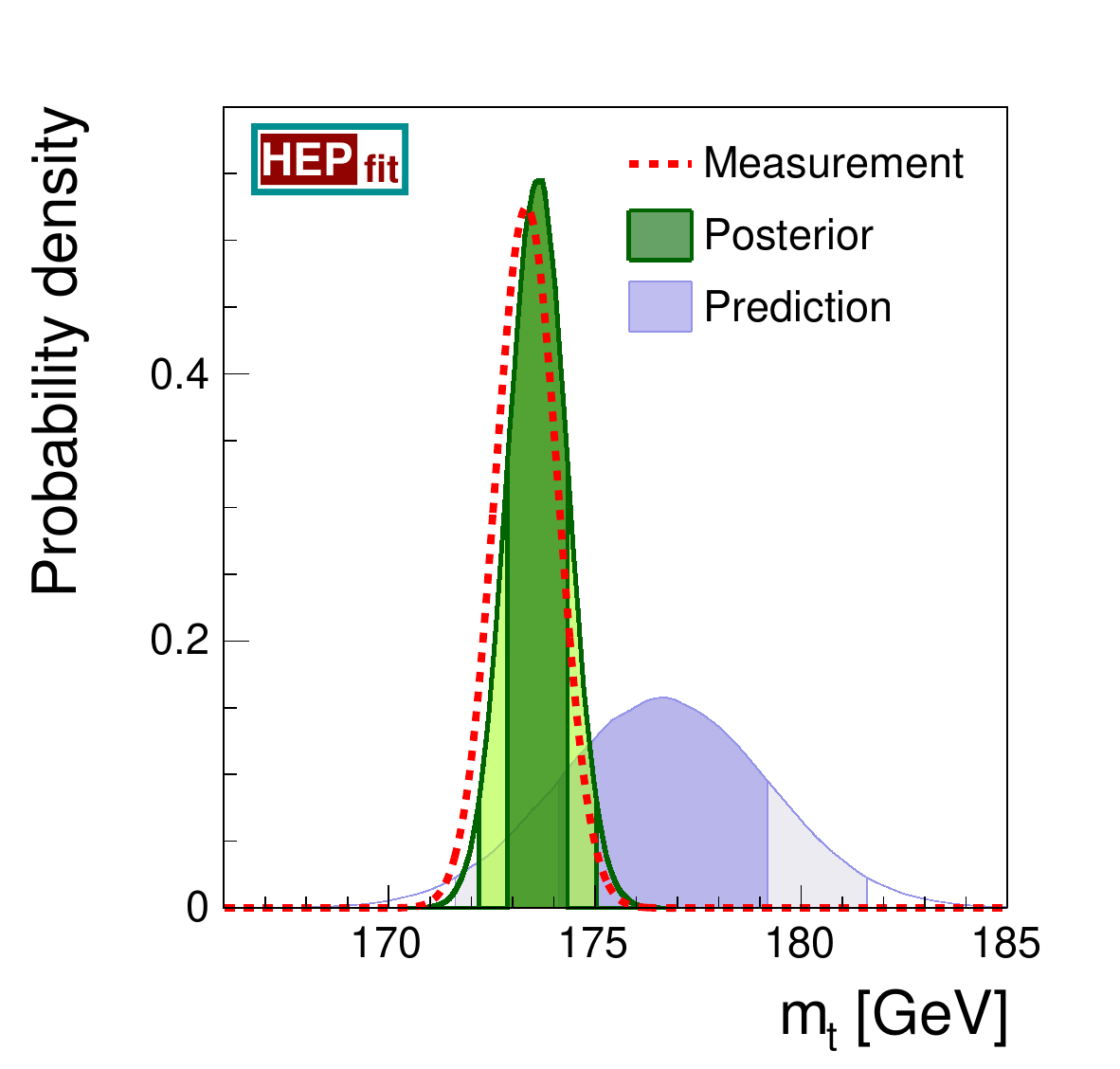}
  \hspace{-7mm}
  \includegraphics[width=.35\textwidth]{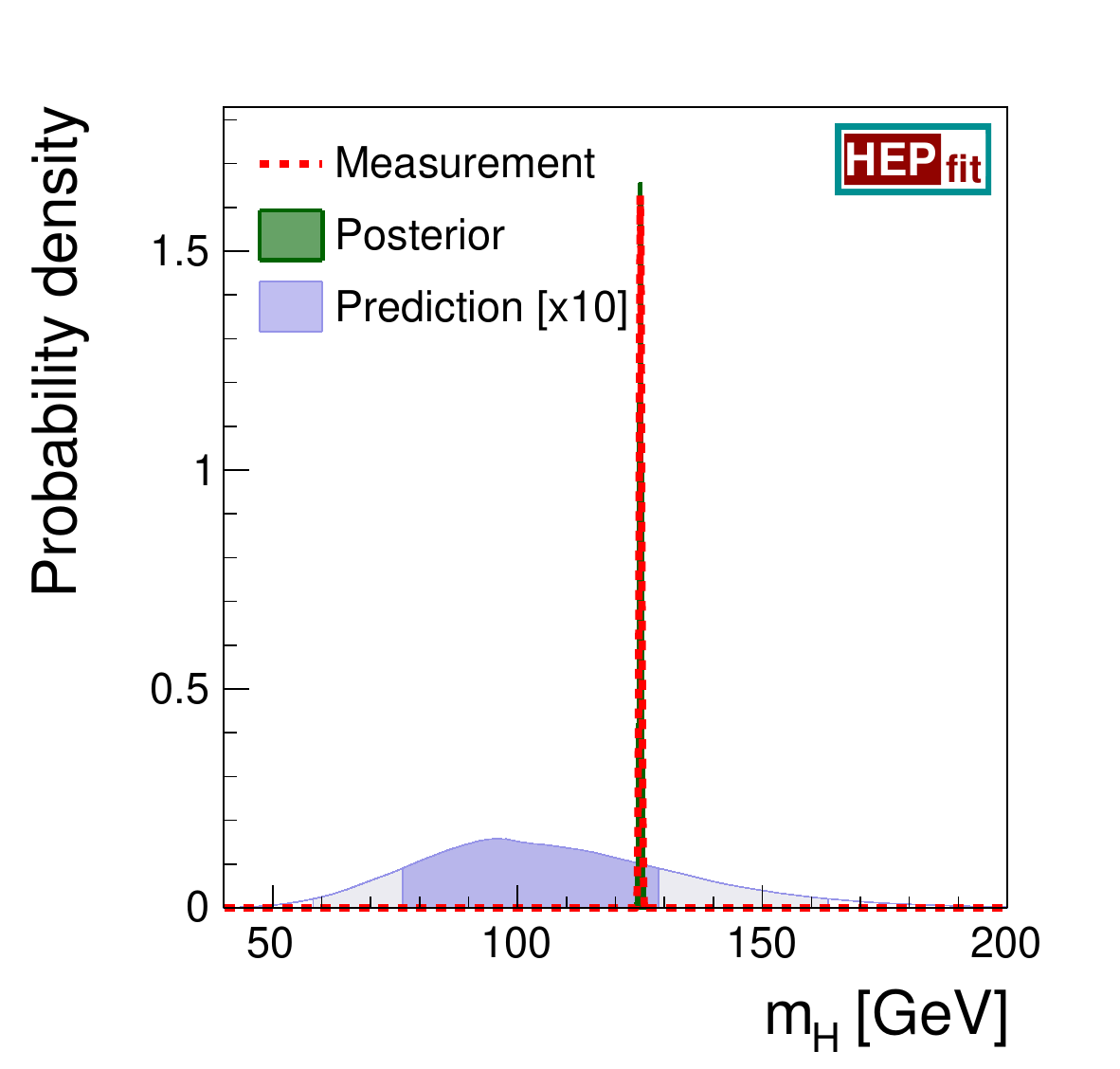}
  \caption{Comparison among the direct measurement, the posterior, and
    the posterior predictive (or indirect) probability distributions
    for the input parameters in the SM fit. The latter is obtained
    from the fit by assuming a flat prior for the parameter under
    consideration. Dark (light)
    regions correspond to $68\%$ ($95\%$) probability ranges.}
  \label{fig:SMinputs}
\end{figure}

In figure~\ref{fig:SMinputs}, we show a comparison of the direct
measurement (\emph{Measurement} in table~\ref{tab:SMfit}), the
posterior probability distribution (\emph{Posterior} in
table~\ref{tab:SMfit}), and the indirect prediction or predictive
posterior probability distribution (\emph{Prediction} in
table~\ref{tab:SMfit}) for the five floating input parameters. These
plots show at a glance the impact of the precision of each input
parameter on the fit, as well as the agreement between the values
preferred by the fit and the direct determinations.

\begin{figure}
  \centering
  \vspace{-1mm}
  \includegraphics[width=.49\textwidth]{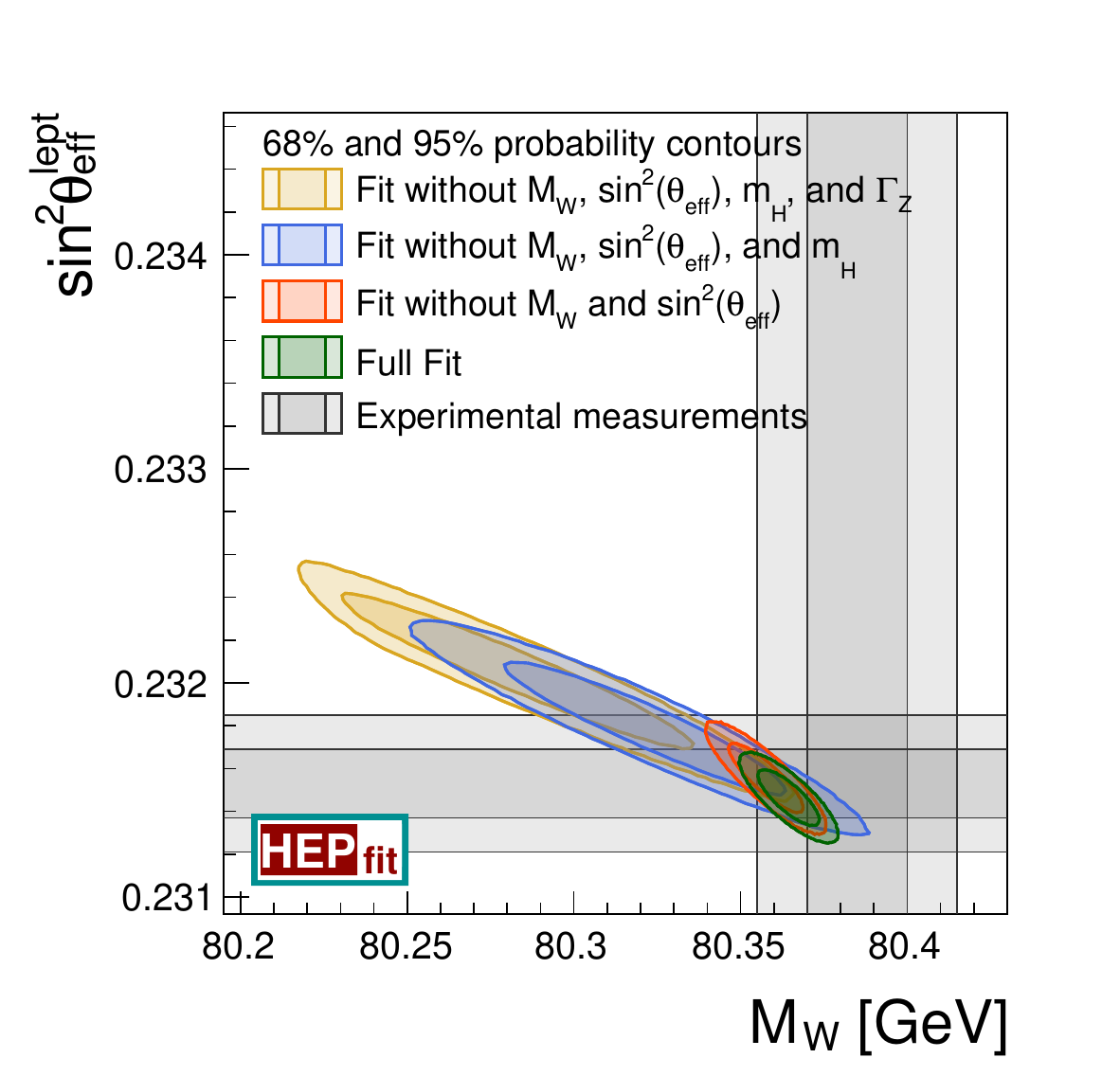} 
  \hspace{-3mm}
  \includegraphics[width=.49\textwidth]{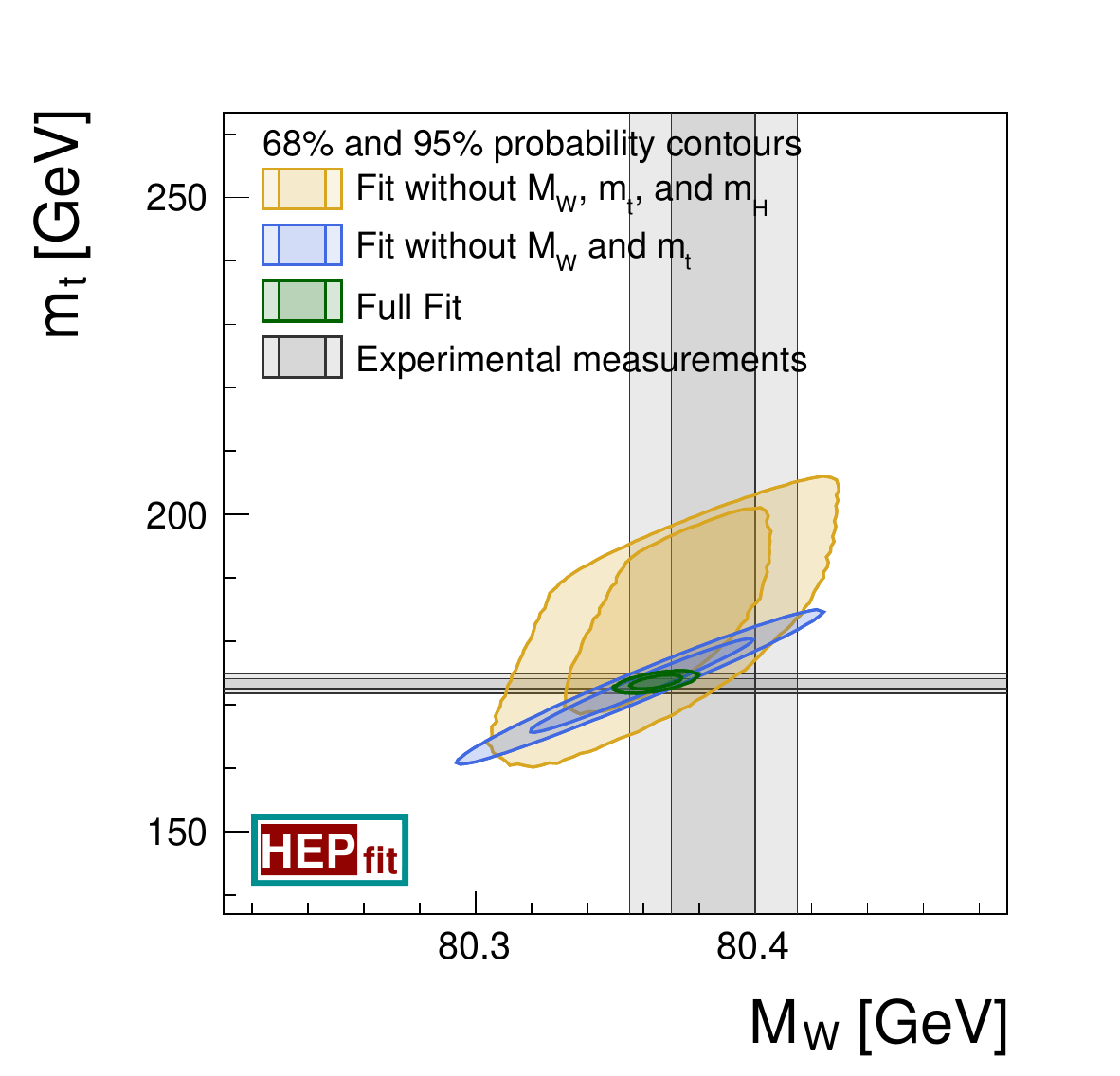} 
  \caption{Left: Comparison of the indirect constraints on $\sin^2{\theta_\mathrm{eff}^\mathrm{lept}}$
   and $M_W$ with the direct experimental measurements. Dark (light)
    regions correspond to $68\%$ ($95\%$) probability. Right: The same for $m_t$ and $M_W$.}
  \label{fig:mw-mt-sin2eff}
\end{figure}

Two of the most important observables in the SM fit are the effective
mixing angle, $\sin^2{\theta_\mathrm{eff}^\mathrm{lept}}$, and the $W$
mass, $M_W$.  In figure~\ref{fig:mw-mt-sin2eff} we show the
consistency of the predictions for these observables with the direct
experimental measurements, their dependence on the top mass, and the
impact of other measurements, such as $m_H$ (varied in the range
$10~\mathrm{GeV} < m_H < 1~\mathrm{TeV}$) and $\Gamma_Z$.

Looking at the pulls in table~\ref{tab:SMfit}, one can notice that there is an overall
agreement between EWPO and SM predictions. Only $A_{\mathrm{FB}}^{0,b}$ shows
some tension between existing measurements and the result of the SM
precision fit. Care must be taken when interpreting this as a possible
hint of NP, for deviations at this level ($\sim 2 \sigma$) are likely to
occur when fitting this many observables. Having this in mind, this anomaly
will be taken into account in exploring possible
parameterizations of NP effects in section~\ref{sec:ew-precision-fit-BSM}.

\begin{table}[tp]
\centering
\begin{tabular}{lcllll} 
\toprule
& Prediction &
\quad\ $\alpha_s$ & 
\ $\Delta\alpha_{\rm had}^{(5)}$ & 
\quad $M_Z$ & 
\quad $m_t$ 
\\
\cmrule 
$M_W$ [GeV] & $ 80.3618 \pm 0.0080 $
& $\pm 0.0008 $ % _AlsMz
& $\pm 0.0060 $ % _dAle5Mz
& $\pm 0.0026 $ % _Mz
& $\pm 0.0046 $ % _mtop
\\ 
$\Gamma_{W}$ [GeV] & $ 2.08849 \pm 0.00079 $
& $\pm 0.00048 $ % _AlsMz
& $\pm 0.00047 $ % _dAle5Mz
& $\pm 0.00021 $ % _Mz
& $\pm 0.00036 $ % _mtop
\\ 
$\Gamma_{Z}$ [GeV] & $ 2.49403 \pm 0.00073 $
& $\pm 0.00059 $ % _AlsMz
& $\pm 0.00031 $ % _dAle5Mz
& $\pm 0.00021 $ % _Mz
& $\pm 0.00017 $ % _mtop
\\ 
$\sigma_{h}^{0}$ [nb] & $ 41.4910 \pm 0.0062 $
& $\pm 0.0059 $ % _AlsMz
& $\pm 0.0005 $ % _dAle5Mz
& $\pm 0.0020 $ % _Mz
& $\pm 0.0005 $ % _mtop
\\ 
$\sin^2\theta_{\rm eff}^{\rm lept}$ & $ 0.23148 \pm 0.00012 $
& $\pm 0.00000 $ % _AlsMz
& $\pm 0.00012 $ % _dAle5Mz
& $\pm 0.00002 $ % _Mz
& $\pm 0.00002 $ % _mtop
\\ 
$P_{\tau}^{\rm pol}=\mathcal{A}_\ell$ & $ 0.14731 \pm 0.00093 $
& $\pm 0.00003 $ % _AlsMz
& $\pm 0.00091 $ % _dAle5Mz
& $\pm 0.00012 $ % _Mz
& $\pm 0.00019 $ % _mtop
\\ 
$\mathcal{A}_c$ & $ 0.66802 \pm 0.00041 $
& $\pm 0.00001 $ % _AlsMz
& $\pm 0.00040 $ % _dAle5Mz
& $\pm 0.00005 $ % _Mz
& $\pm 0.00008 $ % _mtop
\\ 
$\mathcal{A}_b$ & $ 0.934643 \pm 0.000076 $
& $\pm 0.000003 $ % _AlsMz
& $\pm 0.000075 $ % _dAle5Mz
& $\pm 0.000010 $ % _Mz
& $\pm 0.000005 $ % _mtop
\\ 
$A_{\rm FB}^{0, \ell}$ & $ 0.01627 \pm 0.00021 $
& $\pm 0.00001 $ % _AlsMz
& $\pm 0.00020 $ % _dAle5Mz
& $\pm 0.00003 $ % _Mz
& $\pm 0.00004 $ % _mtop
\\ 
$A_{\rm FB}^{0, c}$ & $ 0.07381 \pm 0.00052 $
& $\pm 0.00002 $ % _AlsMz
& $\pm 0.00050 $ % _dAle5Mz
& $\pm 0.00007 $ % _Mz
& $\pm 0.00010 $ % _mtop
\\ 
$A_{\rm FB}^{0, b}$ & $ 0.10326 \pm 0.00067 $
& $\pm 0.00002 $ % _AlsMz
& $\pm 0.00065 $ % _dAle5Mz
& $\pm 0.00008 $ % _Mz
& $\pm 0.00013 $ % _mtop
\\ 
$R^{0}_{\ell}$ & $ 20.7478 \pm 0.0077 $
& $\pm 0.0074 $ % _AlsMz
& $\pm 0.0020 $ % _dAle5Mz
& $\pm 0.0003 $ % _Mz
& $\pm 0.0003 $ % _mtop
\\ 
$R^{0}_{c}$ & $ 0.172222 \pm 0.000026 $
& $\pm 0.000023 $ % _AlsMz
& $\pm 0.000007 $ % _dAle5Mz
& $\pm 0.000001 $ % _Mz
& $\pm 0.000009 $ % _mtop
\\ 
$R^{0}_{b}$ & $ 0.215800 \pm 0.000030 $
& $\pm 0.000013 $ % _AlsMz
& $\pm 0.000004 $ % _dAle5Mz
& $\pm 0.000000 $ % _Mz
& $\pm 0.000026 $ % _mtop
\\ 
\bottomrule
\end{tabular}
\caption{SM predictions computed using the theoretical expressions for
  the EWPO without the corresponding experimental constraints, and 
  individual uncertainties associated with each
  input parameter, except for $m_H$ (see text).} 
\label{tab:SMpred}
\end{table}

In table~\ref{tab:SMpred} we present the full
predictions for all EWPO (computed using the theoretical expressions
used in the fit without the experimental constraints on the
observables) with the breakdown of the parametric uncertainty induced
by $1\sigma$ variations of the input parameters. We do not include
in that table the corresponding column for $m_H$,
since its leading contributions to the EWPO are logarithmic, and hence
its error
does not induce a significant uncertainty in the predictions.  In
several cases, the largest contribution to the parametric errors comes
from the uncertainty in $\Delta \alpha_{\mathrm{had}}^{(5)}(M_Z)$.
This is the dominant source for
$\sin^2{\theta_{\mathrm{eff}}^{\mathrm{lept}}}$ and hence for the
different asymmetries. The uncertainties of $M_W$ and the
pseudo-observables involving decay widths, on the other hand, receive
sizeable contributions from several or all input parameters.  In
particular, with the new PDG value, $\alpha_s(M_Z)$ becomes the
dominant source of uncertainty in all observables involving the
hadronic decay width, with the exception of $R_b^0$, whose error is
controlled by that of $m_t$.

For the sake of comparison, we repeat the fit using the old PDG determination of $\alpha_s(M_Z)$
and report the results in tables~\ref{tab:SMfitOld} and \ref{tab:SMpredOld}. The effect is particularly
visible in all observables involving the hadronic decay width.

\begin{table}[t]
{\footnotesize
\setlength\tabcolsep{5pt}
\begin{center}
\begin{tabular}{lcccccc}
\toprule
& Ref & Measurement  & Posterior & Prediction &\!1D Pull\! &\!nD Pull\! \\
\cmrule  
$\alpha_s(M_Z)$ & \cite{Agashe:2014kda} & $ 0.11850 \pm 0.00050 $ & $ 0.11850 \pm 0.00049 $  & $ 0.1186 \pm 0.0028 $ & 0.1  \\ 
$\Delta\alpha_{\mathrm{had}}^{(5)}(M_Z)$ & \cite{Burkhardt:2011ur} & $ 0.02750 \pm 0.00033 $ & $ 0.02747 \pm 0.00025 $  & $ 0.02743 \pm 0.00038 $ & -0.2  \\ 
\rowcolor[gray]{.85}$M_Z$ [GeV]& \cite{ALEPH:2005ab}& $ 91.1875 \pm 0.0021 $ & $ 91.1879 \pm 0.0021 $  & $ 91.198 \pm 0.011 $ & -0.9 & \\ 
$m_t$ [GeV]& \cite{ATLAS:2014wva} & $ 173.34 \pm 0.76 $ & $ 173.61 \pm 0.73 $  & $ 176.7 \pm 2.5 $ & 1.1  \\ 
$m_H$ [GeV]& \cite{Aad:2015zhl} & $ 125.09 \pm 0.24 $ & $ 125.09 \pm 0.24 $  & $ 102.4 \pm 26.4 $ & -0.6  \\ 
\cmrule  
$M_W$ [GeV]& \cite{Group:2012gb} & $ 80.385 \pm 0.015 $ & $ 80.3641 \pm 0.0060 $  & $ 80.3601 \pm 0.0066 $ & -1.7  \\ 
\cmrule  
$\Gamma_{W}$ [GeV]& \cite{ALEPH:2010aa} & $ 2.085 \pm 0.042 $ & $ 2.08893 \pm 0.00051 $  & $ 2.08893 \pm 0.00051 $ & 0.0  \\ 
\cmrule  
$\sin^2\theta_{\rm eff}^{\rm lept}(Q_{\rm FB}^{\rm had})$\!\!\!\!\! & \cite{ALEPH:2005ab} & $ 0.2324 \pm 0.0012 $ & $ 0.231466 \pm 0.000086 $  & $ 0.231437 \pm 0.000090 $ & -0.8  \\ 
\cmrule  
$P_{\tau}^{\rm pol}=\mathcal{A}_\ell$& \cite{ALEPH:2005ab} & $ 0.1465 \pm 0.0033 $ & $ 0.14746 \pm 0.00068 $  & $ 0.14751 \pm 0.00069 $ & 0.1  \\ 
\cmrule  
\rowcolor[gray]{.85}$\Gamma_{Z}$ [GeV]& \cite{ALEPH:2005ab} & $ 2.4952 \pm 0.0023 $ & $2.49445 \pm 0.00040 $  & $ 2.49439 \pm 0.00041 $ & 0.4 &   \\ 
\rowcolor[gray]{.85}$\sigma_{h}^{0}$ [nb]& \cite{ALEPH:2005ab} & $ 41.540 \pm 0.037 $ & $ 41.4878 \pm 0.0031 $  & $ 41.4880 \pm 0.0032 $ & 1.3&  0.7 \\ 
\rowcolor[gray]{.85}$R^{0}_{\ell}$& \cite{ALEPH:2005ab} & $ 20.767 \pm 0.025 $ & $ 20.7516 \pm 0.0034 $  & $ 20.7513 \pm 0.0035 $ & 0.6&   \\ 
\rowcolor[gray]{.85}$A_{\rm FB}^{0, \ell}$& \cite{ALEPH:2005ab} & $ 0.0171 \pm 0.0010 $ & $ 0.01631 \pm 0.00015 $  & $ 0.01627 \pm 0.00015 $ & 0.9&   \\ 
\cmrule 
\rowcolor[gray]{.75}$\mathcal{A}_{\ell}$ (SLD)& \cite{ALEPH:2005ab} & $ 0.1513 \pm 0.0021 $& $ 0.14746 \pm 0.00068 $  & $ 0.14762 \pm 0.00076 $ & 1.7 &  \\ 
\rowcolor[gray]{.75}$\mathcal{A}_c$& \cite{ALEPH:2005ab} & $ 0.670 \pm 0.027 $ & $ 0.66809 \pm 0.00030 $  & $ 0.66816 \pm 0.00033 $ & 0.03&   \\ 
\rowcolor[gray]{.75}$\mathcal{A}_b$& \cite{ALEPH:2005ab} & $ 0.923 \pm 0.020 $ & $ 0.934648 \pm 0.000058 $  & $ 0.934661 \pm 0.000064 $ & -0.4&   \\ 
\rowcolor[gray]{.75}$A_{\rm FB}^{0, c}$& \cite{ALEPH:2005ab} & $ 0.0707 \pm 0.0035 $ & $ 0.07389 \pm 0.00037 $  & $ 0.07398 \pm 0.00042 $ & -0.9&  1.5   \\ 
\rowcolor[gray]{.75}$A_{\rm FB}^{0, b}$& \cite{ALEPH:2005ab} & $ 0.0992 \pm 0.0016 $ & $ 0.10337 \pm 0.00048 $  & $ 0.10348 \pm 0.00054 $ & -2.5&   \\ 
\rowcolor[gray]{.75}$R^{0}_{c}$& \cite{ALEPH:2005ab} & $ 0.1721 \pm 0.0030 $ & $ 0.172238 \pm 0.000013 $  & $ 0.172239 \pm 0.000013 $ & -0.1&   \\ 
\rowcolor[gray]{.75}$R^{0}_{b}$& \cite{ALEPH:2005ab} & $ 0.21629 \pm 0.00066 $ & $ 0.215784 \pm 0.000025 $  & $ 0.215783 \pm 0.000026 $ & 0.8&   \\ 
\cmrule 
$\sin^2\theta_{\rm eff}^{ee}$& \cite{Aaltonen:2016nuy} & $ 0.23248 \pm 0.00053 $ & \multirow{6}{*}{$0.231466 \pm 0.000086$}  & \multirow{6}{*}{$0.231437 \pm 0.000090$} &  2.1 \\ 
$\sin^2\theta_{\rm eff}^{\mu\mu}$& \cite{Aaltonen:2014loa}& $ 0.2315 \pm 0.0010 $ &   & & 0.1 \\ 
$\sin^2\theta_{\rm eff}^{ee}$& \cite{Abazov:2014jti} & $ 0.23146 \pm 0.00047 $ &  & & 0.2 \\ 
$\sin^2\theta_{\rm eff}^{ee,\mu\mu}$& \cite{Aad:2015uau} & $ 0.2308 \pm 0.0012 $ &  & & -0.5\\ 
$\sin^2\theta_{\rm eff}^{\mu\mu}$& \cite{Chatrchyan:2011ya} & $ 0.2287 \pm 0.0032 $  & & & -0.8\\  
$\sin^2\theta_{\rm eff}^{\mu\mu}$& \cite{Aaij:2015lka} & $ 0.2314 \pm 0.0011 $ &  & & -0.3\\ 
\bottomrule 
\end{tabular}
\end{center}
}
\caption{
  Same as table~\ref{tab:SMfit} using the old PDG determination of $\alpha_s(M_Z)$.} 
\label{tab:SMfitOld}
\end{table}

\begin{table}[!btp]
\centering
\begin{tabular}{lcllll} 
\toprule
& Prediction &
\quad\ $\alpha_s$ & 
\ $\Delta\alpha_{\rm had}^{(5)}$ & 
\quad $M_Z$ & 
\quad $m_t$ 
\\
\cmrule
$M_W$ [GeV] & $ 80.3615 \pm 0.0080 $
& $\pm 0.0003 $ % _AlsMz
& $\pm 0.0060 $ % _dAle5Mz
& $\pm 0.0027 $ % _Mz
& $\pm 0.0046 $ % _mtop
\\ 
$\Gamma_{W}$ [GeV] & $ 2.08872 \pm 0.00066 $
& $\pm 0.00020 $ % _AlsMz
& $\pm 0.00047 $ % _dAle5Mz
& $\pm 0.00021 $ % _Mz
& $\pm 0.00036 $ % _mtop
\\ 
$\Gamma_{Z}$ [GeV] & $ 2.49433 \pm 0.00049 $
& $\pm 0.00025 $ % _AlsMz
& $\pm 0.00031 $ % _dAle5Mz
& $\pm 0.00021 $ % _Mz
& $\pm 0.00017 $ % _mtop
\\ 
$\sigma_{h}^{0}$ [nb] & $ 41.4881 \pm 0.0032 $
& $\pm 0.0024 $ % _AlsMz
& $\pm 0.0005 $ % _dAle5Mz
& $\pm 0.0020 $ % _Mz
& $\pm 0.0005 $ % _mtop
\\ 
$\sin^2\theta_{\rm eff}^{\rm lept}$ & $ 0.23149 \pm 0.00012 $
& $\pm 0.00000 $ % _AlsMz
& $\pm 0.00012 $ % _dAle5Mz
& $\pm 0.00002 $ % _Mz
& $\pm 0.00002 $ % _mtop
\\ 
$P_{\tau}^{\rm pol}=\mathcal{A}_\ell$ & $ 0.14730 \pm 0.00094 $
& $\pm 0.00001 $ % _AlsMz
& $\pm 0.00091 $ % _dAle5Mz
& $\pm 0.00012 $ % _Mz
& $\pm 0.00019 $ % _mtop
\\ 
$\mathcal{A}_c$ & $ 0.66802 \pm 0.00041 $
& $\pm 0.00001 $ % _AlsMz
& $\pm 0.00040 $ % _dAle5Mz
& $\pm 0.00005 $ % _Mz
& $\pm 0.00008 $ % _mtop
\\ 
$\mathcal{A}_b$ & $ 0.934642 \pm 0.000076 $
& $\pm 0.000001 $ % _AlsMz
& $\pm 0.000075 $ % _dAle5Mz
& $\pm 0.000010 $ % _Mz
& $\pm 0.000005 $ % _mtop
\\ 
$A_{\rm FB}^{0, \ell}$ & $ 0.01627 \pm 0.00021 $
& $\pm 0.00000 $ % _AlsMz
& $\pm 0.00020 $ % _dAle5Mz
& $\pm 0.00003 $ % _Mz
& $\pm 0.00004 $ % _mtop
\\ 
$A_{\rm FB}^{0, c}$ & $ 0.07380 \pm 0.00052 $
& $\pm 0.00001 $ % _AlsMz
& $\pm 0.00050 $ % _dAle5Mz
& $\pm 0.00007 $ % _Mz
& $\pm 0.00010 $ % _mtop
\\ 
$A_{\rm FB}^{0, b}$ & $ 0.10325 \pm 0.00067 $
& $\pm 0.00001 $ % _AlsMz
& $\pm 0.00065 $ % _dAle5Mz
& $\pm 0.00008 $ % _Mz
& $\pm 0.00013 $ % _mtop
\\ 
$R^{0}_{\ell}$ & $ 20.7515 \pm 0.0037 $
& $\pm 0.0031 $ % _AlsMz
& $\pm 0.0020 $ % _dAle5Mz
& $\pm 0.0003 $ % _Mz
& $\pm 0.0003 $ % _mtop
\\ 
$R^{0}_{c}$ & $ 0.172234 \pm 0.000015 $
& $\pm 0.000010 $ % _AlsMz
& $\pm 0.000007 $ % _dAle5Mz
& $\pm 0.000001 $ % _Mz
& $\pm 0.000009 $ % _mtop
\\ 
$R^{0}_{b}$ & $ 0.215794 \pm 0.000027 $
& $\pm 0.000006 $ % _AlsMz
& $\pm 0.000004 $ % _dAle5Mz
& $\pm 0.000000 $ % _Mz
& $\pm 0.000026 $ % _mtop
\\ 
\bottomrule
\end{tabular}
\caption{Same as table~\ref{tab:SMpred} using  the old PDG determination of $\alpha_s(M_Z)$.} 
\label{tab:SMpredOld}
\end{table}
%------------------------------------------------------------------------------------------------------------

\FloatBarrier

\section{Electroweak precision fit beyond the Standard Model}
\label{sec:ew-precision-fit-BSM}

We now generalize the SM fit considering different sets of parameters
which account for NP contributions in several extensions of
the SM.

%------------------------------------------------------------------------------------------------------------

\subsection{Non-standard oblique corrections}
\label{subsec:ew-precision-fit-BSM-oblique}

In this section, we use the fit of EWPO to constrain the oblique
parameters $S$, $T$, $U$ introduced in
ref.~\cite{Peskin:1990zt,Peskin:1991sw} and the
$\varepsilon_{1,2,3,b}$ parameters introduced in
refs.~\cite{Altarelli:1990zd,Altarelli:1991fk,Altarelli:1993sz}.

The $S$, $T$, $U$ parameters account for NP effects in the
vacuum-polarization amplitudes of the EW gauge bosons and modify all
EWPO considered here. The explicit dependence of the EWPO on $S$, $T$,
and $U$ can be found in appendix A of ref.~\cite{Ciuchini:2013pca}
where it was also noticed how
the EWPO considered here depend only on the following three specific
combinations of the $S$, $T$, and $U$ parameters (where
$s_W=\sin\theta_W$ and $c_W=\cos\theta_W$),
\begin{eqnarray}
  \label{eq:abc}
  A &=& S - 2c_W^2\, T - \frac{(c_W^2-s_W^2)\,U}{2s_W^2}\,,
  \nonumber \\
  B &=& S - 4c_W^2 s_W^2\, T\,,\\
  C &=& -10(3-8s_W^2)\,S + (63-126s_W^2-40s_W^4)\,T\,. \nonumber
\end{eqnarray}

\begin{figure}[t]
\begin{center}
  \begin{tabular}{c c}
  \includegraphics[width=.45\textwidth]{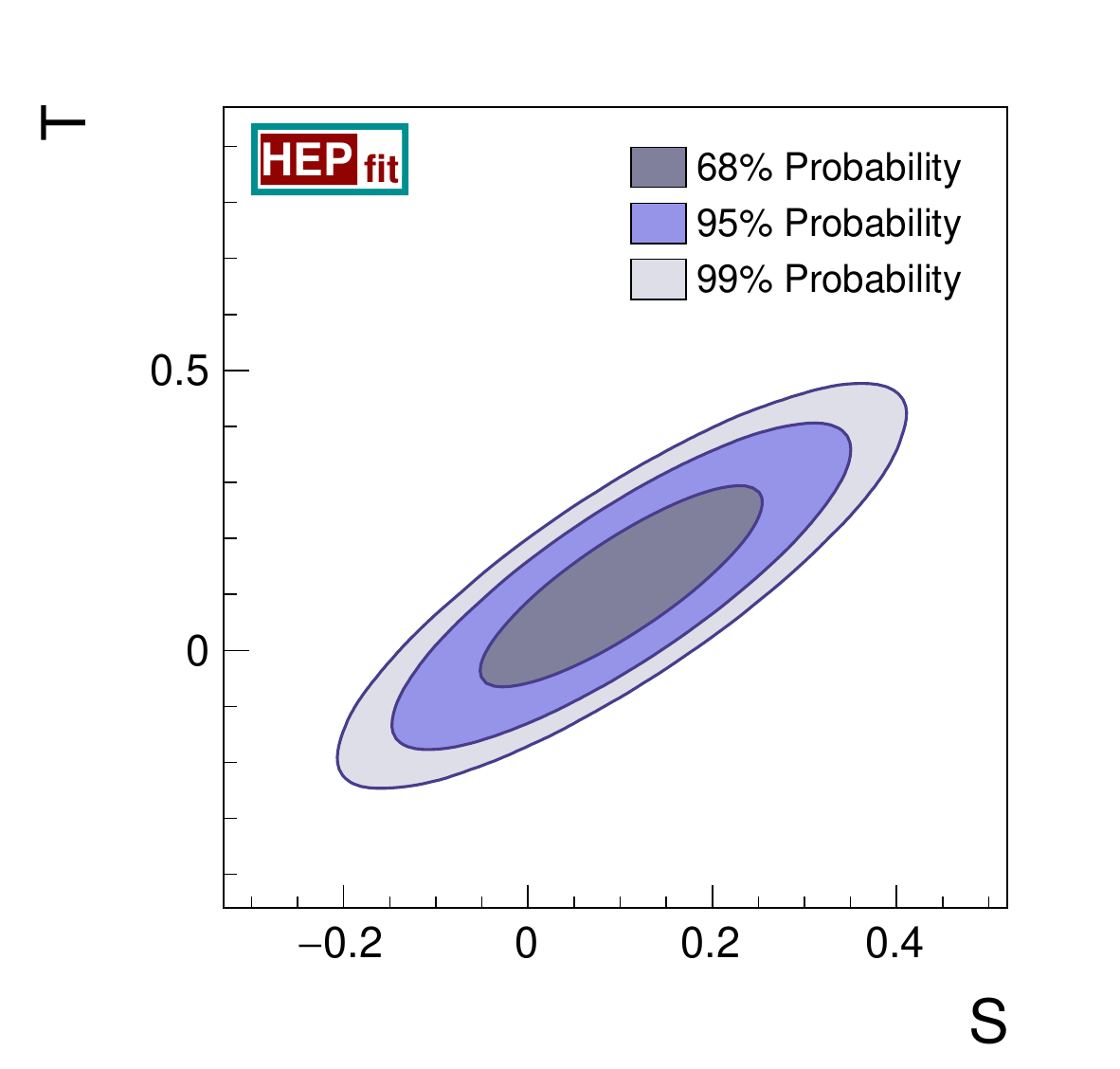} %<----- S vs T
&
  \includegraphics[width=.45\textwidth]{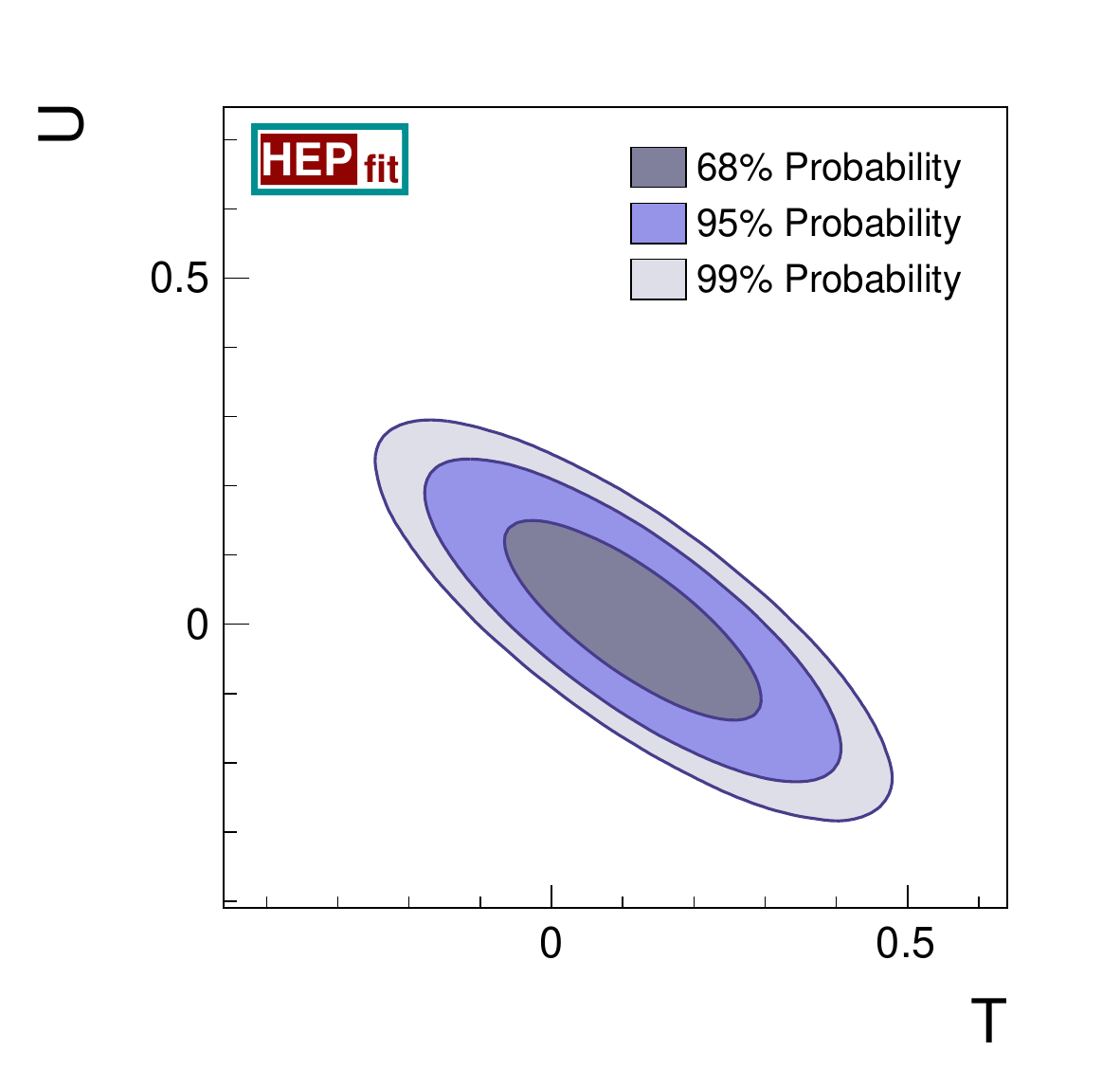}\\ %<----- T vs U
  \includegraphics[width=.45\textwidth]{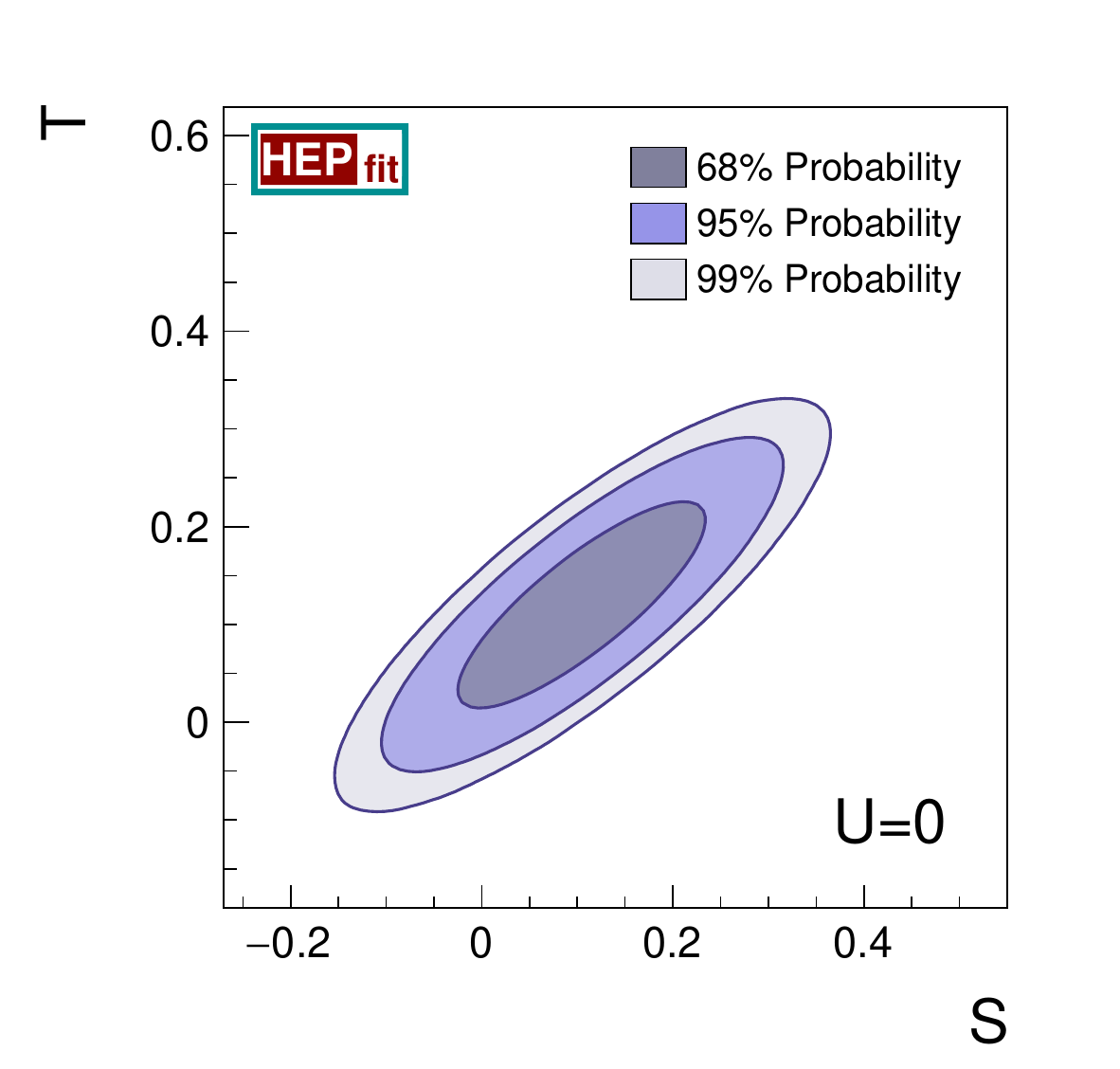} %<----- S vs T (U=0)
&
  \includegraphics[width=.45\textwidth]{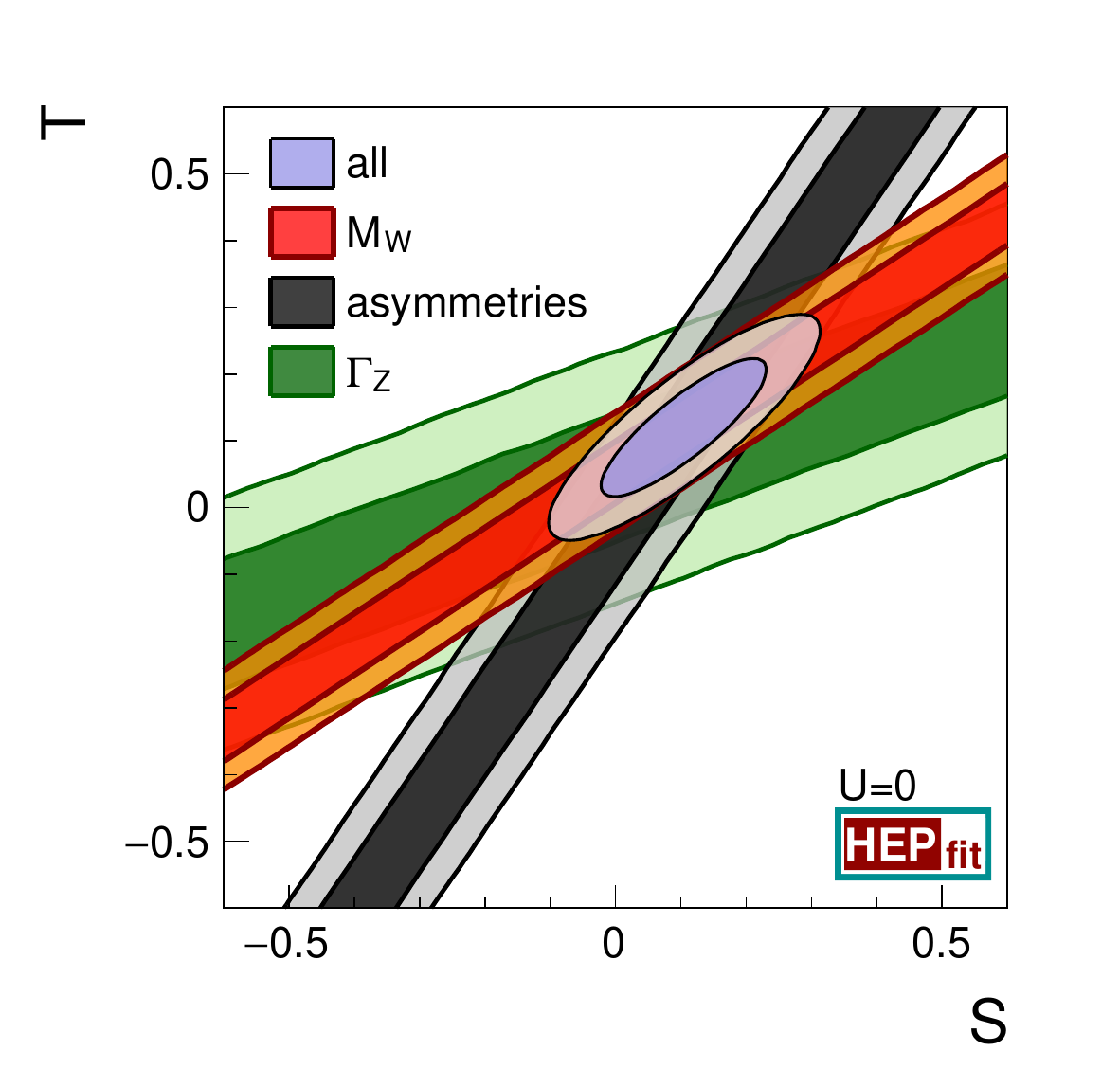} %<----- S vs T (U=0) individual constraints
  \end{tabular}
  \caption{Two-dimensional probability distributions for the oblique
    parameters $S$ and $T$ (upper-left panel), and $T$ and $U$
    (upper-right panel).  From darker to lighter the different regions
    correspond respectively to $68\%$, $95\%$, and $99\%$ probability.  In the lower
    panel we show the two-dimensional distributions for $S$ and $T$
    fixing $U=0$, together with the individual constraints from $M_W$,
    the asymmetry parameters $\sin^2\theta_{\rm eff}^{\rm lept}$,
    $P_\tau^{\rm pol}$, $\mathcal{A}_{f}$, and $A_{\rm FB}^{0,f}$ with
    $f=\ell,c,b$, and $\Gamma_Z$.  In this last plot the dark (light)
    region corresponds to $68\%$ ($95\%$) probability.}
  \label{fig:Oblique}
  \end{center}
\end{figure}

Therefore the extracted values of $S$, $T$, and $U$ are
correlated. For this reason, we give in tables~\ref{tab:STUfit} and
\ref{tab:STfit} the results of the fit together with the correlation
matrix. We also remind the reader that $A$, the only parameter
depending on $U$, describes NP contributions to $M_W$ and $\Gamma_W$,
the parameter $C$ describes NP contributions to $\Gamma_Z$, and NP
contributions to all other EWPO are proportional to $B$.  As
illustrated in figure~\ref{fig:Oblique}, $S$, $T$, and $U$ are
compatible with zero, implying the absence of sizeable oblique
corrections beyond those predicted by the SM. In general,
  in models of new physics with linearly realized electroweak symmetry
  breaking $U$ is largely suppressed relative to $S$ and $T$. For this
  reason we also specify our fit to the case in which $U=0$ and give
  the corresponding results in table~\ref{tab:STfit}, and in the
  bottom plots of figure~\ref{fig:Oblique}. The results of
  table~\ref{tab:STUgLR} later in this section are are also given in
  the $U=0$ scenario.

\begin{table}[h]
\parbox{.45\linewidth}{
\centering
\begin{tabular}{c|c|rrr}
 \hline
 & Result & \multicolumn{3}{c}{Correlation Matrix} \\ 
 \hline 
$S$ & $ 0.09 \pm 0.10 $ & $1.00$ \\ 
$T$ & $ 0.10 \pm 0.12 $ & $0.86$ & $1.00$ \\ 
$U$ & $ 0.01 \pm 0.09 $ & $-0.54$ & $-0.81$ & $1.00$ \\ \hline
 \end{tabular}
\caption{Results of the fit for the oblique parameters $S$, $T$, and $U$.}
\label{tab:STUfit}
}\hfill
\parbox{.45\linewidth}{
\centering
\begin{tabular}{c|c|rr}
 \hline
 & Result & \multicolumn{2}{c}{Correlation Matrix} \\ 
 \hline 
$S$ & $ 0.10 \pm 0.08 $ & $1.00$ \\ 
$T$ & $ 0.12 \pm 0.07 $ & $0.86$ & $1.00$ \\ 
\hline
 \end{tabular}
\caption{Results of the fit for the oblique parameters $S$ and $T$, taking $U=0$.}
\label{tab:STfit}
}
\end{table}

Next we consider the $\varepsilon_{1,2,3,b}$ parameters introduced
in refs.~\cite{Altarelli:1990zd,Altarelli:1991fk,Altarelli:1993sz}.
Unlike the $S$, $T$, and $U$ parameters discussed above, the $\varepsilon_i$
parameters involve SM contributions associated with the top quark and
the Higgs boson, SM flavour non-universal vertex corrections, and
further vacuum-polarization corrections~\cite{Barbieri:2004qk}. Since
the SM is now fully known and there is no need to disentangle top-quark and
Higgs-boson contributions anymore, we separate the genuine NP contribution
from the SM one by introducing
$\delta\varepsilon_i=\varepsilon_i - \varepsilon_{i,\mathrm{SM}}$ for
$i=1,2,3,b$, where $\varepsilon_i$ are the original parameters and
$\varepsilon_{i,\mathrm{SM}}$ contain the SM contribution only.  The
expressions of the EWPO in terms of $\delta\varepsilon_i$ can be found
in ref.~\cite{Ciuchini:2013pca,Ciuchini:2014dea}.

\begin{table}[b]
\centering
\begin{tabular}{c|c|rrrr}
 \hline
 & Result & \multicolumn{4}{c}{Correlation Matrix} \\ 
 \hline 
$\delta \varepsilon_{1}$ & $ ~0.0007 \pm 0.0010 $ & $1.00$ \\ 
$\delta \varepsilon_{2}$ & $ -0.0002 \pm 0.0008 $ & $0.82$ & $1.00$ \\ 
$\delta \varepsilon_{3}$ & $ ~0.0007 \pm 0.0009 $ & $0.87$ & $0.56$ & $1.00$ \\ 
$\delta \varepsilon_{b}$ & $ ~0.0004 \pm 0.0013 $ & $-0.34$ & $-0.32$ & $-0.24$ & $1.00$ \\ 
\hline
 \end{tabular}
\caption{Results of the fit for the $\delta\varepsilon_i$ parameters ($i=1,2,3,b$).}
\label{tab:4deps}
\end{table}

\begin{figure}[t]
  \centering
  \vspace{-1mm}
  \includegraphics[width=.45\textwidth]{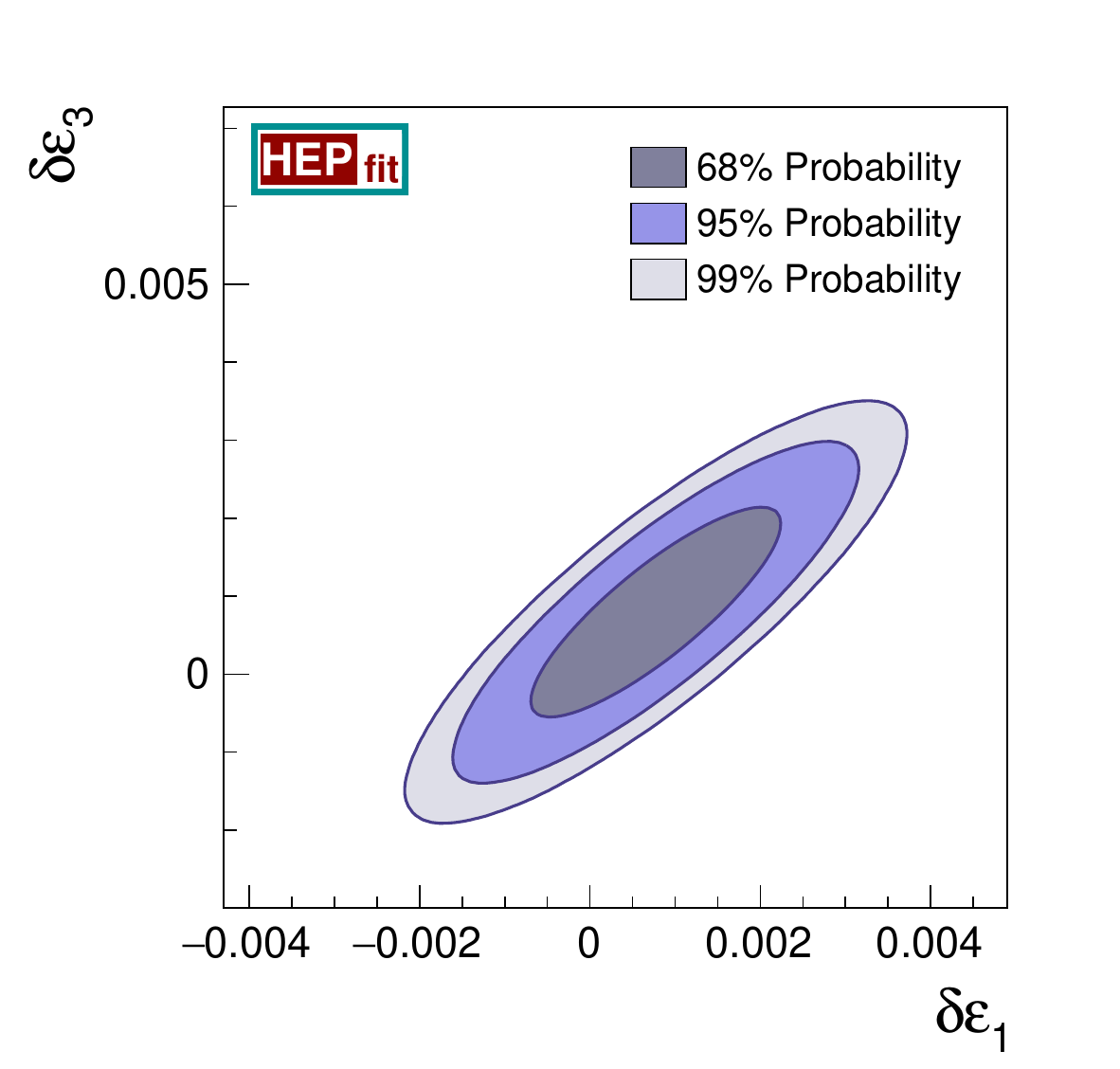}
  \hspace{-3mm}
  \includegraphics[width=.45\textwidth]{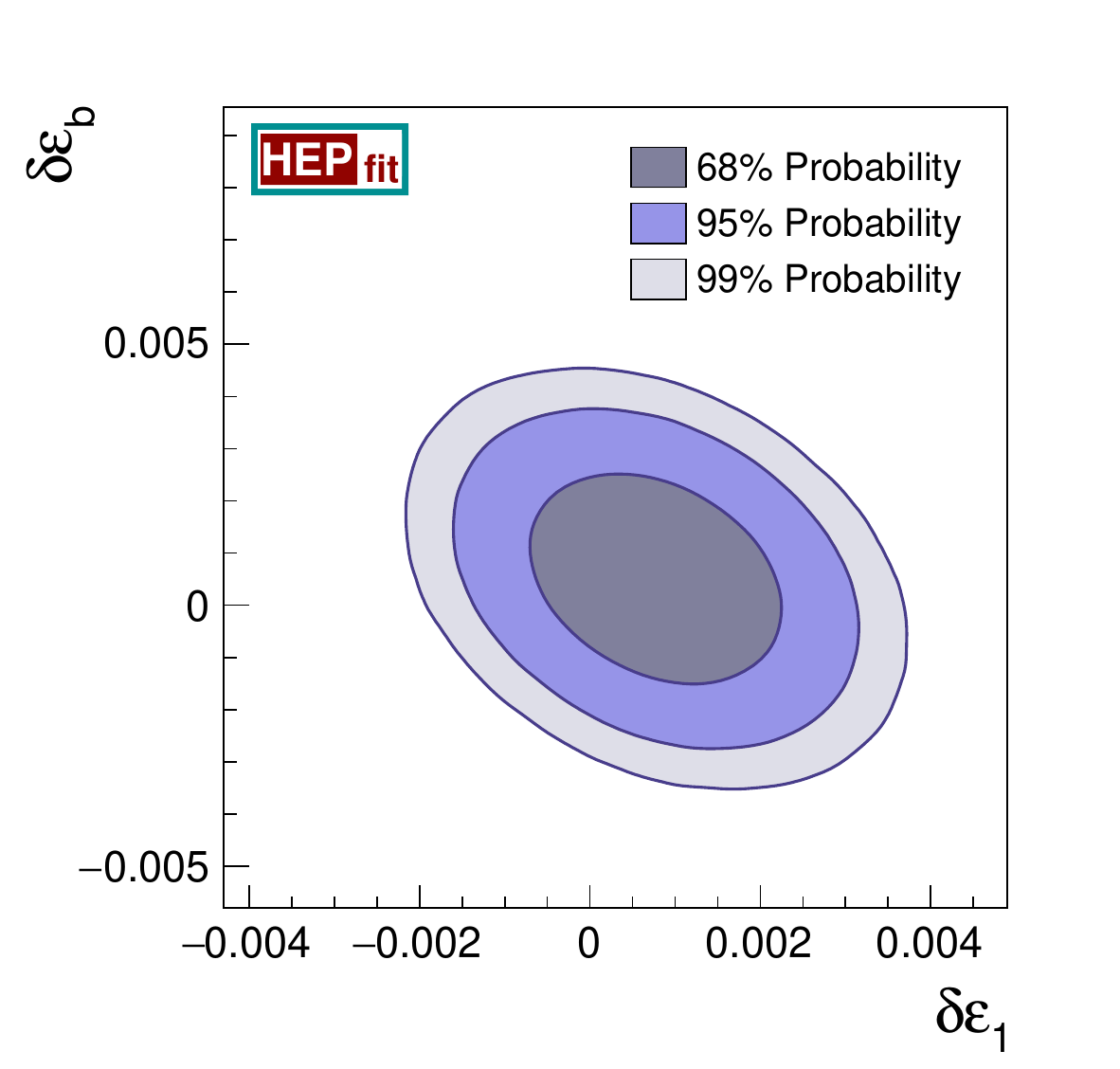}
  \vspace{-2mm}
  \caption{Two-dimensional probability distributions for
    $\delta\varepsilon_1$ and $\delta\varepsilon_3$ (left), and
    $\delta\varepsilon_1$ and $\delta\varepsilon_b$ (right) varying all
    $\delta\varepsilon_i$ parameters.  From darker to
    lighter the different regions correspond to $68\%$, $95\%$, and $99\%$
    probability.\label{fig:Epsilons}}
\end{figure}

The results of our fit for the $\delta\varepsilon_i$ parameters are
summarized in table~\ref{tab:4deps}. Some two-dimensional probability
distributions are plotted in figure~\ref{fig:Epsilons}.  All results
are consistent with the SM. Note that, as mentioned above, the
$\delta \varepsilon_i$ parameters include oblique corrections beyond
those connected to the $S,~T$, and $U$ parameters. More precisely,
\begin{eqnarray}
\delta \varepsilon_1&=&\alpha T-W + 2 X \frac{\sin{\theta_W}}{\cos{\theta_W}}-Y \frac{\sin^2{\theta_W}}{\cos^2{\theta_W}},\\
\delta \varepsilon_2&=&-\frac{\alpha}{4 \sin^2{\theta_W}}U-W + 2 X \frac{\sin{\theta_W}}{\cos{\theta_W}}-V,\\
\delta \varepsilon_3&=&\frac{\alpha}{4 \sin^2{\theta_W}}S-W +  \frac{X}{\sin{\theta_W}\cos{\theta_W}}-Y,
\end{eqnarray}
where $V,~W,~X,~Y$ are part of the extended set of oblique parameters defined in~\cite{Barbieri:2004qk}. With
the results in table~\ref{tab:4deps} and the above equations, one can therefore obtain approximate constraints
on NP scenarios with vanishing contributions to $S,~T$, and/or $U$ but non-zero values of some of the other
parameters ($V$, $W$, $X$, and $Y$).

%------------------------------------------------------------------------------------------------------------

\subsection{Modified $Zb\bar b$ couplings}
\label{subsec:ew-precision-fit-BSM-Zbb}

\begin{figure}[h]
  \centering
  \vspace{-1mm}
  \includegraphics[width=.49\textwidth]{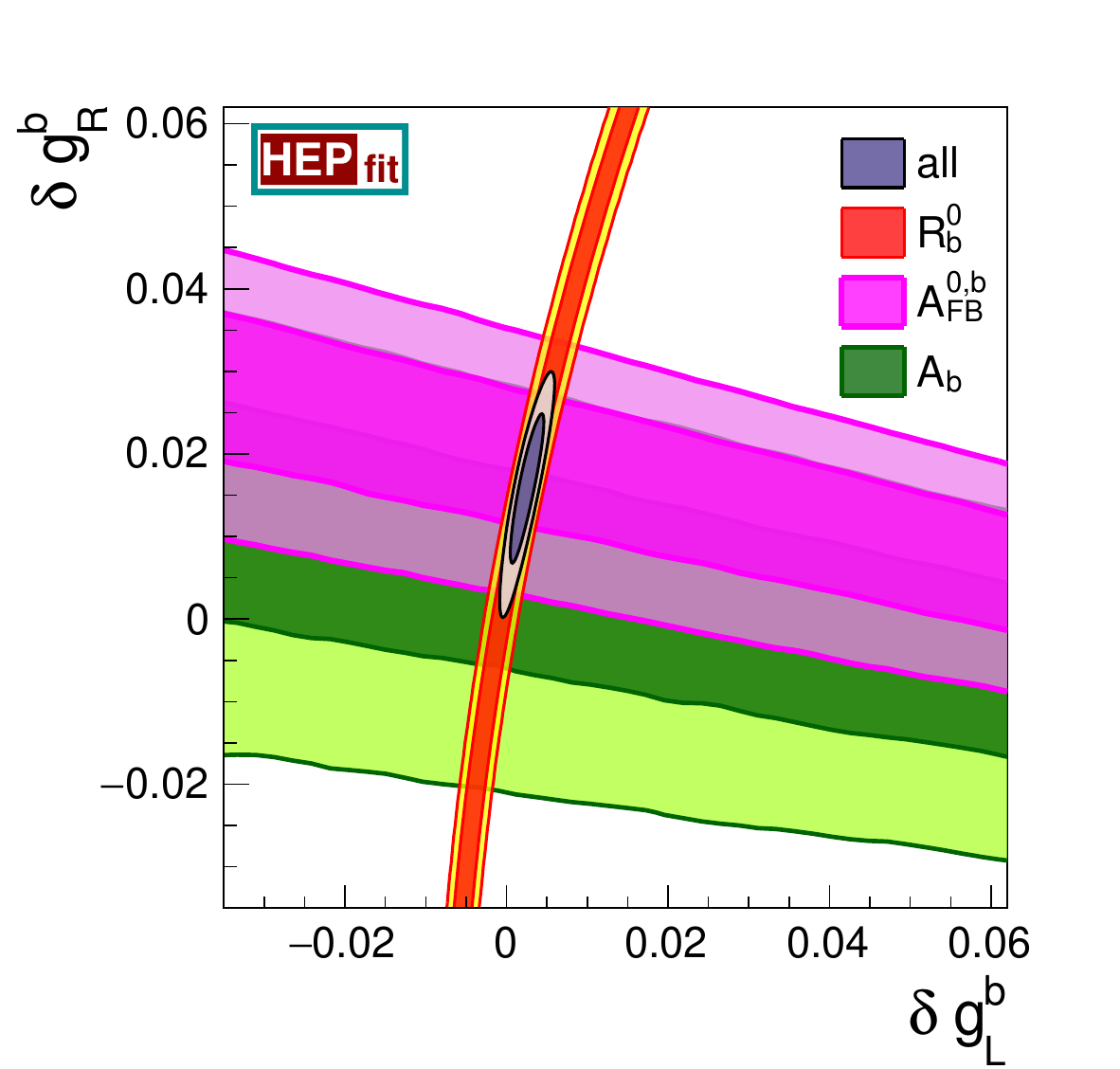}
  \hspace{-3mm}
  \includegraphics[width=.49\textwidth]{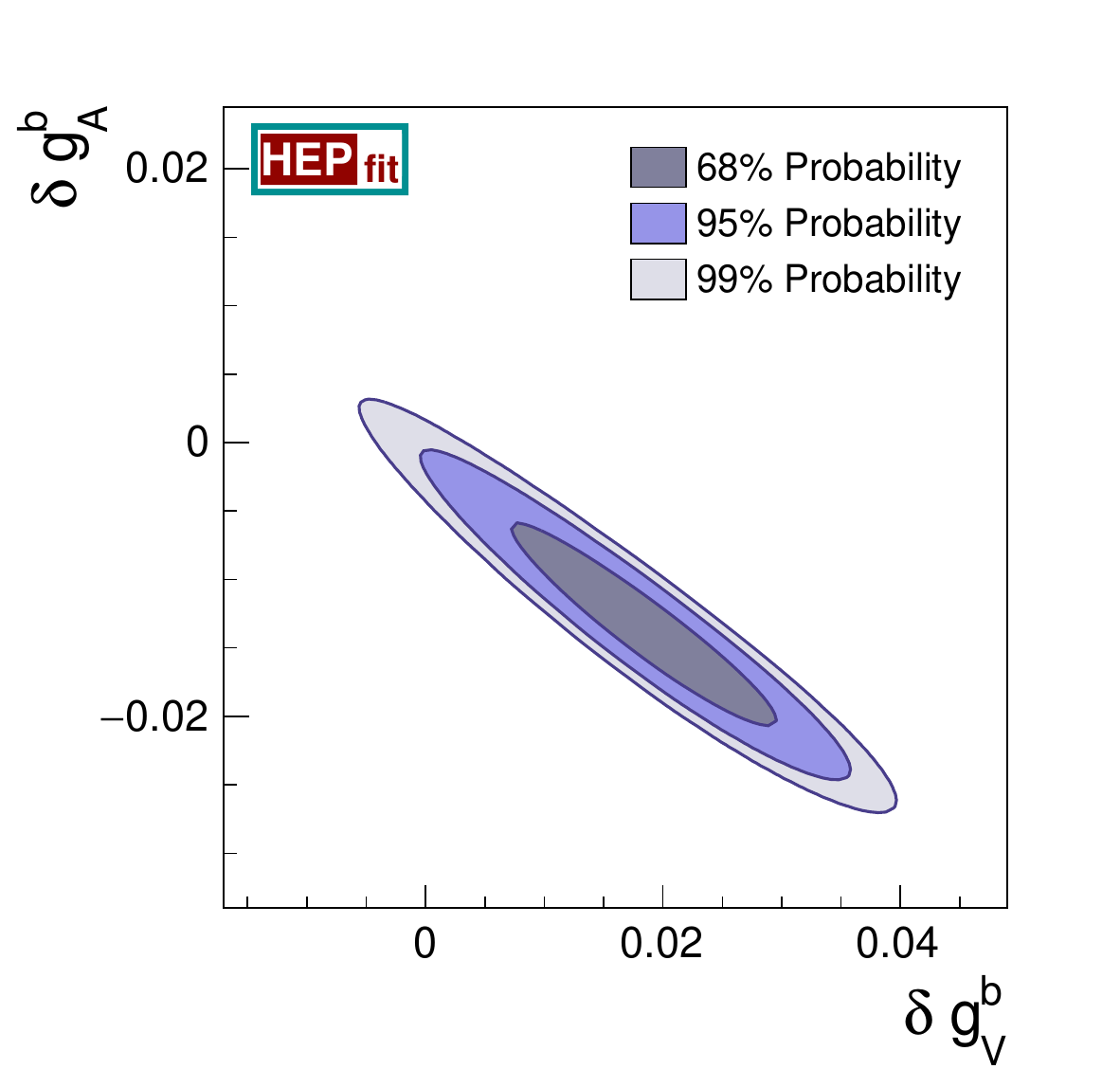}

  \vspace{-2mm}
  \caption{Two-dimensional probability distributions for $\delta
    g_R^b$, $\delta g_L^b$ (left), and $\delta g_V^b$, $\delta g_A^b$
    (right). In the left plot, the dark (light) regions correspond
    to $68\%$ ($95\%$) probability regions.\label{fig:Zbb}}
\end{figure}

Motivated by the apparent discrepancy between the SM prediction for $A_{FB}^{0,b}$ 
and the corresponding experimental result, we also consider here the case where dominant NP contributions
appear in the $Zb\bar{b}$ couplings. We parameterize NP contributions to the
$Zb\bar{b}$ couplings as follows:
\begin{align}
g_i^b 
&= g^b_{i,\mathrm{SM}} + \delta g_i^b\qquad
\mathrm{for}\ \ i=L,\,R\ \ \mathrm{or}\ \ V,\,A\,,
\end{align}
and we present results for both $V$, $A$, and $L$, $R$
couplings. Details on the definitions of these couplings can be found
in ref.~\cite{Ciuchini:2013pca}.
The EW precision fit finds four
solutions for these couplings, but two of them are disfavoured by the
off-peak measurement of the forward-backward asymmetry in
$e^+e^-\to b\bar{b}$~\cite{Choudhury:2001hs}.  In table~\ref{tab:Zbb}
and figure~\ref{fig:Zbb}, we present only the solution closer to the
SM. The observed deviations from zero of the parameters
$ \delta g_i^b$ reflect the deviation from the SM of the measured
value of $A_{\mathrm{FB}}^{0,b}$. While the agreement between 
the SM and $R_b^0$ results in a preferred value of $\delta g_L^b$
consistent with the SM at the $2\sigma$ level, a sizeable contribution
to $\delta g_R^b$ is required to explain the $A_{\mathrm{FB}}^{0,b}$,
and the resulting 95$\%$ probability region in the $\delta g_L^b$-$\delta g_R^b$ 
plane is only marginally compatible with the SM predictions.

\begin{table}[h]
\centering
\begin{tabular}{c|c|rr}
 \hline
 & Result & \multicolumn{2}{c}{Correlation Matrix} \\ 
 \hline 
$\delta g_{R}^{b}$ & $ 0.016 \pm 0.006 $ & $1.00$ \\ 
$\delta g_{L}^{b}$ & $ 0.002 \pm 0.001 $ & $0.90$ & $1.00$ \\ 
\hline
\\
\hline
$\delta g_{V}^{b}$ & $ 0.018 \pm 0.007 $ & $1.00$ \\ 
$\delta g_{A}^{b}$ & $ -0.013 \pm 0.005 $ & $-0.98$ & $1.00$ \\ 
\hline
 \end{tabular}
\caption{Results of the fit for the shifts in the $Zb\bar b$ couplings.}
\label{tab:Zbb}
\end{table}

%------------------------------------------------------------------------------------------------------------

\subsection{Modified $Zb\bar b$ couplings and oblique corrections}
\label{subsec:ew-precision-fit-BSM-Zbb-Oblique}

In several extensions of the SM, oblique corrections and modifications
of the $Zb\bar{b}$ vertex occur simultaneously, possibly affecting
only a specific chirality of the vertex (see for example
refs.~\cite{Grojean:2013qca,Ghosh:2015wiz}). We therefore consider the
following cases: oblique contributions with i) $\delta g_L^b$ and
$\delta g_R^b$, ii)  $\delta g_L^b$ only and iii)  $\delta g_R^b$
only. The corresponding results are presented in table~\ref{tab:STUgLR}. 

\begin{table}
\centering
\begin{tabular}{c|c|rrrr}
\hline
& Result & 
\multicolumn{4}{c}{Correlation Matrix} \\
\hline 
$S$ & $ 0.04 \pm 0.09 $ & $1.00$ \\ 
$T$ & $ 0.08 \pm 0.07 $ & $0.86$ & $1.00$ \\ 
$\delta g_{L}^{b}$ & $ 0.003 \pm 0.001 $ & $-0.24$ & $-0.15$ & $1.00$ \\ 
$\delta g_{R}^{b}$ & $ 0.017 \pm 0.008 $ & $-0.29$ & $-0.22$ & $0.91$ & $1.00$ \\ 
\hline
\multicolumn{6}{c}{$\delta g^b_R = 0$} \\
\hline 
$S$ & $ 0.10 \pm 0.09 $ & $1.00$ \\ 
$T$ & $ 0.12 \pm 0.07 $ & $0.85$ & $1.00$ \\ 
$\delta g_{L}^{b}$ & $ -0.0001 \pm 0.0006 $ & $0.07$ & $0.13$ & $1.00$ \\ 
\hline
\multicolumn{6}{c}{$\delta g^b_L = 0$}\\
\hline 
$S$ & $ 0.08 \pm 0.09 $ & $1.00$ \\ 
$T$ & $ 0.10 \pm 0.07 $ & $0.86$ & $1.00$ \\ 
$\delta g_{R}^{b}$ & $ 0.004 \pm 0.003 $ & $-0.19$ & $-0.21$ & $1.00$ \\ 
\hline
\end{tabular}
\caption{Results of the combined fit of the oblique parameters $S$ and
  $T$, and of the modified $Zb\bar{b}$ couplings, in the case when
  both $\delta g_R^b$ and $\delta g_L^b$ are non zero, and in the case
  in which either $\delta g_R^b=0$ or $\delta g_L^b=0$.}
\label{tab:STUgLR}
\end{table}

%------------------------------------------------------------------------------------------------------------

\section{Constraints on Higgs-boson couplings}
\label{sec:Higgs-coupl-constraints}

In addition to the standard set of EWPO, we have considered all most
recent measurements of Higgs-boson signal strengths, i.e. the ratio
between the measured effective cross section and the corresponding SM
prediction ($\mu\equiv\sigma/\sigma_{\mathrm{SM}}$), taken from
refs.~\cite{Aad:2014eha,Khachatryan:2014ira} for
$H\rightarrow\gamma\gamma$,
refs.~\cite{Aad:2015vsa,Chatrchyan:2014nva} for
$H\rightarrow\tau^+\tau^-$,
refs.~\cite{Aad:2014eva,Chatrchyan:2013mxa,Khachatryan:2014jba} for $H\rightarrow ZZ$,
refs.~\cite{ATLAS:2014aga,Aad:2015ona,Chatrchyan:2013iaa} for
$H\rightarrow W^+W^-$, and
refs.~\cite{Aad:2014xzb,Aad:2015gra,Chatrchyan:2013zna,Khachatryan:2014qaa}
as well as the Tevatron papers ~\cite{Aaltonen:2013ipa,Abazov:2013gmz}
for $H\rightarrow b\bar{b}$.  The Higgs-boson signal strength $\mu$ of
a specific Higgs-search analysis can be calculated as
\begin{equation}
\mu=\sum_i w_ir_i\,\,\,\,\,\,\,\mbox{with}\,\,\,\,\,\,\,
r_i=
\frac{(\sigma\times Br)_i}{(\sigma_{\mathrm{SM}}\times
  Br_{\mathrm{SM}})_i}\,\,\,\,\,\,\,
\mbox{and}\,\,\,\,\,\,\,
w_i=\frac{\epsilon_i (\sigma_{\mathrm{SM}}\times Br_{\mathrm{SM}})_i}
{\sum_j \epsilon_j (\sigma_{\mathrm{SM}}\times Br_{\mathrm{SM}})_j}\,\,\,,
\end{equation}
where $\epsilon_i$ are the experimental efficiencies, and the sums run
over all channels which can contribute to the
final state of the specific analysis.  The SM Higgs-boson production cross
sections (including QCD and, when available, EW corrections) are taken
from ref.~\cite{Heinemeyer:2013tqa} and the SM Higgs-boson decay rates
are taken from ref.~\cite{Contino:2014aaa}.  

In this section, we specialize our discussion to a minimal NP scenario
consisting of an effective theory with only one Higgs boson below the
cutoff scale $\Lambda$. Following
  ref.~\cite{Contino:2010mh}, we assume that custodial
  symmetry is approximately realized, and that the NP scale is sufficiently large
  compared to the energies we are testing, so we can truncate
  the effective Lagrangian at the 2-derivative level. We also assume
  that gauge fields couple to the NP sector via weak gauging, in which case
  the coefficients of operators involving field strengths are loop suppressed
  and we can neglect them.
  Finally, we assume that fermions are only coupled
  to the NP via proto-Yukawa interactions, and we take all corrections
  from NP to be flavor diagonal and universal. This
scenario can be described by a general effective Lagrangian of the
form (see
e.g.~\cite{Giudice:2007fh,Contino:2010mh,Azatov:2012bz,Contino:2013kra}):
\begin{equation}
\mathcal{L}_\mathrm{eff} =
\frac{v^2}{4}{\rm tr}\big(D_\mu\Sigma^\dagger D^\mu\Sigma\big)
\left( 1 + 2\kappa_V\frac{H}{v} + \cdots \right)-m_i\bar f^i_L \left(1+2 \kappa_f\frac{H}{v}
+\cdots\right)f^i_R
+ \cdots,
\label{eq:L_LightHiggs}
\end{equation}
where $v$ is the vacuum expectation value of the Higgs field, 
and the longitudinal components of the $W$ and $Z$ bosons,
$\chi^a(x)$, are described by the two-by-two matrix  
$\Sigma(x) = \exp(i\tau^a\chi^a(x)/v)$, with $\tau^a$ being the Pauli
matrices. 
The deviations in the Higgs-boson couplings to weak gauge-bosons,
$HVV$  ($V=Z,W^\pm$), and to fermions,  $Hf\bar f$, are parameterized by 
the scale factors $\kappa_V$ and $\kappa_f$ respectively,
defined as the ratio between the total Higgs-boson couplings, including
NP effects, and the corresponding couplings in the SM (such that
$\kappa_V=\kappa_f=1$ in the SM).
We only consider the modification of couplings already existing in the SM and,
for loop-induced couplings ($Hgg$, $H\gamma\gamma$, and $HZ\gamma$), we
do not assume NP contributions in loops\footnote{We notice
  that, in the presence of NP, the
relative experimental efficiencies, $\epsilon_i$, will in general be
different from their values in the SM. In particular, the appearance
of new tensor structures in the vertices could modify the kinematic
distributions of the final-state particles, thereby changing the
efficiencies. However, since in this work we only consider rescalings of the SM Higgs-boson
couplings, we will use SM efficiencies $\epsilon_i^{\mathrm{SM}}$ (and hence weight
factors $w_i^{\mathrm{SM}}$) throughout.}.
This class of models is not fully general but it is
more directly constrained by the experimental measurements of
Higgs-boson couplings. It is also the scenario assumed in both ATLAS
and CMS studies of Higgs-boson couplings and allows us to directly
compare to their results, giving us the possibility to test both the {\tt HEPfit}
framework and our correct use of the experimental data for 
Higgs-boson signal-strengths.
 For a detailed description of
the relations between scale factors and the Higgs-boson signal
strengths we refer the reader to ref.~\cite{Heinemeyer:2013tqa}.
%In the presence of NP, the relative experimental efficiencies, $\epsilon_i$, will in general be
%different from their values in the SM. In particular, the appearance
%of new tensor structures in the vertices can modify the kinematic
%distributions of the final-state particles, thereby changing the
%efficiencies. In this work we only consider rescalings of the SM Higgs
%couplings and use the SM weight factors throughout. This assumption is
%justified a posteriori by the overall compatibility of the measurements of
%Higgs-boson properties with the corresponding SM predictions.

\begin{figure}[htb]
  \centering
  \vspace{-1mm}
  \includegraphics[width=.45\textwidth]{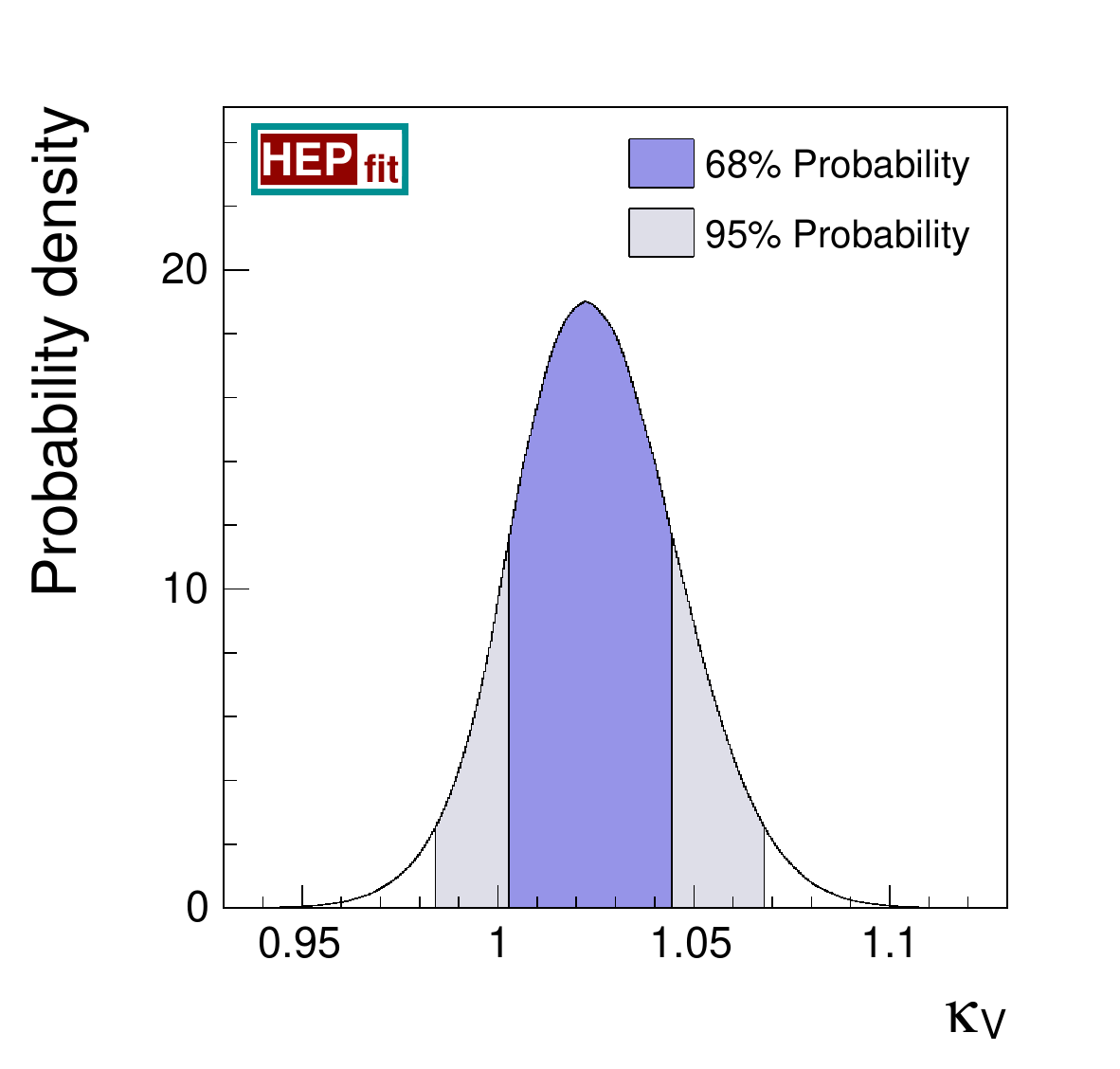}

  \vspace{-2mm}
  \caption{Probability distribution for $\kappa_V$ derived from
    precision EW measurements.  The dark and light regions correspond
    respectively to $68\%$ and $95\%$ probabilities.}
  \label{fig:kV_EWPO}
\end{figure}

In this context we first perform a fit of the EWPO with the only
addition of the scale factor $\kappa_V$. The
only corrections to EWPO are then given by the following
1-loop contributions to the
oblique $S$ and $T$ parameters~\footnote{ Even 
if we assume that custodial symmetry is preserved by the new interactions, there is still
a non-zero contribution to the $T$ parameter. Note that custodial symmetry breaking
is actually parameterized by $\Delta\rho=\alpha T$, which, in this scenario,
is proportional to the square of $U(1)_Y$ gauge coupling, $g^\prime$.
 Indeed $g^\prime$ is one of the parameters that breaks
  custodial symmetry in the SM, and the new effective interactions
  only modify the way custodial symmetry is broken. In the limit of
  $g^\prime\rightarrow 0$ there is no contribution to $T$,
  consistently with the assumption of custodial symmetry in the new physics.}~\cite{Barbieri:2007bh}:
\begin{align}
S &= \frac{1}{12\pi} (1 - \kappa_V^2)
  \ln\bigg(\frac{\Lambda^2}{m_H^2}\bigg)\,,
&
T &= - \frac{3}{16\pi c_W^2} (1 - \kappa_V^2)
  \ln\bigg(\frac{\Lambda^2}{m_H^2}\bigg)\,,
\label{eq:ST}
\end{align}
where we set the cutoff of the effective Lagrangian
to the scale of violation of perturbative unitarity in $WW$ scattering, i.e. 
$\Lambda = 4\pi v/\sqrt{|1-\kappa_V^2|}$. We present the results of the fit for $\kappa_V$ in
table~\ref{tab:kV_EWPO} and fig.~\ref{fig:kV_EWPO}.  

The lower bound on $\kappa_V$ at 95\% corresponds to a cutoff scale
$\Lambda=13$ TeV if $\kappa_V$ is assumed to be smaller than 1,
$\Lambda=8.7$ TeV if $\kappa_V$ is assumed to be larger than 1, and
$\Lambda=8.8$ TeV marginalizing over the sign of $1-\kappa_V$. The fit
disfavours values of $\kappa_V < 1$ ($10\%$ probability), expected for
example in composite Higgs models. This problem can be alleviated by
adding extra contributions to the oblique
parameters~\cite{Grojean:2006nn,Azatov:2013ura,Pich:2012dv,Pich:2013fea}.

\begin{table}[htb]
\centering
\begin{tabular}{c|cc}
\hline
& Result & 95\% Prob.
\\
\hline
$\kappa_V$ & $1.02\pm 0.02$ & $[0.98,\, 1.07]$
\\
\hline
\end{tabular}
\caption{Results of the fit for the scale factor $\kappa_V$ at 68\% and
  95\% probabilities.
\label{tab:kV_EWPO}}
\end{table}

\begin{figure}[t]
  \centering
  \vspace{-1mm}
  \includegraphics[width=.44\textwidth]{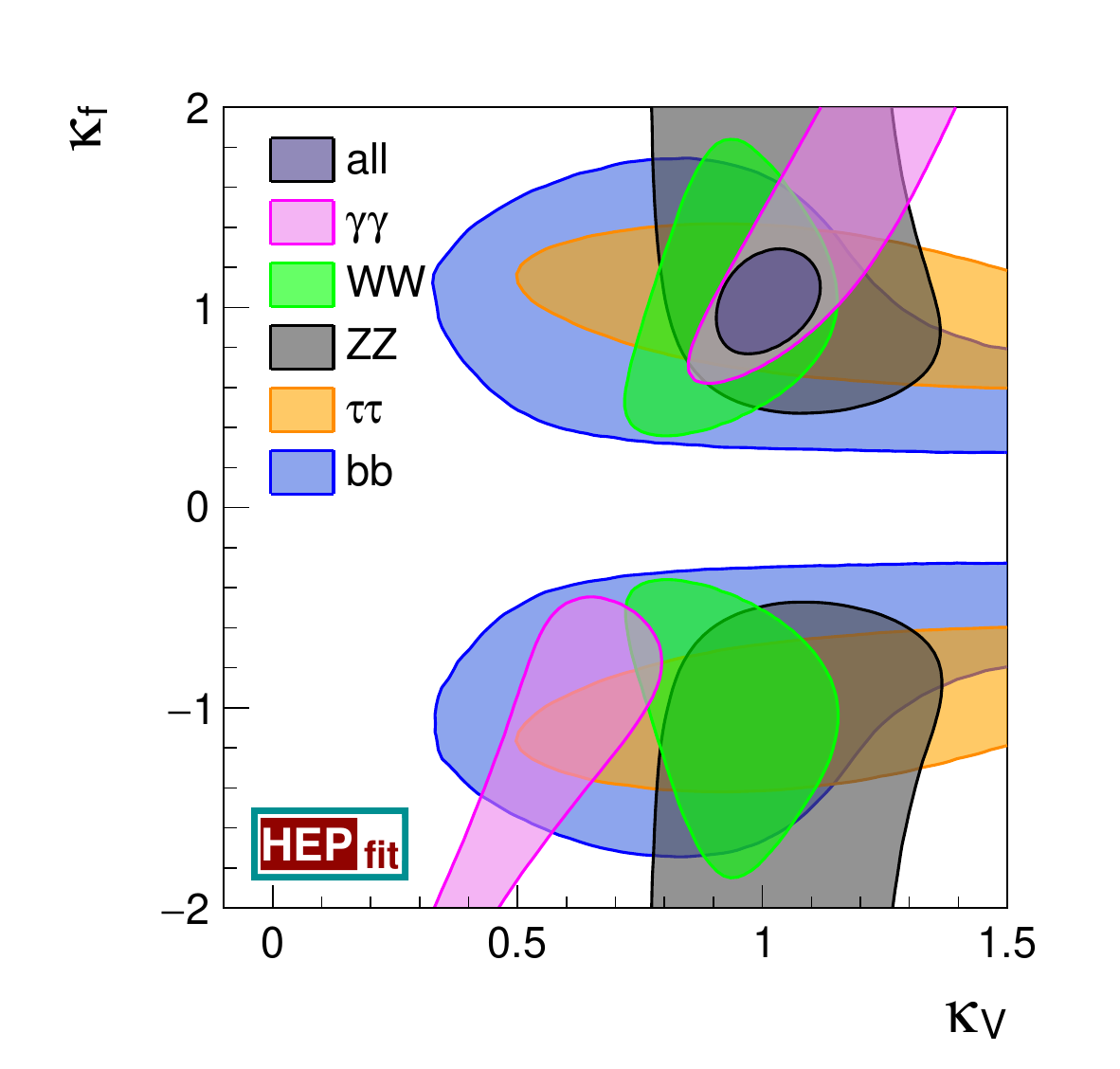}
  \hspace{-3mm}
  \includegraphics[width=.44\textwidth]{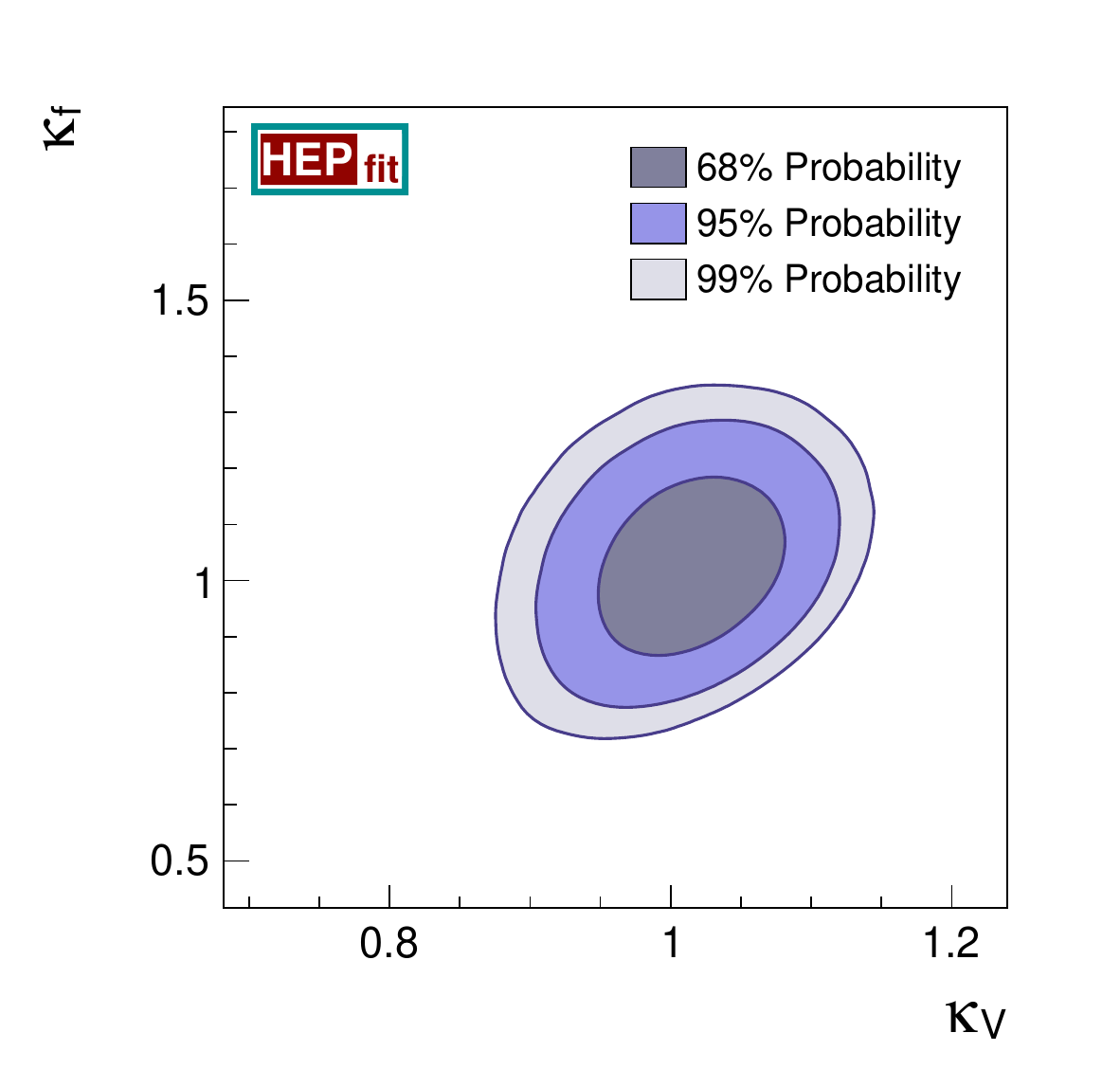}
  \vspace{-2mm}
  \caption{
      Left: constraints from individual channels at
      95\% probability.
      Right: two-dimensional probability distributions for $\kappa_V$
      and $\kappa_f$ at 68\%, 95\%, and 99\% (darker to
      lighter), obtained from the fit to the Higgs-boson signal
      strengths.
      \label{fig:kV_kf}}
\end{figure}

\begin{figure}[t!]
  \centering
  \vspace{-2mm}
  \includegraphics[width=.44\textwidth]{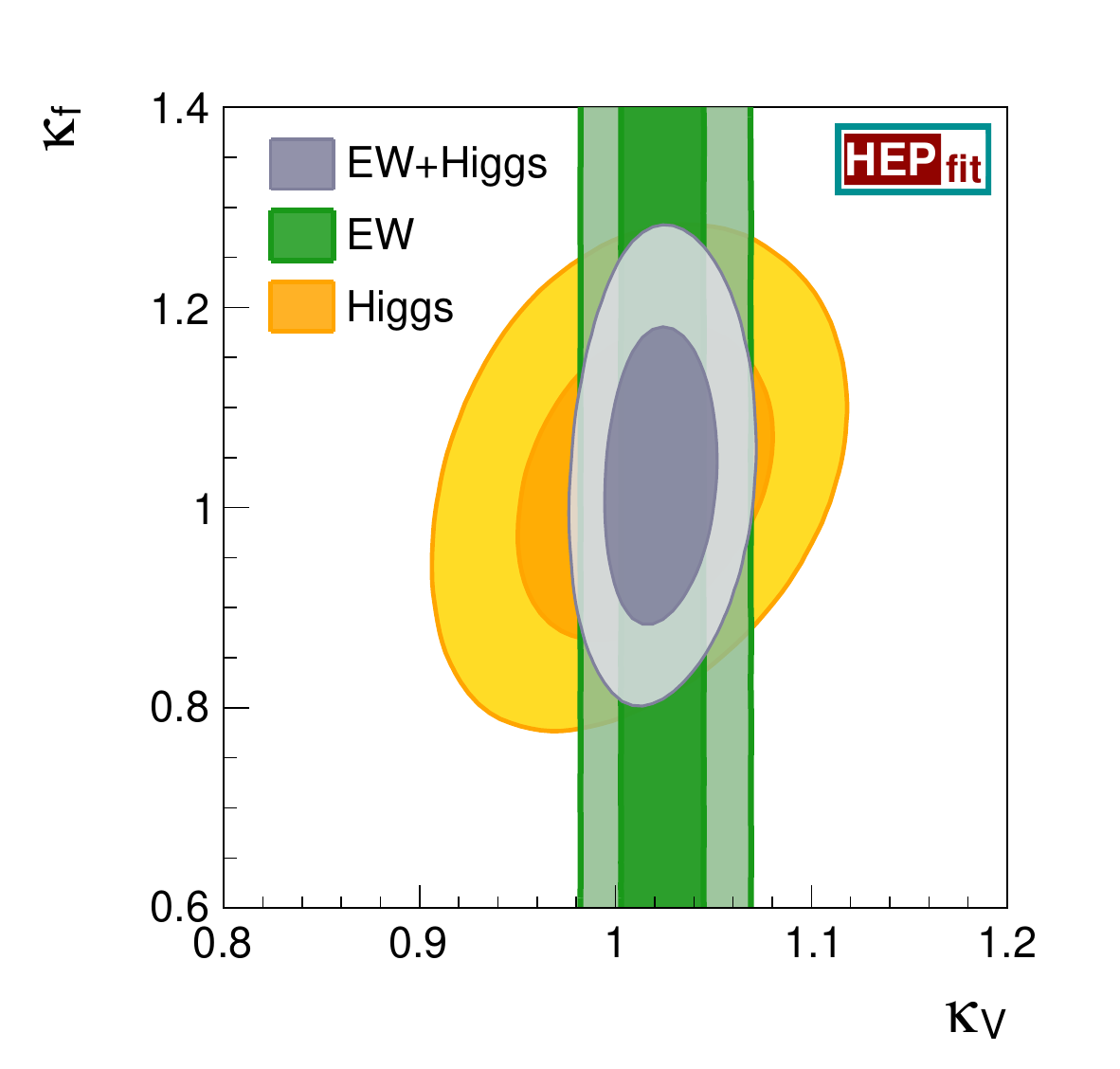}
  \vspace{-3mm}
  \caption{
    Two-dimensional $68\%$ (dark) and $95\%$ (light) probability
    contours for $\kappa_V$ and $\kappa_f$ (from darker to lighter),
    obtained from the fit to the Higgs-boson signal strengths and the
    EWPO.
    \label{fig:kV_kf-EW}}
\end{figure}

The two-dimensional probability distributions for $\kappa_V$ and
$\kappa_f$ obtained from the fit to Higgs-boson signal strengths are
summarized in table~\ref{tab:kV_kf} and shown in 
figure~\ref{fig:kV_kf}.
The
left panel of of figure~\ref{fig:kV_kf} shows the 95\%
probability contours obtained from a fit including only
each individual channel (e.g. $H\rightarrow \gamma \gamma$), 
as well as the result from the global fit. 
Since both production cross sections and decay rates depend on the
modified couplings via products of the form $\kappa_i\kappa_j$,
theoretical predictions are symmetric under the simultaneous exchange
$\{\kappa_V,\ \kappa_f\} \leftrightarrow \{-\kappa_V,\
-\kappa_f\}$. We therefore restrict the parameter space to positive
$\kappa_V$ only.  Note also that, when performing the global fit to
all channels, the region with negative $\kappa_f$ is not populated
even at $99\%$ probability, so that we only show positive values of
$\kappa_f$ in the right-hand-side plot of figure ~\ref{fig:kV_kf}. The
effect of performing a combined fit of both Higgs-boson signal strengths and
EWPO is summarized in table~\ref{tab:kV_kf-EW} and illustrated in
figure~\ref{fig:kV_kf-EW} (note that in tables~\ref{tab:kV_kf}
and~\ref{tab:kV_kf-EW} we only show the results corresponding to the
SM-like solution, i.e. $\kappa_{V,f}>0$). It is interesting to notice
that the constraint on $\kappa_V$ from EWPO is stronger than the one
obtained from the Higgs-boson signal strengths alone.
\begin{table}[t]
\centering
\begin{tabular}{c|c|c|rr}
 \hline
 & Result & 95\% Prob. & \multicolumn{2}{c}{Correlation Matrix} \\ 
 \hline 
$\kappa_{V}$ & $ 1.01 \pm 0.04 $ & $[ 0.93, 1.10]$ & $1.00$ \\ 
$\kappa_{f}$ & $ 1.03 \pm 0.10 $ & $[ 0.83, 1.23]$ & $0.31$ & $1.00$ \\ 
\hline
 \end{tabular}
\caption{
SM-like solution in the fit of $\kappa_V$ and $\kappa_f$  to the Higgs-boson signal strengths.}
\label{tab:kV_kf}
\end{table}

\begin{table}[t!]
\centering
\begin{tabular}{c|c|c|rr}
 \hline
 & Result & 95\% Prob. & \multicolumn{2}{c}{Correlation Matrix} \\ 
 \hline 
$\kappa_{V}$ & $ 1.02 \pm 0.02 $  & $[ 0.99, 1.06]$ & $1.00$ \\ 
$\kappa_{f}$ & $ 1.03 \pm 0.10 $  & $[ 0.85, 1.23]$ & $0.14$ & $1.00$ \\ 
\hline
 \end{tabular}
\caption{
Same as table~\protect\ref{tab:kV_kf} but considering both the Higgs-boson signal strengths and the EWPO.\label{tab:kV_kf-EW}}
\end{table}

\begin{figure}[t!]
  \centering
  \vspace{-1mm}
  \hspace*{-5mm}
  \begin{tabular}{lll}
  \includegraphics[width=.35\textwidth]{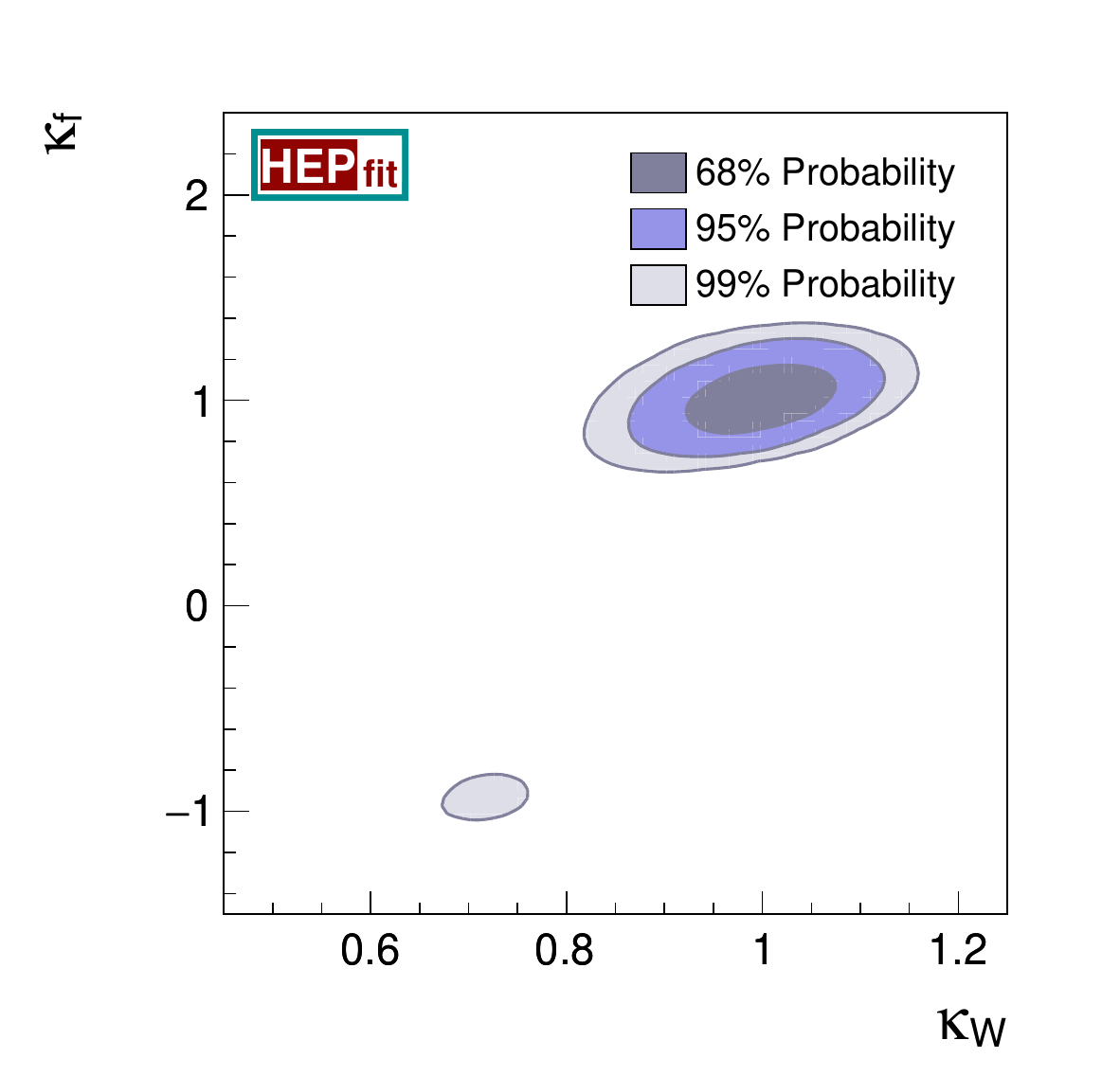}
  &\hspace{-7mm}
  \includegraphics[width=.35\textwidth]{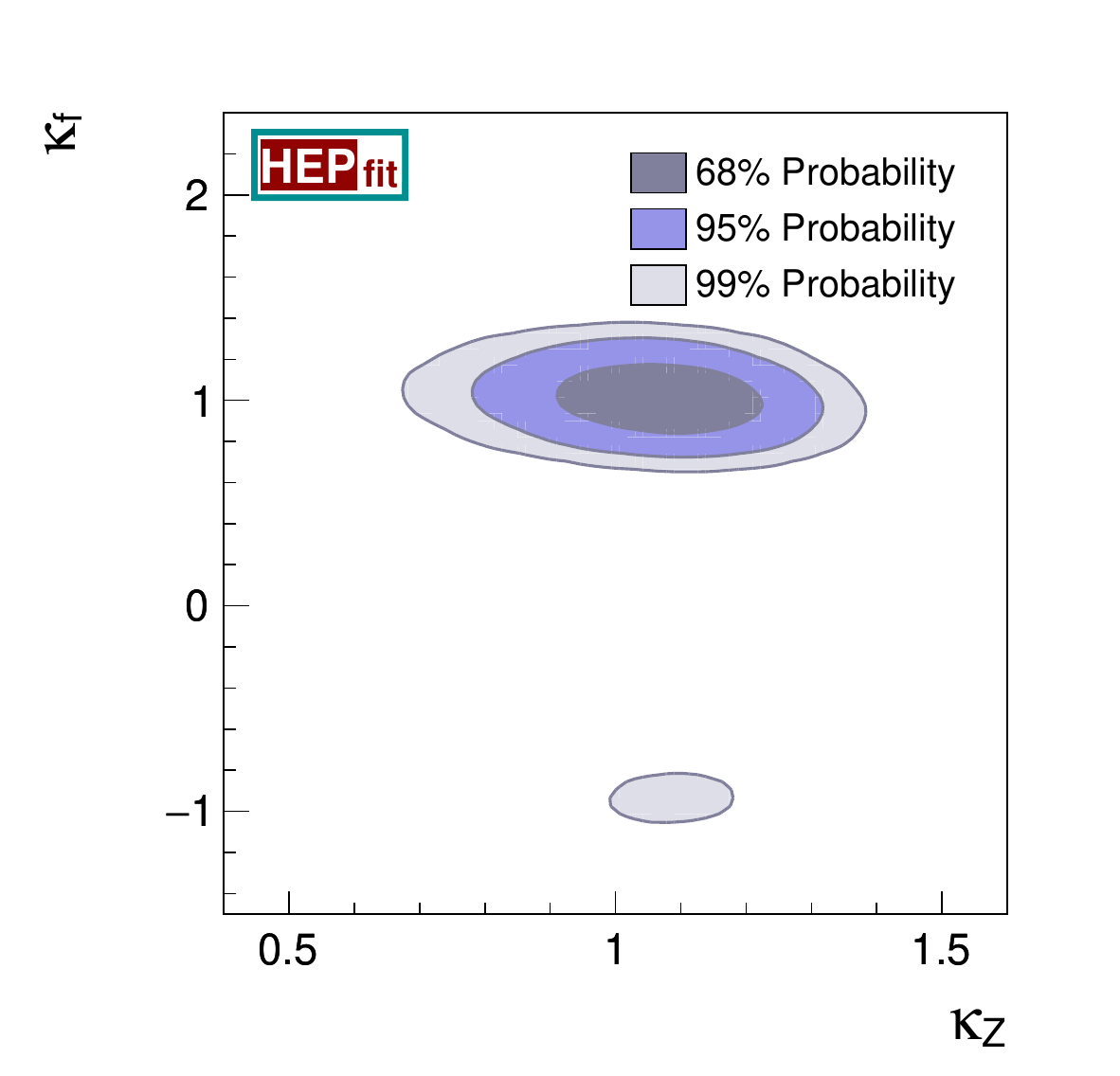}
  &\hspace{-7mm}
  \includegraphics[width=.35\textwidth]{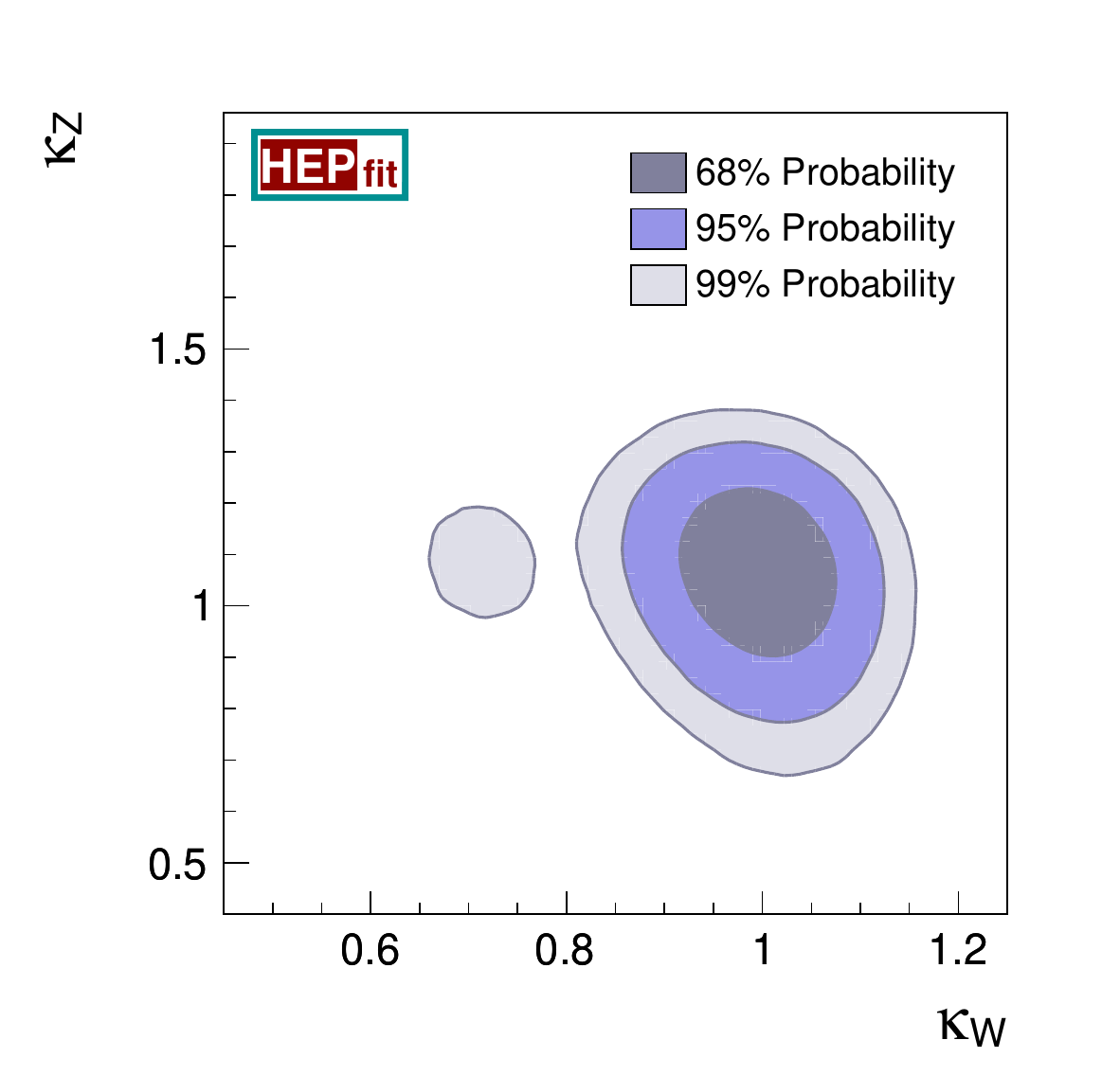}
  \end{tabular}
  
  \vspace{-2mm}
  \caption{
    Two-dimensional probability distributions for $\kappa_W$ and
    $\kappa_f$ (left), for $\kappa_Z$ and $\kappa_f$ (center), and for
    $\kappa_W$ and $\kappa_Z$ (right) at $68\%$, $95\%$, and $99\%$
    (darker to lighter), obtained from the fit to the Higgs-boson
    signal strengths. Note that a small region with $\kappa_f<0$
    is still allowed at $99\%$ probability.
    \label{fig:kW_kZ_kf}}
\end{figure}

We then lift the assumption of custodial symmetry and rescale the
$HZZ$ and $HW^+W^-$ couplings independently, introducing two
parameters $\kappa_Z$ and $\kappa_W$, while keeping a unique rescaling
factor for all fermionic couplings, $\kappa_f$. We obtain the results
summarized in table~\ref{tab:kW_kZ_kf} and the corresponding
probability distributions shown in figure~\ref{fig:kW_kZ_kf}, which
are consistent with custodial symmetry. We notice that theoretical
predictions are symmetric under the exchanges
$\{\kappa_W,\ \kappa_f\} \leftrightarrow \{-\kappa_W,\ -\kappa_f\}$
and/or $\kappa_Z \leftrightarrow -\kappa_Z$, where $\kappa_Z$ can flip
the sign independent of $\kappa_W$, since the interference between the
$W$ and $Z$ contributions to the vector-boson fusion cross section is
negligible. Hence we have considered only the parameter space where
both $\kappa_W$ and $\kappa_Z$ are positive. In this case, we ignore
EWPO in the fit, since setting $\kappa_W\neq\kappa_Z$ generates power
divergences in the oblique corrections, indicating that the detailed
information on the UV theory is necessary for calculating the oblique
corrections.
\begin{table}[t]
\setlength{\tabcolsep}{4pt}
\centering
\begin{tabular}{c|c|c|rrr}
 \hline
 & Result & 95\% Prob. & \multicolumn{3}{c}{Correlation Matrix} \\ 
 \hline 
$\kappa_{W}$ & $ 1.00 \pm 0.05 $  & $[ 0.89, 1.10]$ & $1.00$ \\ 
$\kappa_{Z}$ & $ 1.07 \pm 0.11 $  & $[ 0.85, 1.27]$ & $-0.17$ & $1.00$ \\ 
$\kappa_{f}$ & $ 1.01 \pm 0.11 $  & $[ 0.80, 1.22]$ & $0.41$ & $-0.14$ & $1.00$ \\ 
\hline
 \end{tabular}
\caption{
SM-like solution in the fit of $\kappa_W$, $\kappa_Z$, and $\kappa_f$ to the Higgs-boson signal strengths.\label{tab:kW_kZ_kf}}
\end{table}

\begin{figure}[t!]
  \centering
  \hspace*{-5mm}
  \begin{tabular}{lll}
  \includegraphics[width=.35\textwidth]{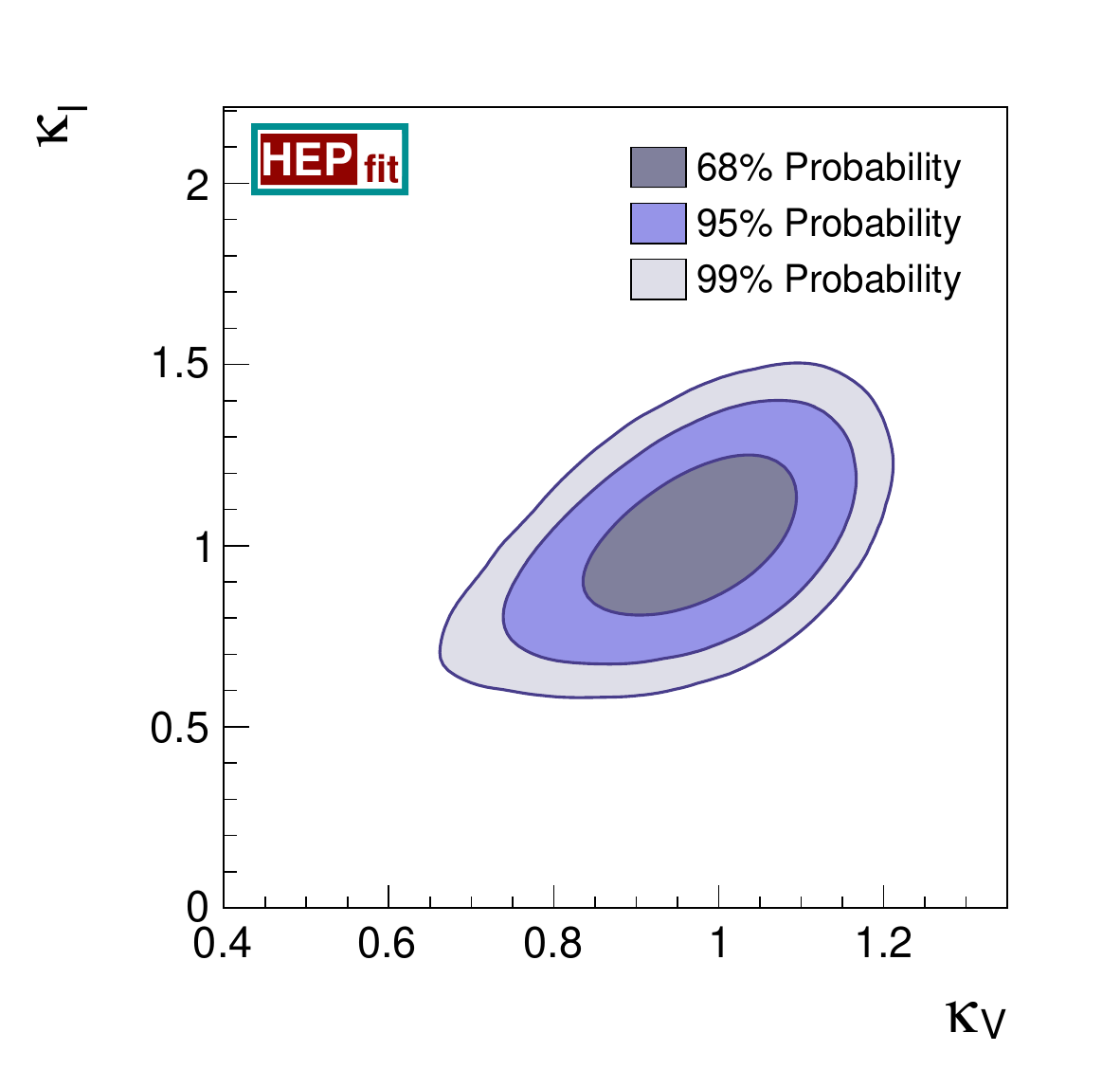}
  &\hspace{-7mm}
  \includegraphics[width=.35\textwidth]{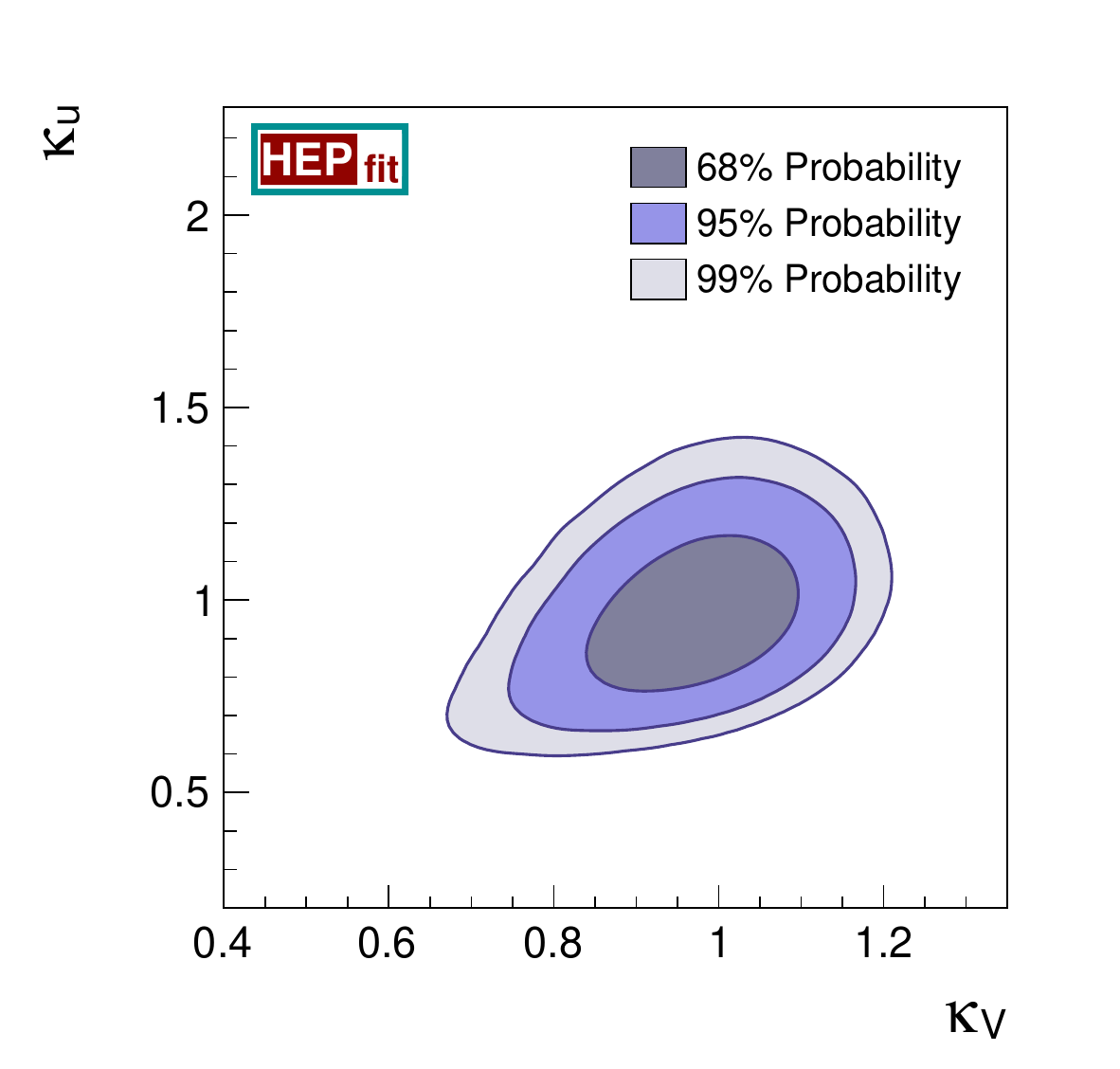}
  &\hspace{-7mm}
  \includegraphics[width=.35\textwidth]{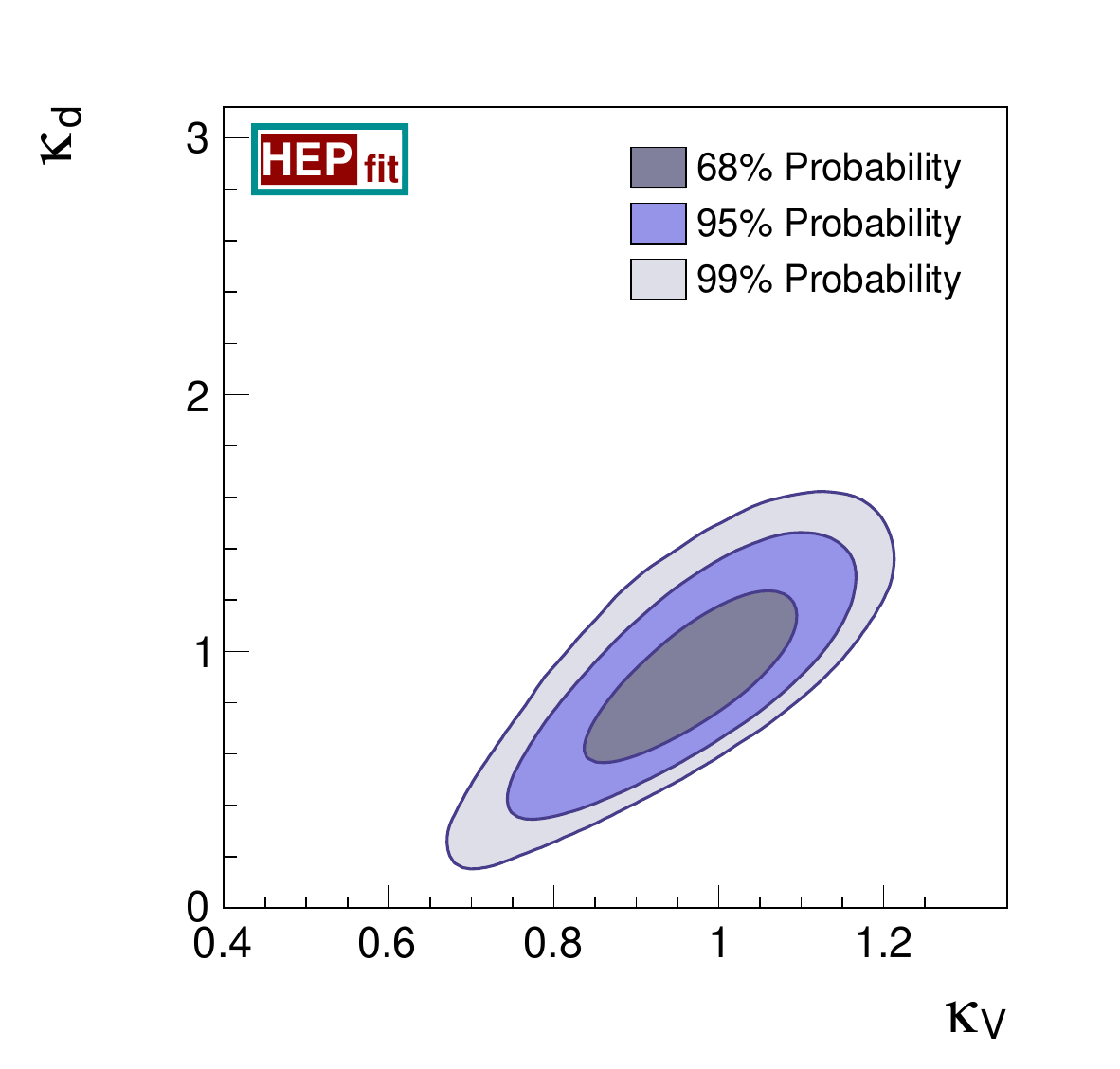}
  \\
  \includegraphics[width=.35\textwidth]{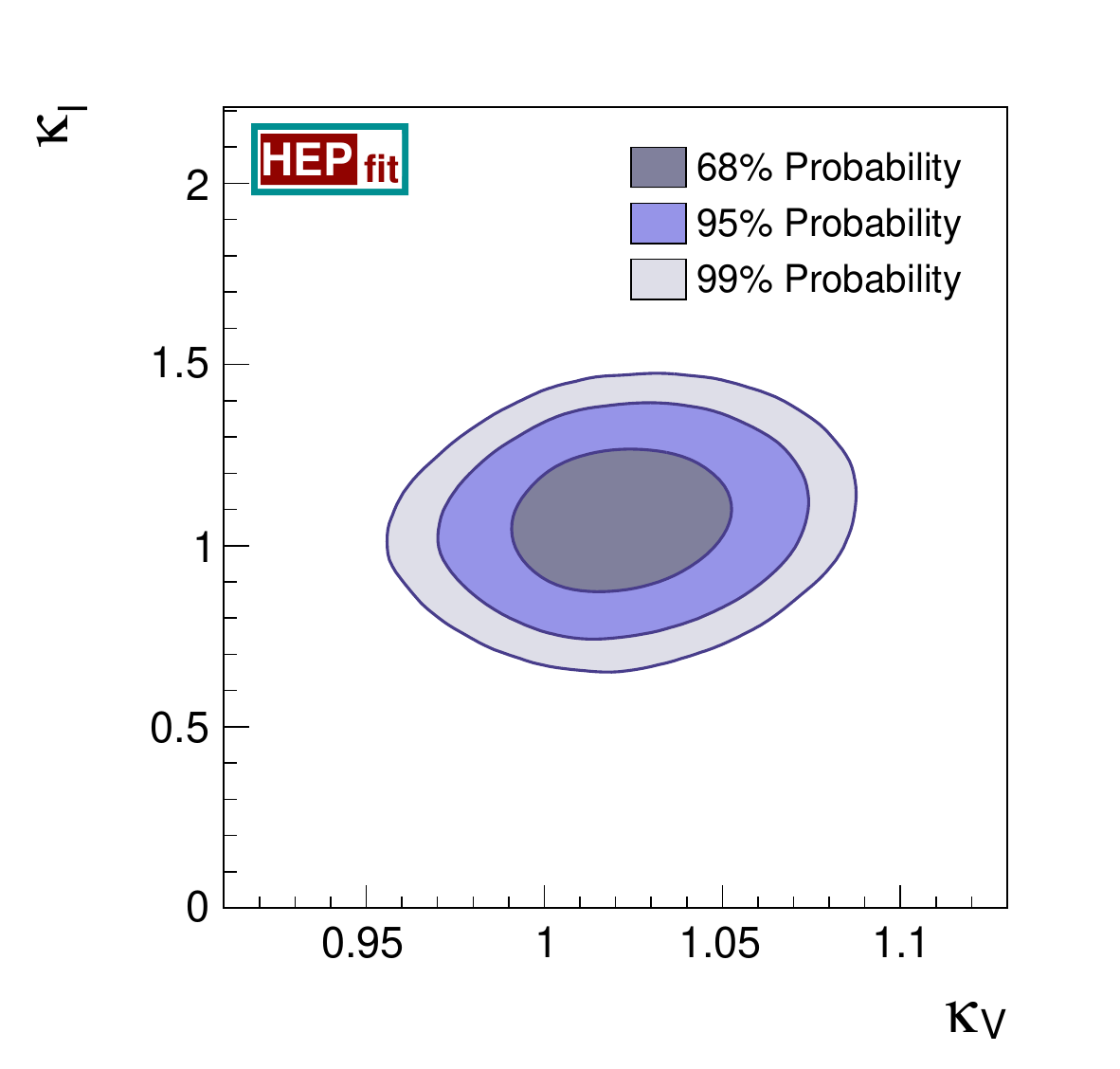}
  &\hspace{-7mm}
  \includegraphics[width=.35\textwidth]{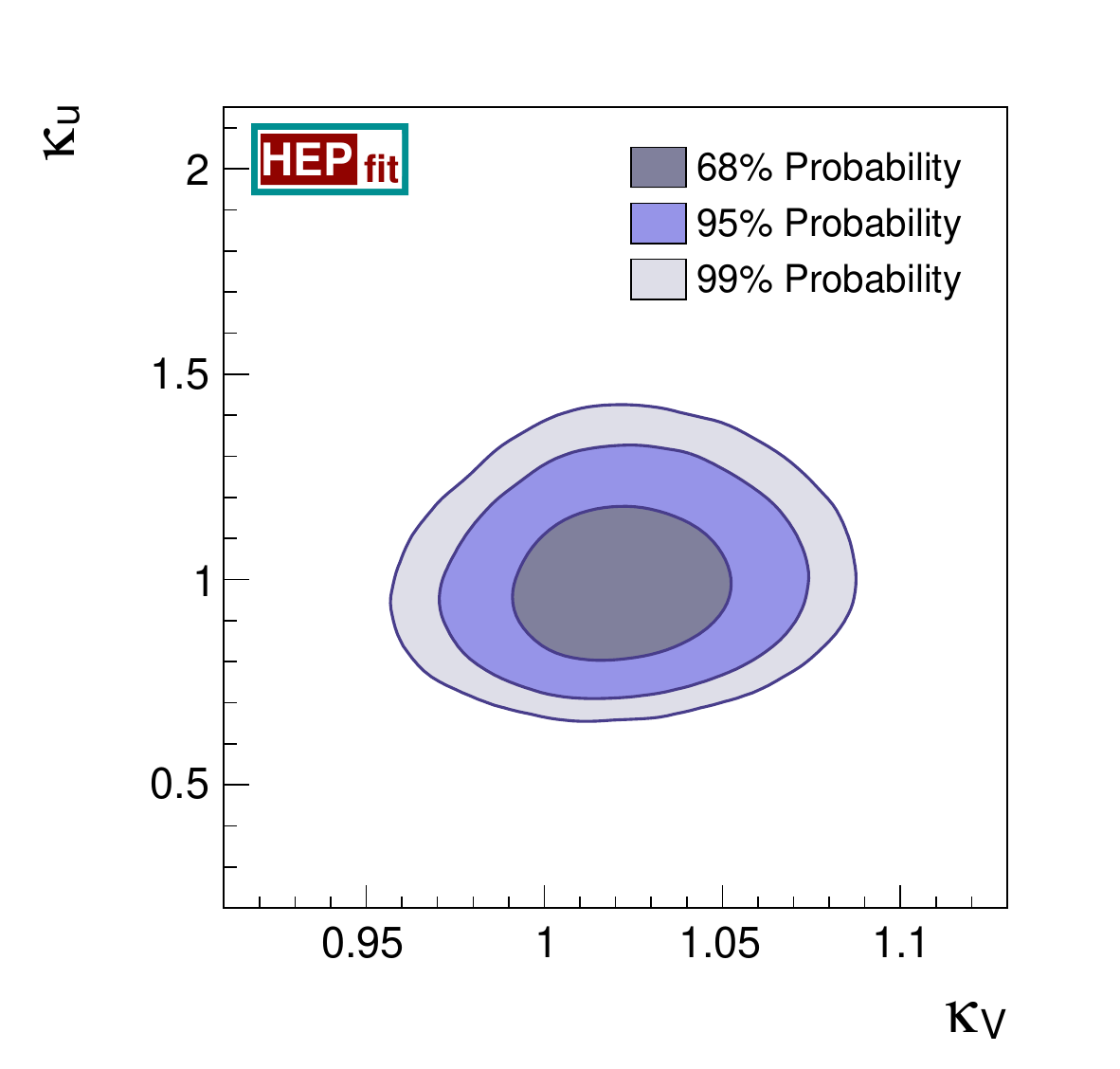}
  &\hspace{-7mm}
  \includegraphics[width=.35\textwidth]{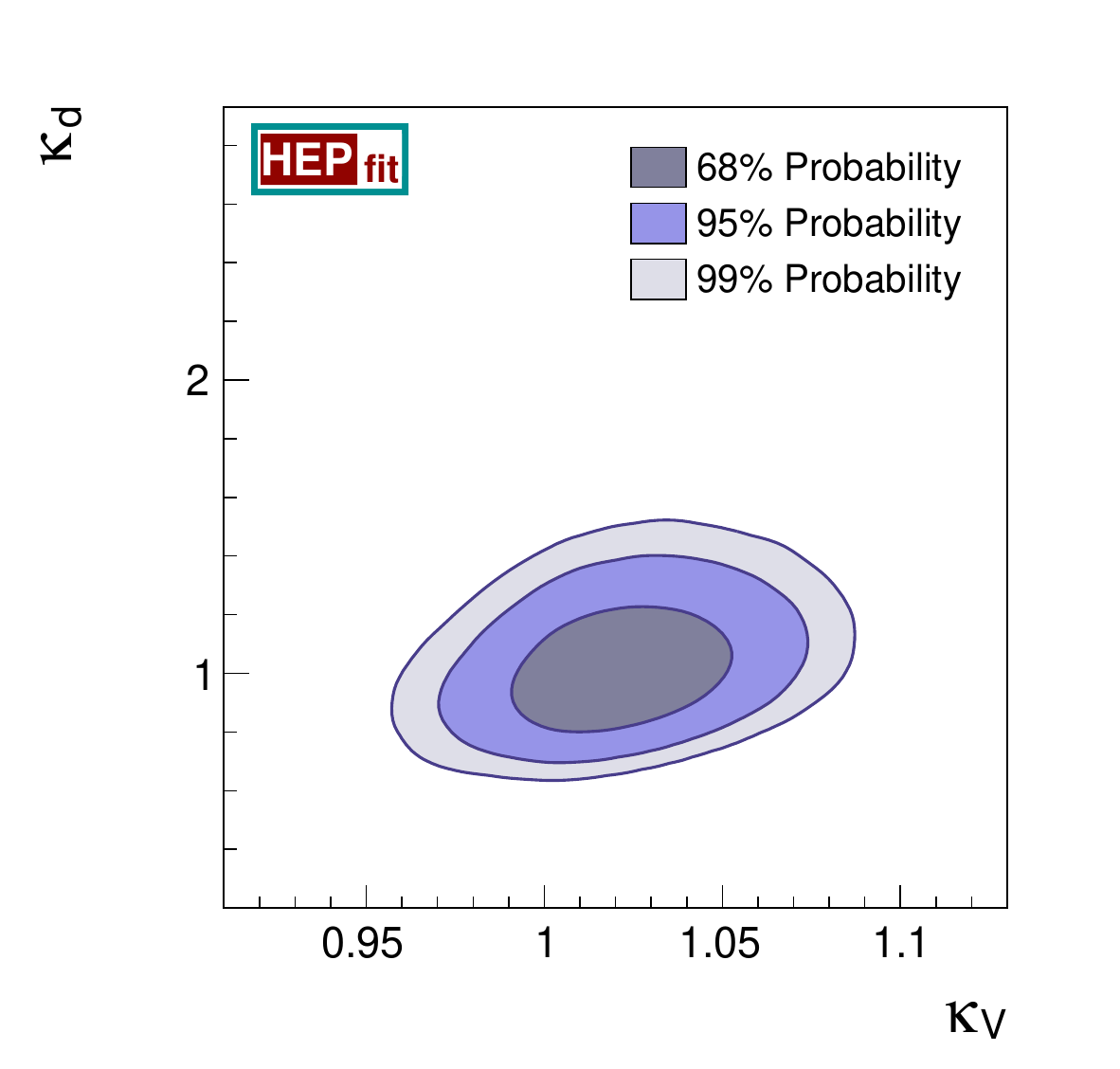}
  \end{tabular}
  
  \vspace{-2mm}
  \caption{
    Two-dimensional probability distributions for $\kappa_V$ and
    $\kappa_\ell$, for $\kappa_V$ and $\kappa_u$, and for $\kappa_V$
    and $\kappa_d$, at $68\%$, $95\%$, and $99\%$ (darker to
    lighter), obtained from the fit to the Higgs-boson signal
    strengths only (top plots) or the combination of Higgs-boson
    signal strengths and EWPO (bottom
    plots).\label{fig:kV_kl_ku_kd-EW}}
\end{figure}

We also consider the case in which we only lift fermion universality
and introduce different rescaling factors for charged leptons
($\kappa_\ell$), up-type quarks ($\kappa_u$), and down-type quarks
($\kappa_d$), while keeping a unique parameter $\kappa_V$ for both
$HVV$ couplings. In this case, from the Higgs-boson signal strengths
we obtain the constraints on the scale factors presented in
table~\ref{tab:kV_kl_ku_kd} and in the top plots of
figure~\ref{fig:kV_kl_ku_kd-EW}. By adding the EWPO to the fit, the
constraints become stronger, as shown in
table~\ref{tab:kV_kl_ku_kd-EW} and in the bottom plots of
figure~\ref{fig:kV_kl_ku_kd-EW}.  In this case, the Higgs-boson signal
strengths are approximately symmetric under the exchanges
$\kappa_\ell \leftrightarrow -\kappa_\ell$,
$\kappa_d \leftrightarrow -\kappa_d$ and/or
$\{\kappa_V,\ \kappa_u\} \leftrightarrow \{-\kappa_V,\ -\kappa_u\}$.
These approximate symmetries follow from the small effect of the
interference between tau and/or bottom-quark loops with top-quark/$W$
loops in the Higgs-boson decay into two photons, as well as the
relatively small interference between bottom- and top-quark loops in
gluon-fusion, for $\left|\kappa_{V,u,d,\ell}\right|\sim 1$. Moreover,
we find that negative values of $\kappa_u$ are disfavoured in the
fit. Hence, in figure~\ref{fig:kV_kl_ku_kd-EW} we consider only the
parameter space where all $\kappa$'s are positive. Again, the results
on table~\ref{tab:kW_kZ_kf} correspond to the SM-like solution,
i.e. $\kappa_{V,u,d,\ell}>0$.

\begin{table}[t]
\setlength{\tabcolsep}{3pt}
\centering
\begin{tabular}{c|c|c|rrrr}
 \hline
 & Result & 95\% Prob. & \multicolumn{4}{c}{Correlation Matrix} \\ 
 \hline 
$\kappa_{V}$ & $ 0.97 \pm 0.08 $  & $[ 0.80, 1.13]$ & $1.00$ \\ 
$\kappa_{\ell}$ & $ 1.01 \pm 0.14 $  & $[ 0.73, 1.30]$ & $0.54$ & $1.00$ \\ 
$\kappa_{u}$ & $ 0.97 \pm 0.13 $  & $[ 0.73, 1.25]$ & $0.42$ & $0.41$ & $1.00$ \\ 
$\kappa_{d}$ & $ 0.91 \pm 0.21 $  & $[ 0.48, 1.35]$ & $0.81$ & $0.61$ & $0.77$ & $1.00$ \\ 
\hline
 \end{tabular}
\caption{
SM-like solution in the fit of $\kappa_V$, $\kappa_\ell$, $\kappa_u$, and $\kappa_d$ to the Higgs-boson signal strengths.\label{tab:kV_kl_ku_kd}}
\end{table}

\begin{table}[t!]
\setlength{\tabcolsep}{3pt}
\centering
\begin{tabular}{c|c|c|rrrr}
 \hline
 & Result & 95\% Prob. & \multicolumn{4}{c}{Correlation Matrix} \\ 
 \hline 
$\kappa_{V}$ & $ 1.02 \pm 0.02 $  & $[ 0.98, 1.06]$ & $1.00$ \\ 
$\kappa_{\ell}$ & $ 1.07 \pm 0.12 $  & $[ 0.82, 1.32]$ & $0.15$ & $1.00$ \\ 
$\kappa_{u}$ & $ 1.01 \pm 0.12 $  & $[ 0.79, 1.27]$ & $0.10$ & $0.24$ & $1.00$ \\ 
$\kappa_{d}$ & $ 1.01 \pm 0.13 $  & $[ 0.76, 1.30]$ & $0.31$ & $0.38$ & $0.78$ & $1.00$ \\ 
\hline
 \end{tabular}
\caption{
Same as table~\protect\ref{tab:kV_kl_ku_kd},  but considering both the
Higgs-boson signal strengths and the EWPO. \label{tab:kV_kl_ku_kd-EW}}
\end{table}

Finally, we consider the case in which both the assumptions of
custodial symmetry and fermion universality are lifted, and perform a
five-parameter fit of $\kappa_W$, $\kappa_Z$, $\kappa_\ell$,
$\kappa_u$, and $\kappa_d$ reported in table \ref{tab:kW_kZ_kl_ku_kd}.
Following the previous discussion, we restrict all the parameters but
$\kappa_u$ (which has an important interference with $\kappa_W$ in $H
\to \gamma\gamma$) to be positive. 
\begin{table}[t!]
\setlength{\tabcolsep}{3pt}
\centering
\begin{tabular}{c|c|c|rrrrr}
 \hline
 & Result & 95\% Prob. & \multicolumn{5}{c}{Correlation Matrix} \\ 
 \hline 
$\kappa_{W}$ & $ 0.94 \pm 0.10 $  & $[ 0.73, 1.13]$ & $1.00$ \\ 
$\kappa_{Z}$ & $ 1.03 \pm 0.13 $  & $[ 0.77, 1.28]$ & $0.34$ & $1.00$ \\
$\kappa_{\ell}$ & $ 1.02 \pm 0.15 $  & $[ 0.73, 1.33]$ & $0.55$ & $0.22$ & $1.00$ \\
$\kappa_{u}$ & $ 0.95 \pm 0.13 $  & $[-0.96, -0.72] \cup [ 0.68, 1.28]$ & $0.49$ &$0.04$ & $0.44$ & $1.00$ \\ 
$\kappa_{d}$ & $  0.91 \pm 0.22 $  & $[ 0.46, 1.36]$ & $0.81$ & $0.36$ & $0.62$ & $0.78$ & $1.00$ \\ 
\hline
 \end{tabular}
\caption{
Results of the simultaneous fit of $\kappa_W$, $\kappa_Z$,
$\kappa_\ell$, $\kappa_u$, and $\kappa_d$, considering only 
Higgs-boson signal strengths. \label{tab:kW_kZ_kl_ku_kd}}
\end{table}

The results presented in this section agree with the recent LHC
combination of Higgs couplings in ref.~\cite{Khachatryan:2016vau},
taking into account that the coupling to down quarks in our analysis
also includes the Tevatron measurements. See also
refs.~\cite{Falkowski:2013dza,Ellis:2013lra,Djouadi:2013qya,Belanger:2013xza,Chpoi:2013wga,Bechtle:2014ewa,Bergstrom:2014vla,Corbett:2015ksa}
for other recent Higgs couplings analyses.

%------------------------------------------------------------------------------------------------------------

\section{Expected sensitivities at future lepton colliders}
\label{sec:future}

Future lepton colliders represent an opportunity to reach the ultimate
precision both on EWPO and Higgs-boson couplings. In this work, we
assess the impact of this improvement in precision by considering the
following proposed $e^+e^-$ colliders: the Future Circular Collider
(FCCee) project at CERN~\cite{Gomez-Ceballos:2013zzn}, the
International Linear Collider (ILC) in
Japan~\cite{Barklow:2015tja,Fujii:2015jha}, and the Circular Electron
Positron Collider (CepC) in China~\cite{CepCreport}.  For completeness
in the comparison we also consider the improvements in the
measurements of EWPO and Higgs-boson signal strengths expected at the
High Luminosity LHC (HL-LHC)
\cite{CMS:2013xfa,ATL-PHYS-PUB-2013-014,ATL-PHYS-PUB-2014-011,ATL-PHYS-PUB-2014-016}.  In
this section we describe the different physics scenarios we will
consider, and estimate the improvements they offer in terms of
sensitivity to the different NP models described in
sections~\ref{sec:ew-precision-fit-BSM} and
\ref{sec:Higgs-coupl-constraints}, comparing the results with those
obtained using current data. See
refs.~\cite{Erler:2000jg,Freitas:2013xga,Fan:2014vta,Baak:2014ora,Ge:2016zro} for
earlier analyses of this kind.

Across its years of operation, the FCCee design includes running at
the $Z$ pole, and at the $WW$, $HZ$, and $t \bar t$ production
thresholds, with the possibility of a dedicated run at center-of-mass
energy $\sqrt{s}\gtrsim 350$~GeV to explore the top-quark
couplings. Compared to other options for future $e^+e^-$ colliders,
the FCCee also offers the largest integrated luminosity and allows to
assess an optimistic best-case scenario.  The expected performance of
the FCCee machine is documented in
refs.~\cite{Gomez-Ceballos:2013zzn,AzziFCCweek}, and summarized in
table~\ref{tab:FCCee}. The values of integrated luminosity presented
there are a useful baseline for our study. Further improvements in
performance are under consideration, including an increase in
center-of-mass energy. Within the context of our analyses, these
improvements would further reduce the statistical uncertainties. On
the other hand, since the precision on the observables considered in
our study will be mainly dominated by the systematic uncertainties,
our conclusions would still hold to a large extent.

\begin{table}[htb]
\begin{center}
\begin{tabular}{|c|c|c|c|c|c|}
\hline
FCCee               & $Z$ pole & $WW$ & $HZ$ & $t \bar t$  & Above $t \bar t$ \\
                          &                & threshold & threshold & threshold & threshold \\
\hline
$\sqrt{s}$ [GeV] & 90 & 160 & 240 & 350 & $>350$ \\ 
${\cal L}~[\mathrm{ab}^{-1}/\mathrm{year}]$ & 88 & 15 & 3.5 & 1.0 & 1.0 \\ 
Years of operation & 0.3 / 2.5 & 1 & 3 & 0.5 & 3 \\ 
\hline
Events & $10^{12} / 10^{13}$ & $10^{8}$ & $2 \times 10^{6}$ & $2.1 \times 10^{5}$ & $7.5 \times 10^{4}$ \\ 
\hline
\end{tabular}
\end{center}
\caption{\label{tab:FCCee} Expected performances of the FCCee machine, taken from ref.~\cite{AzziFCCweek}}
\end{table}

The ILC project consists of a linear $e^+e^-$ collider optimized for
Higgs-boson and top-quark precision measurements, and would initially
run at energies $\sqrt{s}=250,~350$, and $500$
GeV~\cite{Barklow:2015tja}.  The current proposed scenarios would
involve approximately 20 years of operation, including a luminosity
upgrade. There is also the possibility of extending the energy reach
of the machine up to 1 TeV, and we include this in our list of physics
scenarios. The energy and luminosity settings of the Higgs-boson runs
that we study in this work are given in
table~\ref{tab:ILC}~\cite{Dawson:2013bba}.  Improved measurements of
the properties of the $Z$ lineshape at $\sqrt{s}\approx 91$ GeV, on
the other hand, would require a machine upgrade from the Technical
Design Report to achieve an optimal luminosity
performance~\cite{Barklow:2015tja}.  We therefore do not consider this
scenario here. As far as EWPO are concerned, we only include the
improvements in the Higgs-boson, top-quark, and $W$ masses, where the
latter is obtained from the measurements of
$e^+ e^- \rightarrow W^+ W^-$ above threshold with a target overall
uncertainty at the level of approximately $3$ MeV.

\begin{table}[htb]
\begin{center}
\begin{tabular}{|c|ccc|ccc|}
\hline
ILC   & \multicolumn{2}{c}{~~~~~~~~~Phase 1}&& \multicolumn{2}{c}{~~~~~~~~~Phase 2 }& \\
  & \multicolumn{2}{c}{ }&& \multicolumn{2}{c}{~~~~~~~~~(Luminosity upgrade)}& \\
\hline
$\sqrt{s}$ [GeV] & 250 & 500 & 1000 & 250 & 500 &100 \\ 
$\int {\cal L}~\!dt~[\mathrm{ab}^{-1}]$ & 0.25 & 0.5 & 1 & 1.15 & 1.6 & 2.5 \\ 
$\int dt$ ($10^7$ s) & 3 & 3 & 3 & 3 & 3 & 3 \\ 
\hline
\end{tabular}
\end{center}
\caption{\label{tab:ILC} Expected performances of the ILC machine, taken from ref.~\cite{Dawson:2013bba}}
\end{table}

Finally, the CepC project is designed as a Higgs-boson and/or $Z$
factory~\cite{CepCreport}. Running at $\sqrt{s}\approx 240$ GeV the
CepC would produce about $10^6$ Higgs-boson particles, allowing
measurements of its couplings at the percent level or better. During
the $\sqrt{s}\approx 91$ GeV run, on the other hand, up to $10^{11}$
$Z$ bosons could be produced, improving the sensitivity to the $Z$
couplings to the $10^{-4}$ level. With this statistics, the overall
uncertainty for most observables is expected to be dominated by
systematic effects.  For the run at the $Z$-pole energy, we will
assume a total integrated luminosity larger than 150 fb$^{-1}$,
necessary to achieve the expected precision for all the different EWPO
in table 4.1 of ref.~\cite{CepCreport}.  As in the case of the ILC, an
improved measurement of the $W$ mass is possible at center-of-mass
energies above the $W^+ W^-$ production threshold. For the
$\sqrt{s}=250$~GeV run a direct $M_W$ measurement is expected with a
similar uncertainty of approximately 3 MeV.

% The expected experimental uncertainties on the different EWPO and
% Higgs-boson signal strengths at the future colliders introduced
% above are summarized in tables~\ref{table:EWPOfuture} and
% \ref{table:Higgsfuture}, where we also include projections for the
% HL-LHC~\cite{CMS:2013xfa,ATL-PHYS-PUB-2013-014,ATL-PHYS-PUB-2014-011,ATL-PHYS-PUB-2014-016}.

The expected experimental uncertainties on the different EWPO at the
future colliders introduced above are summarized in
table~\ref{table:EWPOfuture}~\cite{Gomez-Ceballos:2013zzn,Fujii:2015jha,CepCreport,Baak:2013fwa}.
When no input is provided for
FCCee~\cite{Gomez-Ceballos:2013zzn,Baak:2013fwa}, we adopted
conservative numbers, depending on the experimental observable. The
corresponding information for the expected accuracies in Higgs-boson
signal strengths are summarized in table~\ref{table:Higgsfuture}.  In
both tables~\ref{table:EWPOfuture} and \ref{table:Higgsfuture}, we
have also included projections for the
HL-LHC~\cite{CMS:2013xfa,ATL-PHYS-PUB-2013-014,ATL-PHYS-PUB-2014-011,ATL-PHYS-PUB-2014-016}.

On the theory side, while the theoretical uncertainties associated to
unknown higher-order corrections to EWPO in perturbation theory are
subdominant compared with current experimental errors, this is no
longer the case when we take into account the projected future
experimental precision summarized in table~\ref{table:EWPOfuture}.
The present theoretical uncertainties for the most relevant EWPO are
shown in table~\ref{table:EWPOfutureErrors}, where we compare them to
the corresponding current and future experimental errors. It is clear
that we need to improve SM calculations in order for theoretical
uncertainties in the predictions of EWPO not to become a limiting
factor at future experiments. The future projected theoretical errors
in table~\ref{table:EWPOfutureErrors} assume that the complete
${\cal O}(\alpha\alpha_s^2)$ corrections, the fermionic
${\cal O}(\alpha^2\alpha_s^2)$ and ${\cal O}(\alpha^3)$ corrections,
and the leading 4-loop corrections entering via the $\rho$ parameter
in the different observables will become
available~\cite{Freitas:2013xga, Freitas:2014owa, Freitas:2016sty}.
There are other sources of theoretical uncertainties not considered in
the previous discussion.  First, as explained in
section~\ref{sec:ew-precision-fit-SM}, the parametric uncertainties on
the theoretical predictions for the different EWPO receive important
contributions from the current errors in the experimental measurements
of $\Delta \alpha_{\mathrm{had}}^{(5)}(M_Z)$ and $\alpha_s(M_Z)$ (see
table~\ref{tab:SMpred}).  Apart from the experimental improvements
summarized in table~\ref{table:EWPOfuture}, we also assume in all
future scenarios that a measurement of
$\Delta \alpha_{\mathrm{had}}^{(5)}$ is possible with a precision of
$\pm 5\times 10^{-5}$. Such an improvement is expected to be within
the reach of ongoing and future experiments measuring the
$e^+ e^- \rightarrow \mathrm{hadrons}$ cross section. This requires
measuring the ratio $R$ of the hadronic to the muonic $e^+e^-$ cross
sections with a relative uncertainty of
$1\%$~\cite{Asner:2008nq}. Likewise, for the strong coupling
constant at the $Z$ pole, we use future lattice QCD projections,
which estimate an uncertainty
$\delta \alpha_s(M_Z)=\pm 0.0002$~\cite{LubiczFantasy}.  Another
observable which suffers of additional theoretical uncertainties is
the top-quark mass.  At $e^+ e^-$ colliders the top-quark mass can be
extracted by reconstructing the $t\bar{t}$ production cross section in
a scan around the production threshold. From the shape of the
differential cross section one can derive the top-quark mass in
different theoretically well-defined schemes, e.g. the
potential-subtracted (PS) top-quark mass~\cite{Beneke:1998rk}, or the
so-called $1S$ top-quark mass~\cite{Hoang:1999zc}.
In both schemes the top-quark mass can be extracted with a theoretical uncertainty $\lesssim 50$ MeV~\cite{Beneke:2015kwa,Beneke:2015lwa}, 
to be added to the projected experimental uncertainties shown in table~\ref{table:EWPOfuture}. 
The relation between the PS or $1S$ top-quark mass and the $\overline{\mathrm{MS}}$ top-quark mass has been calculated to 4 loops in 
perturbative QCD~\cite{Marquard:2015qpa}, and introduces an additional uncertainty of 
approximately $\sim 20$~MeV ($\sim 10$~MeV) in the translation from the PS ($1S$) mass. In our fits we will
assume a combined uncertainty in the top-quark mass of 50 MeV for both
the ILC and FCCee-$t\bar{t}$ scenarios. 

\begin{table}[h]
\centering
{\footnotesize
\begin{tabular}{lcccccc}
\toprule
& {Current} & {HL-LHC}  & {ILC}& \multicolumn{2}{c}{{FCCee}} & {CepC} \\
& {Data} & & & &(Run)& \\
\cmrule
${\alpha_s(M_Z)}$ &
  $ 0.1179 {\pm} 0.0012$ & % Current
  $ $ & % HLLHC
  $ $ & % ILC
  $ $& % FCC  
  $ $ % CepC
\\
${\Delta\alpha_{\rm had}^{(5)}(M_Z)}$ &
  $ 0.02750{\pm} 0.00033$ & % Current
  $ $ & % HLLHC
  $ $ & % ILC 
  $ $ && % FCC 
  $ $ % CepC
\\
${M_Z}$ [GeV] &
  $ 91.1875{\pm} 0.0021$ & % Current
  $ $ & % HLLHC
  $ $ & % ILC 
  $ {\pm} 0.0001$ &\tiny{(FCCee-${Z}$)}& % FCC 
  $ {\pm} 0.0005$ % CepC
\\
${m_t}$ [GeV] &
  $ 173.34{\pm} 0.76$ & % Current
  $ {\pm} 0.6$ & % HLLHC
  $ {\pm} 0.017$ & % ILC
  $ {\pm} 0.014$ &\tiny{(FCCee-${t\bar{t}}$)}& % FCC  
  $ $ % CepC
\\
${m_H}$ [GeV] &
  $ 125.09{\pm} 0.24$ & % Current
  $ {\pm} 0.05$ & % HLLHC
  $ {\pm} 0.015$ & % ILC 
  $ {\pm} 0.007$ & \tiny{(FCCee-${HZ}$)}& % FCC 
  $ {\pm} 0.0059$ % CepC
\\
\cmrule
${M_W}$ [GeV] &
  $ 80.385{\pm} 0.015$ & % Current
  $ {\pm} 0.011$ & % HLLHC
  $ {\pm} 0.0024$ & % ILC
  $ {\pm} 0.001$ &\tiny{(FCCee-${WW}$)}& % FCC  
  $ {\pm} 0.003$ % CepC
\\
${\Gamma_W}$ [GeV] &
  $ 2.085{\pm} 0.042$ & % Current
  $ $ & % HLLHC
  $ $ & % ILC
  $ {\pm} 0.005$ &\tiny{(FCCee-${WW}$)}& % FCC  
  $ $ % CepC
\\
${\Gamma_{Z}}$ [GeV] &
  $ 2.4952{\pm} 0.0023$ & % Current
  $ $ & % HLLHC
  $ $ & % ILC
  $ {\pm} 0.0001$ &\tiny{(FCCee-${Z}$)}& % FCC
  $ {\pm} 0.0005$ % CepC
\\
${\sigma_{h}^{0}}$ [nb] &
  $ 41.540{\pm} 0.037$ & % Current
  $ $ & % HLLHC
  $ $ & % ILC
  $ {\pm} 0.025$ &\tiny{(FCCee-${Z}$)}& % FCC
  $ $ % CepC
\\
%${\sin^2\theta_{\rm eff}^{\rm lept}(Q_{\rm FB}^{\rm had})}$ &
${\sin^2\theta_{\rm eff}^{\rm lept}}$ &
  $ 0.2324{\pm} 0.0012$ & % Current
  $ $ & % HLLHC
  $ $ & % ILC
  $ {\pm} 0.0001$ &\tiny{(FCCee-${Z}$)}& % FCC
  $ {\pm} 0.000023$ % CepC
\\
${P_\tau^{\rm pol}}$ &
  $ 0.1465{\pm} 0.0033$ & % Current
  $ $ & % HLLHC
  $ $ & % ILC
  $ {\pm} 0.0002$ &\tiny{(FCCee-${Z}$)}& % FCC
  $ $ % CepC
\\
${\mathcal{A}_\ell}$ &
  $ 0.1513{\pm} 0.0021$ & % Current
  $ $ & % HLLHC
  $ $ & % ILC
  $ {\pm} 0.000021$ &\tiny{(FCCee-${Z}$ [pol])}& % FCC
  % CepC
\\
${\mathcal{A}_{c}}$ &
  $ 0.670{\pm} 0.027$ & % Current
  $ $ & % HLLHC
  $ $ & % ILC
  $ {\pm} 0.01$ &\tiny{(FCCee-${Z}$ [pol])}& % FCC
  $ $ % CepC
\\
${\mathcal{A}_{b}}$ &
  $ 0.923{\pm} 0.020$ & % Current
  $ $ & % HLLHC
  $ $ & % ILC
  $ {\pm} 0.007$ &\tiny{(FCCee-${Z}$ [pol])}& % FCC
  $ $ % CepC
\\
${A_{\rm FB}^{0,\ell}}$ &
  $ 0.0171{\pm} 0.0010$ & % Current
  $ $ & % HLLHC
  $ $ & % ILC
  $ {\pm} 0.0001$ &\tiny{(FCCee-${Z}$)}& % FCC
  $ {\pm} 0.0010$ % CepC
\\
${A_{\rm FB}^{0,c}}$ &
  $ 0.0707{\pm} 0.0035$ & % Current
  $ $ & % HLLHC
  $ $ & % ILC
  $ {\pm} 0.0003$ &\tiny{(FCCee-${Z}$)}& % FCC
  $ $ % CepC
\\
${A_{\rm FB}^{0,b}}$ &
  $ 0.0992{\pm} 0.0016$ & % Current
  $ $ & % HLLHC
  $ $ & % ILC
  $ {\pm} 0.0001$ &\tiny{(FCCee-${Z}$)}& % FCC
  $ {\pm} 0.00014$ % CepC
\\
${R^{0}_{\ell}}$ &
  $ 20.767{\pm} 0.025$ & % Current
  $ $ & % HLLHC
  $ $ & % ILC
  $ {\pm} 0.001$ &\tiny{(FCCee-${Z}$)} & % FCC
  $ {\pm} 0.007$ % CepC
\\
${R^{0}_{c}}$ &
  $ 0.1721{\pm} 0.0030$ & % Current
  $ $ & % HLLHC
  $ $ & % ILC
  $ {\pm} 0.0003$ &\tiny{(FCCee-${Z}$)}& % FCC
  $ $ % CepC
\\
${R^{0}_{b}}$ &
  $ 0.21629{\pm} 0.00066$ & % Current
  $ $ & % HLLHC
  $ $ & % ILC
  $ {\pm} 0.00006$ &\tiny{(FCCee-${Z}$)}& % FCC
  $ {\pm} 0.00018$ % CepC
\\
\bottomrule
\end{tabular}
}
\caption{Expected experimental sensitivities to the different EWPO at
  future colliders. Apart from the improvements quoted in this table,
  we also assume that future measurements of $\Delta
  \alpha_{\mathrm{had}}^{(5)}(M_Z)$ and $\alpha_S(M_Z)$, whose errors dominate in the
  parametric uncertainties of the theoretical predictions, are possible with
  an error of approximately $\pm 5\times 10^{-5}$ and $\pm 0.0002$, respectively. 
  This assumption is particularly relevant for the FCCee and CepC fits, where the experimental precision for the bulk of electroweak precision measurements will be largely improved.
\label{table:EWPOfuture}
}
\end{table}

\begin{table}[h]
\centering
\setlength\tabcolsep{5pt}
{\scriptsize
\begin{tabular}{lccccccccccc}
\toprule
& {Current} & {HL-LHC} & \multicolumn{6}{c}{{ILC}} & {FCCee} & {CepC}\\
& & & \multicolumn{3}{c}{{Phase 1}} & \multicolumn{3}{c}{{Phase 2}} &  &  \\
& & & {250} & {500} & {1000} & {250} & {500} & {1000} &  &  \\
\cmrule
${H\rightarrow b\bar{b}}$ &
  $ {\gtrsim 23\%}$ & % Current
  $ {5\mbox{-}36\%}$ & % HLLHC
  $ {1.2\%}$ & % ILC Phase 1 250
  $ {1.8\mbox{-}28\%}$ & % ILC Phase 1 500
  $ {0.3\mbox{-}6\%}$ & % ILC Phase 1 1000
  $ {0.56\%}$ & % ILC Phase 2 250
  $ {0.37\mbox{-}16\%}$ & % ILC Phase 2 500
  $ {0.3\mbox{-}3.8\%}$ & % ILC Phase 2 1000
  $ {0.2\mbox{-}0.6\%}$ & % FCC
  $ {0.28\%}$ % CepC
\\
${H\rightarrow c\bar{c}}$ &
  $ $ & % Current
  $ $ & % HLLHC
  $ {8.3\%}$ & % ILC Phase 1 250
  $ {6.2\mbox{-}13\%}$ & % ILC Phase 1 500
  $ {3.1\%}$ & % ILC Phase 1 1000
  $ {3.9\%}$ & % ILC Phase 2 250
  $ {3.5\mbox{-}7.2\%}$ & % ILC Phase 2 500
  $ {2\%}$ & % ILC Phase 2 1000
  $ {1.2\% }$ & % FCC
  $ {2.2\%}$ % CepC
\\
${H\rightarrow gg}$ &
  $ $ & % Current
  $ $ & % HLLHC
  $ {7\%}$ & % ILC Phase 1 250
  $ {4.1\mbox{-}11\%}$ & % ILC Phase 1 500
  $ {2.3\%}$ & % ILC Phase 1 1000
  $ {3.3\%}$ & % ILC Phase 2 250
  $ {2.3\mbox{-}6\%}$ & % ILC Phase 2 500
  $ {1.4\%}$ & % ILC Phase 2 1000
  $ {1.4\%}$ & % FCC
  $ {1.6\%}$ % CepC
\\
${H\rightarrow W W}$ &
  $ {\gtrsim 15\%}$& % Current
  $ {4\mbox{-}11 \%}$& % HLLHC
  $ {6.4\%}$ & % ILC Phase 1 250
  $ {2.4\mbox{-}9.2\%}$ & % ILC Phase 1 500
  $ {1.6\%}$ & % ILC Phase 1 1000
  $ {3\%}$ & % ILC Phase 2 250
  $ {1.3\mbox{-}5.1\%}$ & % ILC Phase 2 500
  $ {1\%}$ & % ILC Phase 2 1000
  $ {0.9\%}$ & % FCC
  $ {1.5\%}$ % CepC
\\
${H\rightarrow \tau\tau}$ &
  $ {\gtrsim 25 \%}$ & % Current
  $ {5\mbox{-}15\%}$ & % HLLHC
  $ {4.2\%}$ & % ILC Phase 1 250
  $ {5.4\mbox{-}9\%}$ & % ILC Phase 1 500
  $ {3.1\%}$ & % ILC Phase 1 1000
  $ {2\%}$ & % ILC Phase 2 250
  $ {3\mbox{-}5\%}$ & % ILC Phase 2 500
  $ {2\%}$ & % ILC Phase 2 1000
  $ {0.7\%}$ &% FCC
  $ {1.2\%}$ % CepC
\\
${H\rightarrow ZZ}$ &
  $ {\gtrsim 24\%}$ & % Current
  $ {4\mbox{-}17 \%}$ & % HLLHC
  $ {19\%}$ & % ILC Phase 1 250
  $ {8.2\mbox{-}25\%}$ & % ILC Phase 1 500
  $ {4.1\%}$ & % ILC Phase 1 1000
  $ {8.8\%}$ & % ILC Phase 2 250
  $ {4.6\mbox{-}14\%}$ & % ILC Phase 2 500
  $ {2.6\%}$ & % ILC Phase 2 1000
  $ {3.1\%}$ & % FCC
  $ {4.3\%}$ % CepC
\\
${H\rightarrow \gamma\gamma}$ &
  $ {\gtrsim 20 \%}$ & % Current
  $ {4\mbox{-}28 \%}$ & % HLLHC
  $ {38\%}$ & % ILC Phase 1 250
  $ {20\mbox{-}38\%}$ & % ILC Phase 1 500
  $ {7\%}$ & % ILC Phase 1 1000
  $ {16\%}$ & % ILC Phase 2 250
  $ {13\mbox{-}19\%}$ & % ILC Phase 2 500
  $ {5.4\%}$ & % ILC Phase 2 1000
  $ {3.0\%}$ & % FCC
  $ {9\%}$ % CepC
\\
${H\rightarrow Z\gamma}$ &
  $ $ & % Current
  $ {10\mbox{-}27\%}$ & % HLLHC
  $ $ & % ILC Phase 1 250
  $ $ & % ILC Phase 1 500
  $ $ & % ILC Phase 1 1000
  $ $ & % ILC Phase 2 250
  $ $ & % ILC Phase 2 500
  $ $ & % ILC Phase 2 1000
  $ $ & % FCC
  $ $ % CepC
\\
${H\rightarrow \mu\mu}$ &
  $ $ & % Current
  $ {14\mbox{-}23\%}$ & % HLLHC
  $ {}$ & % ILC Phase 1 250
  $ {}$ & % ILC Phase 1 500
  $ {31\%}$ & % ILC Phase 1 1000
  $ $ & % ILC Phase 2 250
  $ $ & % ILC Phase 2 500
  $ {20\%}$ & % ILC Phase 2 1000
  $ 13{\%}$ & % FCC
  $ {17\%}$ % CepC
\\
\bottomrule
\end{tabular}
}
\caption{Future expected sensitivity to Higgs-boson observables at
  various future colliders considered in this study.
\label{table:Higgsfuture}}
\end{table}

\begin{table}[h]
\centering
{\footnotesize
\begin{tabular}{lcccccc}
\toprule
{}& {Current} & {Future}  & {Current} & {ILC} & {FCC-ee} & {CepC} \\
{Observable}& {Th. Error} & {Th. Error}  & {Exp. Error} & { } & { } & { } \\
\cmrule
${M_W~[\mt{MeV}]}$ &
  $ 4$ & % Current Th. Error
  $ 1$ & % Future Th. Error
  $ 15$ & % Current Exp. Error
  $ 3-4$ & % ILC
  $ 1$ & % FCC
  $ 3$ % CepC
\\
${\sin^2{\theta_{\mt{eff}}^{\mt{lept}}}~[10^{-5}]}$ &
  $ 4.5$ & % Current Th. Error
  $ 1.5$ & % Future Th. Error
  $ 16$ & % Current Exp. Error
  $ $ & % ILC
  $ 0.6$ & % FCC
  $ 2.3$ % CepC
\\
${\Gamma_Z~[\mt{MeV}]}$ &
  $ 0.5$ & % Current Th. Error
  $ 0.2$ & % Future Th. Error
  $ 2.3$ & % Current Exp. Error
  $ $ & % ILC
  $ 0.1$ & % FCC
  $ 0.5$ % CepC
\\
${R_b^0~[10^{-5}]}$ &
  $ 15$ & % Current Th. Error
  $ 10$ & % Future Th. Error
  $ 66$ & % Current Exp. Error
  $ $ & % ILC
  $ 6$ & % FCC
  $ 17$ % CepC
\\
\bottomrule
\end{tabular}
}
\caption{Projected theoretical uncertainty for the different EWPO and
  comparison with the corresponding experimental sensitivity at
  various future colliders considered in this study.
\label{table:EWPOfutureErrors}
}
\end{table}

In what follows we estimate the sensitivity to the different new
physics scenarios at the above-mentioned future experiments. To do so,
we  assume that the
future experimental measurements will be fully compatible with the SM
predictions. In particular, we use the following reference values
of the SM input parameters (see column \textit{Posterior} in
table~\ref{tab:SMfit}),
\begin{equation}
\begin{split}
m_H=125.09~\mathrm{GeV},~~m_t=173.61~\mathrm{GeV},~~M_Z=91.1879~\mathrm{GeV},\\
\alpha_s(M_Z)=0.1180~~\mbox{and}~~\Delta \alpha_{\mathrm{had}}^{(5)}(M_Z)=0.02747,~~~~~~~~~~~ 
\end{split}
\label{eq:sm-param-values}
\end{equation}
and take as errors the ones given in Tables~\ref{table:EWPOfuture} and
\ref{table:Higgsfuture}.  In our analysis we assume that the
theoretical calculations necessary to match the experimental precision
will be available, and in our fits we use the future projected
uncertainties in table~\ref{table:EWPOfutureErrors}.  To illustrate
the impact of theoretical uncertainties, we also consider
another scenario where, as in the current EWPO  fit, theoretical
uncertainties are subdominant and are neglected in the analysis.  In
this scenario we also assume that the only uncertainty affecting
the top-quark mass parameter is the one given in
table~\ref{table:EWPOfuture}.

\begin{figure}
  \centering
  \vspace{-1mm}
  \includegraphics[width=.5\textwidth]{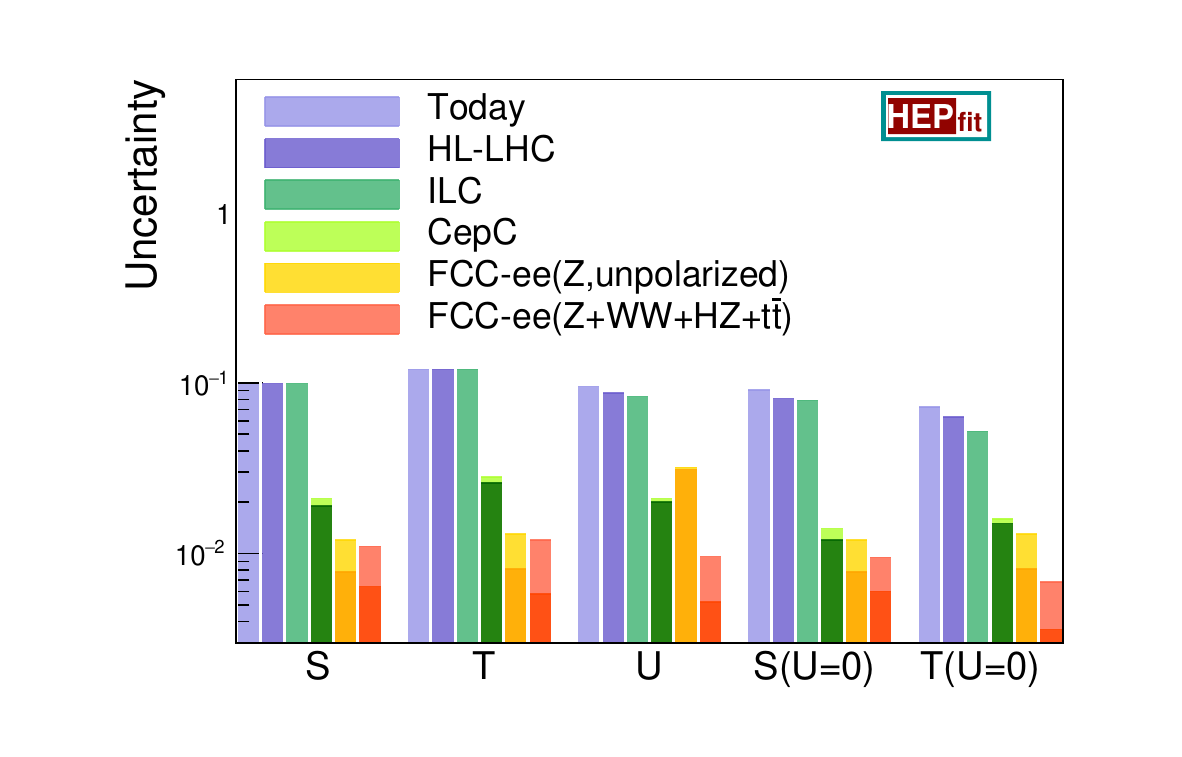} 
  \hspace{-3mm}
  \includegraphics[width=.5\textwidth]{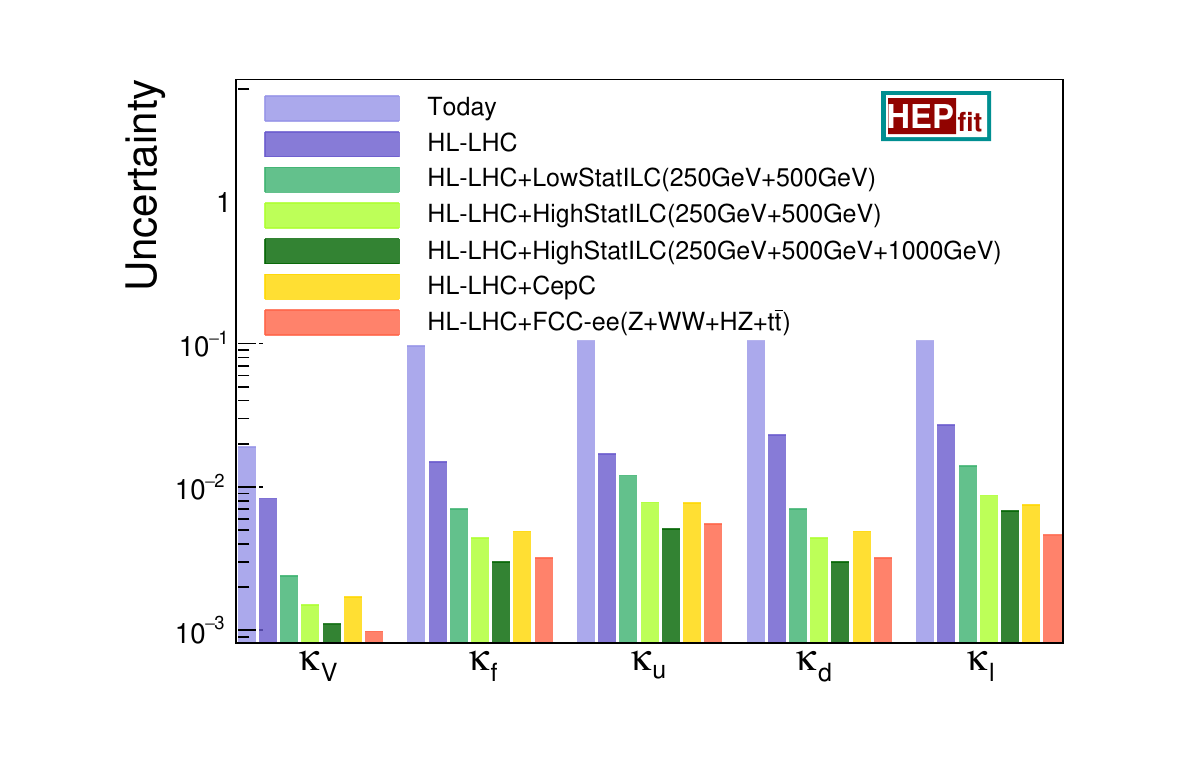}
  \caption{Comparison of expected sensitivities to oblique parameters
    (left) and modified Higgs couplings (right) from future collider
    experiments. Different shades of the same colour correspond to
    results including or neglecting the future theoretical
    uncertainties.~\label{fig:futureErrorBars}}
\end{figure}

With these settings we have performed fits to the main NP scenarios
studied in sections~\ref{sec:ew-precision-fit-BSM} and
\ref{sec:Higgs-coupl-constraints}, and compared the results with those
obtained in a fit assuming the errors of current data.\footnote{For
  consistency in the comparison, in this fit we also set the central
  values to the SM predictions summarized in
  eq.~(\ref{eq:sm-param-values}).} The results of the fits to EWPO
only are summarized in table~\ref{table_summary_EWPO}, while those
from the fits to EWPO plus Higgs-boson observables are reported in
table~\ref{table_summary_EWPOHiggs}.  In these tables we illustrate
the sensitivity to each NP parameter introduced in
sections~\ref{sec:ew-precision-fit-BSM} and
\ref{sec:Higgs-coupl-constraints} by showing the 1-$\sigma$ uncertainty 
on the corresponding parameter from the fit. A comparison of the
projected sensitivity on EW parameters and Higgs coupling constants
for various future colliders is shown in fig.~\ref{fig:futureErrorBars}.

From the results in table~\ref{table_summary_EWPO}
we observe how the FCCee, with dedicated runs aimed at improving the measurements of the different EWPO, 
offers the best performance in terms of constraints on NP. We show the results obtained with the $Z$-pole
runs, with and without polarization, and also show the effect of
adding the improved measurement of the $W$ mass ($WW$ column) as well
as the sensitivity reached after the completion of the whole FCCee
program ($t\bar{t}$ column).  Several things are apparent from this
table. The first one is that, for the NP models considered here, the
use of polarized beams at the FCCee would have only a minor impact on
the constraining power of the machine. Looking into the results for
the different models we observe how, as expected, the major
improvement in sensitivity comes from the more precise properties of
the $Z$ lineshape. After this first run, one can still achieve notable
improvements in the sensitivity to the $U$ parameter
($\delta \varepsilon_{2}$) from the measurement of $M_W$ (notice that
this is essentially the only EWPO that depends on $U$). Likewise, the
sensitivity to $\kappa_V$ can be reduced by a factor of $\sim 2$ with
the measurement of $m_t$. This can be understood from
eq.~(\ref{eq:ST}), the lower-right panel of fig.~\ref{fig:Oblique}, and
the positive correlation between $M_W$ and $m_t$.

In general, the FCCee program would improve current constraints by
about an order of magnitude.  The CepC also offers good prospects to
obtain more stringent NP constraints from EWPO. However, given the
information currently available about the machine performance, the
CepC bounds would only be a factor of approximately $4$-$5$ better
than the bounds derived from current EWPO. Notice also that the
current physics program lacks a dedicated run to improve the
measurement of the top-quark mass, which plays a significant role in
some cases as explained above. In fact, at the ILC, even without a
dedicated run at the $Z$ pole, the precise determinations of $M_W$ and
$m_t$ are enough to reach the same sensitivity to $\kappa_V$ as at the
CepC.

In table~\ref{table_summary_EWPO} we have also illustrated the effect
of the theoretical uncertainties in the results of electroweak fits
with the information from future $e^+ e^-$ colliders. In this table,
the results in the columns with grey background have been
computed using the projected theoretical uncertainties, while such
uncertainties have been neglected in the columns with white
background. As one expects from looking at
table~\ref{table:EWPOfutureErrors}, the effect of the future
theoretical uncertainties on the CepC results are mild, but they are
clearly non-negligible compared to the FCCee precision. Indeed, in the
case of the FCCee, theoretical uncertainties can reduce the
sensitivity to NP in some cases by up to a factor of $2$ compared to
cases in which the theoretical errors are subdominant.

\begin{sidewaystable}
\centering
{
\begin{tabular}{llccC{0.59cm}C{0.59cm}C{0.59cm}C{0.59cm}C{0.59cm}C{0.59cm}C{0.59cm}C{0.59cm}C{0.59cm}C{0.59cm}C{0.59cm}C{0.59cm}}
\toprule
&& {Current} & {HL-LHC} & \multicolumn{2}{c}{{ILC}}  & \multicolumn{8}{c}{{FCCee}} & \multicolumn{2}{c}{{CepC}} \\
&&  &  & & & \multicolumn{2}{c}{{${Z}$ (no pol)}} & \multicolumn{2}{c}{{${Z}$ (pol)}} & \multicolumn{2}{c}{{${WW}$}} & \multicolumn{2}{c}{{${t\bar{t}}$}} & &\\
\cmrule
${\Delta S}$\!\!\!\!&\!\!\!\!${[\times 10^{-3}]}$ &
  $ 100$ & % Current
  $ 99$& % HLLHC
\colcell   $ 99$ & % ILC
  $ 99$ & % ILC (no th unc)
\colcell $ 12$ & % FCC Z nopol
  $ 7.8$ & % FCC Z nopol (no th unc)
\colcell $ 11$ & % FCC Z pol
  $ 6.4$ & % FCC Z pol (no th unc)
\colcell $ 11$ & % FCC WW
  $ 6.4$ & % FCC WW (no th unc)
\colcell $ 11$ & % FCC tt
  $ 6.3$ & % FCC tt (no th unc)
\colcell $ 21$ & % CepC
  $ 19$ % CepC (no th unc)
\\
${\Delta T}$\!\!\!\!&\!\!\!\!${[\times 10^{-3}]}$ &
  $ 120$ & % Current
  $ 120$& % HLLHC
\colcell   $ 120$ & % ILC 
  $ 120$ & % ILC (no th unc) 
\colcell $ 13$ & % FCC Z nopol
  $ 8.1$ & % FCC Z nopol (no th unc)
\colcell $ 13$ & % FCC Z pol
  $ 7.9$ & % FCC Z pol (no th unc)
\colcell $ 13$ & % FCC WW
  $ 7.9$ & % FCC WW (no th unc)
\colcell $ 12$ & % FCC tt
  $ 5.8$ & % FCC tt (no th unc)
\colcell $ 28$ & % CepC
  $ 26$ % CepC (no th unc)
\\
${\Delta U}$\!\!\!\!&\!\!\!\!${[\times 10^{-3}]}$ &
  $ 95$ & % Current
  $ 87$& % HLLHC
\colcell   $ 83$ & % ILC  
  $ 82$ & % ILC (no th unc)
\colcell $ 32$ & % FCC Z nopol
  $ 31$ & % FCC Z nopol (no th unc)
\colcell $ 32$ & % FCC Z pol
  $ 31$ & % FCC Z pol (no th unc)
\colcell $ 9.8$ & % FCC WW
  $ 5.4$ & % FCC WW (no th unc)
\colcell $ 9.6$ & % FCC tt
  $ 5.2$ & % FCC tt (no th unc)
\colcell $ 21$ & % CepC
  $ 20$ % CepC (no th unc)
\\
\cmrule %----------------------------------------------------------------------------------------------------
${\Delta S}$\!\!\!\!&\!\!\!\!${[\times 10^{-3}]}$ &
  $ 91$ & % Current
  $ 81$& % HLLHC
\colcell   $ 79$ & % ILC  
  $ 79$ & % ILC (no th unc)
\colcell $ 12$ & % FCC Z nopol
  $ 7.8$ & % FCC Z nopol (no th unc)
\colcell $ 11$ & % FCC Z pol
  $ 6.4$ & % FCC Z pol (no th unc)
\colcell $ 9.5$ & % FCC WW
  $ 6.1$ & % FCC WW (no th unc)
\colcell $ 9.5$ & % FCC tt
  $ 6$ & % FCC tt (no th unc)
\colcell $ 14$ & % CepC
  $ 12$ % CepC (no th unc)
\\
${\Delta T}$\!\!\!\!&\!\!\!\!${[\times 10^{-3}]}$ &
  $ 72$ & % Current
  $ 63$& % HLLHC
\colcell   $ 52$ & % ILC  
  $ 52$ & % ILC (no th unc)
\colcell $ 13$ & % FCC Z nopol
  $ 8.1$ & % FCC Z nopol (no th unc)
\colcell $ 13$ & % FCC Z pol
  $ 7.9$ & % FCC Z pol (no th unc)
\colcell $ 10$ & % FCC WW
  $ 7.4$ & % FCC WW (no th unc)
\colcell $ 6.8$ & % FCC tt
  $ 3.6$ & % FCC tt (no th unc)
\colcell $ 16$ & % CepC
  $ 15$ % CepC (no th unc)
\\
\multicolumn{2}{c}{${(U=0)}$}&
  $ $ & % Current
  $ $& % HLLHC
\colcell   $ $ & % ILC  
  $ $ & % ILC (no th unc)
\colcell $ $ & % FCC Z nopol
  $ $ & % FCC Z nopol (no th unc)
\colcell $ $ & % FCC Z pol
  $ $ & % FCC Z pol (no th unc)
\colcell $ $ & % FCC WW
  $ $ & % FCC WW (no th unc)
\colcell $ $ & % FCC tt
  $ $ & % FCC tt (no th unc)
\colcell $ $ & % CepC
  $ $ % CepC (no th unc)
\\
\cmrule %----------------------------------------------------------------------------------------------------
${\Delta \varepsilon_{1}^{\mbox{\tiny NP}}}$\!\!\!\!&\!\!\!\!${[\times 10^{-5}]} $ &
  $ 96$ & % Current
  $ 96$& % HLLHC
\colcell   $ 96$ & % ILC  
  $ 95$ & % ILC (no th unc)
\colcell $ 11$ & % FCC Z nopol
  $ 7.3$ & % FCC Z nopol (no th unc)
\colcell $ 11$ & % FCC Z pol
  $ 7.2$ & % FCC Z pol (no th unc)
\colcell $ 11$ & % FCC WW
  $ 7.2$ & % FCC WW (no th unc)
\colcell $ 9.5$ & % FCC tt
  $ 4.7$ & % FCC tt (no th unc)
\colcell $ 25$ & % CepC
  $ 23$ % CepC (no th unc)
\\
${\Delta \varepsilon_2^{\mbox{\tiny NP}}}$\!\!\!\!&\!\!\!\!${[\times 10^{-5}]}$ &
  $ 86$ & % Current
  $ 81$& % HLLHC
\colcell   $ 77$ & % ILC  
  $ 76$ & % ILC (no th unc)
\colcell $ 29$ & % FCC Z nopol
  $ 28$ & % FCC Z nopol (no th unc)
\colcell $ 28$ & % FCC Z pol
  $ 28$ & % FCC Z pol (no th unc)
\colcell $ 8.6$ & % FCC WW
  $ 4.8$ & % FCC WW (no th unc)
\colcell $ 8.5$ & % FCC tt
  $ 4.7$ & % FCC tt (no th unc)
\colcell $ 21$ & % CepC
  $ 19$ % CepC (no th unc)
\\
${\Delta \varepsilon_3^{\mbox{\tiny NP}}}$\!\!\!\!&\!\!\!\!${[\times 10^{-5}]}$ &
  $ 91$ & % Current
  $ 87$& % HLLHC
\colcell   $ 88$ & % ILC  
  $ 87$ & % ILC (no th unc)
\colcell $ 9.9$ & % FCC Z nopol
  $ 6.6$ & % FCC Z nopol (no th unc)
\colcell $ 9.3$ & % FCC Z pol
  $ 5.5$ & % FCC Z pol (no th unc)
\colcell $ 9.2$ & % FCC WW
  $ 5.5$ & % FCC WW (no th unc)
\colcell $ 9.3$ & % FCC tt
  $ 5.5$ & % FCC tt (no th unc)
\colcell $ 20$ & % CepC
  $ 18$ % CepC (no th unc)
\\
${\Delta \varepsilon_b^{\mbox{\tiny NP}}}$\!\!\!\!&\!\!\!\!${[\times 10^{-5}]}$ &
  $ 130$ & % Current
  $ 130$& % HLLHC
\colcell   $ 130$ & % ILC  
  $ 130$ & % ILC (no th unc)
\colcell $ 15$ & % FCC Z nopol
  $ 12$ & % FCC Z nopol (no th unc)
\colcell $ 15$ & % FCC Z pol
  $ 12$ & % FCC Z pol (no th unc)
\colcell $ 15$ & % FCC WW
  $ 12$ & % FCC WW (no th unc)
\colcell $ 14$ & % FCC tt
  $ 11$ & % FCC tt (no th unc)
\colcell $ 41$ & % CepC
  $ 37$ % CepC (no th unc)
\\
\cmrule %----------------------------------------------------------------------------------------------------
\!\!${\Delta \delta g_{L}^b}$\!\!&\!\!\!\!${[\times 10^{-4}]}$ &
  $ 14 $ & % Current
  $ 14$& % HLLHC
\colcell   $ 14$ & % ILC  
  $ 14$ & % ILC (no th unc)
\colcell $ 1.5$ & % FCC Z nopol
  $ 1.3$ & % FCC Z nopol (no th unc)
\colcell $ 1.2$ & % FCC Z pol
  $ 1.1$ & % FCC Z pol (no th unc)
\colcell $ 1.2$ & % FCC WW
  $ 1.1$ & % FCC WW (no th unc)
\colcell $ 1.2$ & % FCC tt
  $ 1.1$ & % FCC tt (no th unc)
\colcell $ 2.4$ & % CepC
  $ 2.2$ % CepC (no th unc)
\\
\!\!${\Delta \delta g_{R}^b}$\!\!&\!\!\!\!${[\times 10^{-4}]}$ &
  $ 72 $ & % Current
  $ 70$& % HLLHC
\colcell   $ 70$ & % ILC  
  $ 70$ & % ILC (no th unc)
\colcell $ 7.1$ & % FCC Z nopol
  $ 6.6$ & % FCC Z nopol (no th unc)
\colcell $ 5.3$ & % FCC Z pol
  $ 5.3$ & % FCC Z pol (no th unc)
\colcell $ 5.3$ & % FCC WW
  $ 5.3$ & % FCC WW (no th unc)
\colcell $ 5.3$ & % FCC tt
  $ 5.3$ & % FCC tt (no th unc)
\colcell $ 8.9$ & % CepC
  $ 8.6$ % CepC (no th unc)
\\
\cmrule %----------------------------------------------------------------------------------------------------
${\Delta \kappa_{V}}$\!\!\!\!&\!\!\!\!${[\times 10^{-3}]}$ &
  $ 22$ & % Current
  $ 14$& % HLLHC
\colcell   $ 4.5$ & % ILC  
  $ 4.4$ & % ILC (no th unc)
\colcell $ 4.6$ & % FCC Z nopol
  $ 3.9$ & % FCC Z nopol (no th unc)
\colcell $ 4.4$ & % FCC Z pol
  $ 3.7$ & % FCC Z pol (no th unc)
\colcell $ 4.1$ & % FCC WW
  $ 3.7$ & % FCC WW (no th unc)
\colcell $ 1.8$ & % FCC tt
  $ 1.3$ & % FCC tt (no th unc)
\colcell $ 5$ & % CepC
  $ 4.7$ % CepC (no th unc)
\\
\bottomrule
\end{tabular}
}
\caption{Comparison of the current and expected sensitivities,
  $\Delta$, to the different NP scenarios at the future colliders
  considered in this study. In this table, the future projections for
  the sensitivity to $\kappa_V$ has been computed considering only the
  improvements in EWPO. (See table~\ref{table_summary_EWPOHiggs} for
  the projections using EWPO and Higgs-boson observables.) For the
  case of future lepton colliders we quote results that also include the
  expected future theoretical errors given in
  table~\ref{table:EWPOfutureErrors} (dark background), as well as
  results in which the theoretical errors have been neglected (white
  background). \label{table_summary_EWPO}}
\end{sidewaystable}

\begin{sidewaystable}
\centering
{
\begin{tabular}{llccC{1.5cm}C{1.5cm}C{1.5cm}C{1.5cm}C{1.5cm}}
\toprule
&& {Current} & {HL-LHC} & {ILC${{\tiny\begin{array}{c}250\\500\end{array}}}$} & {ILC${{\tiny\begin{array}{c}250\\500\end{array}}}$} & {ILC\!${{\tiny\begin{array}{c}250\\500\\1000\end{array}}}$} & {FCCee} & {CepC} \\
&& & & {\scriptsize Low stats}& {\scriptsize High stats} & {\scriptsize High stats} & & \\
\cmrule
${\Delta \kappa_{V}}$\!\!\!\!&\!\!\!\!${[\times 10^{-4}]}$ &
  $  190$ & % Current
  $ 83$& % HLLHC
  $ 24$ & % ILC 250/500 LS
  $ 15$ & % ILC 250/500 HS
  $ 11$ & % ILC 250/500/1000 HS
  $ 9.7~(8.6)$ & % FCC
  $ 17$ % CepC
\\
${\Delta \kappa_{f}}$\!\!\!\!&\!\!\!\!${[\times 10^{-4}]}$ &
  $  960$ & % Current
  $ 150$& % HLLHC
  $ 70$ & % ILC 250/500 LS
  $ 44$ & % ILC 250/500 HS
  $ 30$ & % ILC 250/500/1000 HS  
  $ 31$ & % FCC
  $ 49$ % CepC
\\
\cmrule %----------------------------------------------------------------------------------------------------
${\Delta \kappa_{W}}$\!\!\!\!&\!\!\!\!${[\times 10^{-4}]}$ &
  $ 510$ & % Current
  $ 120$ & % HLLHC
  $ 33$ & % ILC 250/500 LS
  $ 19$ & % ILC 250/500 HS
  $ 12$ & % ILC 250/500/1000 HS 
  $ 29$ & % FCC
  $ 100$ % CepC
\\
${\Delta \kappa_{Z}}$\!\!\!\!&\!\!\!\!${[\times 10^{-4}]}$ &
  $  1100$ & % Current
  $  150$& % HLLHC
  $ 46$ & % ILC 250/500 LS
  $ 25$ & % ILC 250/500 HS
  $ 23$ & % ILC 250/500/1000 HS 
  $ 12$ & % FCC
  $ 18$ % CepC
\\
${\Delta \kappa_{f}}$\!\!\!\!&\!\!\!\!${[\times 10^{-4}]}$ &
  $ 1100$ & % Current
  $ 160$& % HLLHC
  $ 71$ & % ILC 250/500 LS
  $ 45$ & % ILC 250/500 HS
  $ 30$ & % ILC 250/500/1000 HS 
  $ 39$ & % FCC
  $ 100$ % CepC
\\
\cmrule %----------------------------------------------------------------------------------------------------
${\Delta \kappa_{V}}$\!\!\!\!&\!\!\!\!${[\times 10^{-4}]}$ &
  $ 210$ & % Current
  $ 120$& % HLLHC
  $ 25~(24)$ & % ILC 250/500 LS
  $ 16$ & % ILC 250/500 HS
  $ 12$ & % ILC 250/500/1000 HS  
  $ 10~(8.8)$ & % FCC
  $ 17$ % CepC
\\
${\Delta \kappa_{u}}$\!\!\!\!&\!\!\!\!${[\times 10^{-4}]}$ &
  $  1200$ & % Current
  $ 170$& % HLLHC
  $ 120$ & % ILC 250/500 LS
  $ 78$ & % ILC 250/500 HS
  $ 51$ & % ILC 250/500/1000 HS  
  $ 55$ & % FCC
  $ 77$ % CepC
\\
${\Delta \kappa_{d}}$\!\!\!\!&\!\!\!\!${[\times 10^{-4}]}$ &
  $  1400$ & % Current
  $ 230$& % HLLHC
  $ 70$ & % ILC 250/500 LS
  $ 44$ & % ILC 250/500 HS
  $ 30$ & % ILC 250/500/1000 HS 
  $ 32 (31)$ & % FCC
  $ 49$ % CepC
\\
${\Delta \kappa_{\ell}}$\!\!\!\!&\!\!\!\!${[\times 10^{-4}]}$ &
  $  1300$ & % Current
  $ 270$& % HLLHC
  $ 140$ & % ILC 250/500 LS
  $ 87$ & % ILC 250/500 HS
  $ 68$ & % ILC 250/500/1000 HS
  $ 46$ & % FCC
  $ 75$ % CepC
\\
\bottomrule
\end{tabular}
}
\caption{ Comparison of the current and expected sensitivities,
  $\Delta$, to modified Higgs-boson couplings at the future colliders
  considered in this study. Results have been obtained using the
  improvements in both the Higgs-boson signal strengths and the EWPO
  (for $\kappa_V$). The latter are only included in the scenarios with
  $\kappa_W=\kappa_Z=\kappa_V$. The EWPO used in this table have been
  computed including the projected future theoretical uncertainties in
  the precision observables. Neglecting these only seems to have a
  noticeable effect on the FCCee results, as this is the only case
  where the EWPO determination of $\kappa_V$ is precise enough to have
  an impact in the combination with Higgs-boson data. In this case we
  quote in parenthesis the results obtained assuming subdominant
  theoretical uncertainties in the EWPO.  The projections for the
  sensitivities to Higgs-boson observables at future lepton colliders
  (ILC, FCCee and CepC) also include the information from the expected
  improvement in the Higgs-boson signal strengths at the
  HL-LHC. \label{table_summary_EWPOHiggs}}
\end{sidewaystable}

Finally, in table~\ref{table_summary_EWPOHiggs}, we show the level of
sensitivity to modified Higgs-boson couplings achievable at the
various future colliders considered in this study, in the different
scenarios explained in section~\ref{sec:Higgs-coupl-constraints}.  In
this case, let us emphasize however that, even before any future
lepton collider, the HL-LHC will provide much better determinations of
the Higgs-boson properties compared to what has been so far obtained
with current data. Using the fermion-universal custodial-symmetric
scenario as a reference, i.e. $\kappa_W=\kappa_Z\equiv \kappa_V$ and
$\kappa_u=\kappa_d=\kappa_\ell\equiv \kappa_f$, the HL-LHC would be
twice as sensitive to deviations from $\kappa_V=1$, and up to $6$-$7$
times as sensitive to deviations from $\kappa_f=1$. These results
would be further improved at lepton colliders by a factor of 9 (5) for
$\kappa_V$ ($\kappa_f$). The much larger gain in sensitivity for
$\kappa_f$ than for $\kappa_V$ can be understood by noticing that the
measurement of Higg-boson couplings to vector bosons will be
systematic dominated within the current LHC program, while the
measurement of Higgs-boson couplings to fermions will need the full
HL-LHC luminosity for the systematic uncertainty to be comparable to
the statistical one.  Focusing on the results obtained at the
different lepton colliders we observe how, assuming custodial
symmetry, the FCCee would offer a somewhat better performance than
CepC in terms of measuring the Higgs-boson couplings to both vector
bosons and fermions (in part because of the more precise determination
of $\kappa_V$ via EWPO).  At the ILC the results indicate that, again
assuming custodial symmetry, the initial phase would not be enough to
match the FCCee or CepC precision. Matching the CepC would be possible
after a luminosity upgrade even in the absence of a dedicated run at
$\sqrt{s}=1$ TeV. Including such a run in the physics program would
make the ILC performances comparable to the FCCee physics reach for
this scenario. For the scenario with $\kappa_Z\not =\kappa_W$ we
observe that, while the CepC Higgs-boson run will only explore
center-of-mass energies $\sqrt{s}\approx 240$ GeV, where Higgs-boson
production occurs mostly via $ZH$ associated production, running at
the FCCee with $\sqrt{s}=350$~GeV or at the ILC with
$\sqrt{s}=500$~GeV or $\sqrt{s}=1000$~GeV gives also access to
$W$-boson fusion production (as well as $t\bar{t}H$ associated
production in the ILC case). This results in a FCCee (ILC)
determination of $\kappa_W$ approximately 3 (10) times more precise
than at the CepC.

%------------------------------------------------------------------------------------------------------------
\FloatBarrier

\section{Conclusions}
\label{sec:conclusions}

With the discovery of a SM-like Higgs boson during Run I of the LHC,
the possibility of using EW and Higgs-boson precision measurements as
a portal to NP has become a reality. Through the steady
improvement of both theoretical and experimental accuracies,
electroweak and Higgs-boson precision physics could lead us a long way
towards determining the UV completion of the SM and the more
fundamental origin of the spontaneously-broken realization of the
electroweak symmetry.

Indirect searches for NP are indeed as important as ever in
the physics program of Run II of the LHC: they will probe physics at
inaccessible high scales and provide clues on the nature of new
particles.  In this context, it is very valuable, if not essential, to
provide a complete and consistent framework in which all available
experimental data, from precision measurements of electroweak
observables and Higgs-boson couplings to flavour-physics results, can be
analyzed to constrain the theory in a statistically significant
way. 

The study presented in this paper illustrates how this can be achieved
in the context of the {\tt HEPfit} package, and provides results for a
global fit of EWPO and Higgs-boson signal-strength measurements
obtained from LHC Run I data collected at 7 and 8 TeV.  

At the moment, the constraints derived for Higgs-boson couplings to SM gauge bosons
and fermions are overall compatible with SM predictions within the
current accuracy. From the results of
section~\ref{sec:Higgs-coupl-constraints} we see that the combined study
of EWPO and Higgs-boson observables can provide more stringent
constraints on Higgs-boson couplings. We can foresee that the higher
statistics expected in Run II of the LHC will offer the possibility to
isolate potential NP effects from global fits of SM precision
observables.  This will become even more crucial at the HL-LHC and at
a future generation of $e^+e^-$ colliders (FCCee, ILC, CepC) where
very high experimental precision for EWPO and Higgs-boson couplings will be
achievable. We have dedicated a section of this paper to a
 study of the sensitivity of different future experimental facilities 
to NP effects, and have determined at what point more accurate
theoretical predictions will be needed (see also ref.~\cite{deBlas:2016nqo}).

Finally, we notice that deviations from the SM predictions of EWPO and Higgs-boson
couplings constitute indirect evidence of new physics that still
need to be interpreted in terms of specific physical degrees of freedom.
A more refined theoretical approach, which entails a
generalization of the SM Lagrangian to systematically include
all effective interactions generated by the presence of NP at the UV scale,
will then be necessary in order to explore the nature of such
deviations. In a following paper~\cite{hepfit:2015daje} we
will explore the possibility of using an effective field theory
approach to build a model-independent study of NP effects in Higgs-boson
couplings and use the {\tt HEPfit} framework to combine it with a fit
of all available electroweak precision data.

%------------------------------------------------------------------------------------------------------------

\acknowledgments

We thank P. Janot for pointing out an omission in the list of FCCee
measurements taken as input. M.~C. is associated to the Dipartimento
di Matematica e Fisica, Universit\`a di Roma Tre, and E.~F. and
L.~S. are associated to the Dipartimento di Fisica, Universit\`a di
Roma ``La Sapienza''.  The research leading to these results has
received funding from the European Research Council under the European
Union's Seventh Framework Programme (FP/2007-2013) / grants n.~267985
and n.~279972.  The work of L.~R. is supported in part by the
U.S. Department of Energy under grant DE-SC0010102, and by the
National Science Foundation under Grant No. NSF PHY11-25915.
L.~R. would like to thanks the Kavli Institute for Theoretical Physics
(UCSB) for hospitality while this work was being completed.
%------------------------------------------------------------------------------------------------------------

\bibliographystyle{JHEP}
\bibliography{hepfit}

\providecommand{\href}[2]{#2}\begingroup\raggedright\begin{thebibliography}{100}

\bibitem{hepfit}
{\tt HEPfit} Collaboration, \textit{{\tt HEPfit}: a Code for the Combination of
  Indirect and Direct Constraints on High Energy Physics Models}, in
  preparation.

\bibitem{hepfitsite}
{\tt HEPfit} Collaboration, {\url{http://hepfit.roma1.infn.it}}.

\bibitem{Ciuchini:2013pca}
M.~Ciuchini, E.~Franco, S.~Mishima, and L.~Silvestrini, {\it {Electroweak
  Precision Observables, New Physics and the Nature of a 126 GeV Higgs Boson}},
   {\em JHEP} {\bf 08} (2013) 106, [\href{http://arxiv.org/abs/1306.4644}{{\tt
  arXiv:1306.4644}}].

\bibitem{deBlas:2014ula}
J.~de~Blas, M.~Ciuchini, E.~Franco, D.~Ghosh, S.~Mishima, M.~Pierini, L.~Reina,
  and L.~Silvestrini, {\it {Global Bayesian Analysis of the Higgs-boson
  Couplings}},  in {\em {International Conference on High Energy Physics 2014
  (ICHEP 2014) Valencia, Spain, July 2-9, 2014}}.
\newblock \href{http://arxiv.org/abs/1410.4204}{{\tt arXiv:1410.4204}}.

\bibitem{Ciuchini:2014dea}
M.~Ciuchini, E.~Franco, S.~Mishima, M.~Pierini, L.~Reina, and L.~Silvestrini,
  {\it {Update of the electroweak precision fit, interplay with Higgs-boson
  signal strengths and model-independent constraints on new physics}},  in {\em
  {International Conference on High Energy Physics 2014 (ICHEP 2014) Valencia,
  Spain, July 2-9, 2014}}.
\newblock \href{http://arxiv.org/abs/1410.6940}{{\tt arXiv:1410.6940}}.

\bibitem{Reina:2015yuh}
L.~Reina, J.~de~Blas, M.~Ciuchini, E.~Franco, D.~Ghosh, S.~Mishima, M.~Pierini,
  and L.~Silvestrini, {\it {Precision constraints on non-standard Higgs-boson
  couplings with HEPfit}},  {\em PoS} {\bf EPS-HEP2015} (2015) 187.

\bibitem{hepfit:LP15}
J.~de~Blas, M.~Ciuchini, E.~Franco, D.~Ghosh, S.~Mishima, M.~Pierini, L.~Reina,
  and L.~Silvestrini, {\it Updates on fits to electroweak parameters},  in {\em
  27th International Symposium on Lepton Photon Interactions at High Energy
  (LP15), Ljubljana, Slovenia, August 17-22, 2015}.
\newblock To appear in the Proceedings.

\bibitem{hepfit:2015daje}
J.~de~Blas, M.~Ciuchini, E.~Franco, D.~Ghosh, S.~Mishima, M.~Pierini, L.~Reina,
  and L.~Silvestrini.
\newblock {in preparation}.

\bibitem{Baak:2014ora}
{\bf Gfitter Group} Collaboration, M.~Baak, J.~C{\'u}th, J.~Haller, A.~Hoecker,
  R.~Kogler, K.~M{\"o}nig, M.~Schott, and J.~Stelzer, {\it {The global
  electroweak fit at NNLO and prospects for the LHC and ILC}},  {\em Eur. Phys.
  J.} {\bf C74} (2014) 3046, [\href{http://arxiv.org/abs/1407.3792}{{\tt
  arXiv:1407.3792}}].

\bibitem{Agashe:2014kda}
{\bf Particle Data Group} Collaboration, K.~A. Olive et~al., {\it {Review of
  Particle Physics}},  {\em Chin. Phys.} {\bf C38} (2014) 090001 and 2015
  update.

\bibitem{Caldwell:2008fw}
A.~Caldwell, D.~Kollar, and K.~Kroninger, {\it {BAT: The Bayesian Analysis
  Toolkit}},  {\em Comput. Phys. Commun.} {\bf 180} (2009) 2197--2209,
  [\href{http://arxiv.org/abs/0808.2552}{{\tt arXiv:0808.2552}}].

\bibitem{Akhundov:2013ons}
A.~Akhundov, A.~Arbuzov, S.~Riemann, and T.~Riemann, {\it {The ZFITTER
  project}},  {\em Phys. Part. Nucl.} {\bf 45} (2014), no.~3 529--549,
  [\href{http://arxiv.org/abs/1302.1395}{{\tt arXiv:1302.1395}}].

\bibitem{Burkhardt:2011ur}
H.~Burkhardt and B.~Pietrzyk, {\it {Recent BES measurements and the hadronic
  contribution to the QED vacuum polarization}},  {\em Phys. Rev.} {\bf D84}
  (2011) 037502, [\href{http://arxiv.org/abs/1106.2991}{{\tt
  arXiv:1106.2991}}].

\bibitem{ALEPH:2005ab}
{\bf SLD Electroweak Group, DELPHI, ALEPH, SLD, SLD Heavy Flavour Group, OPAL,
  LEP Electroweak Working Group, L3} Collaboration, S.~Schael et~al., {\it
  {Precision electroweak measurements on the $Z$ resonance}},  {\em Phys.
  Rept.} {\bf 427} (2006) 257--454,
  [\href{http://arxiv.org/abs/hep-ex/0509008}{{\tt hep-ex/0509008}}].

\bibitem{ATLAS:2014wva}
{\bf ATLAS, CDF, CMS, D0} Collaboration, {\it {First combination of Tevatron
  and LHC measurements of the top-quark mass}},
  \href{http://arxiv.org/abs/1403.4427}{{\tt arXiv:1403.4427}}.

\bibitem{Aad:2015zhl}
{\bf ATLAS, CMS} Collaboration, G.~Aad et~al., {\it {Combined Measurement of
  the Higgs Boson Mass in $pp$ Collisions at $\sqrt{s}=7$ and 8 TeV with the
  ATLAS and CMS Experiments}},  {\em Phys. Rev. Lett.} {\bf 114} (2015) 191803,
  [\href{http://arxiv.org/abs/1503.07589}{{\tt arXiv:1503.07589}}].

\bibitem{Group:2012gb}
{\bf CDF, D0} Collaboration, T.~E.~W. Group, {\it {2012 Update of the
  Combination of CDF and D0 Results for the Mass of the W Boson}},
  \href{http://arxiv.org/abs/1204.0042}{{\tt arXiv:1204.0042}}.

\bibitem{ALEPH:2010aa}
{\bf Tevatron Electroweak Working Group, CDF, DELPHI, SLD Electroweak and Heavy
  Flavour Groups, ALEPH, LEP Electroweak Working Group, SLD, OPAL, D0, L3}
  Collaboration, L.~E.~W. Group, {\it {Precision Electroweak Measurements and
  Constraints on the Standard Model}},
  \href{http://arxiv.org/abs/1012.2367}{{\tt arXiv:1012.2367}}.

\bibitem{Aaltonen:2016nuy}
{\bf CDF} Collaboration, T.~A. Aaltonen et~al., {\it {Measurement of
  $\sin^2\theta^{\rm lept}_{\rm eff}$ using $e^+e^-$ pairs from $\gamma^*/Z$
  bosons produced in $p\bar{p}$ collisions at a center-of-momentum energy of
  1.96 TeV}},  \href{http://arxiv.org/abs/1605.02719}{{\tt arXiv:1605.02719}}.

\bibitem{Aaltonen:2014loa}
{\bf CDF} Collaboration, T.~A. Aaltonen et~al., {\it {Indirect measurement of
  $\sin^2 \theta_W$ (or $M_W$) using $\mu^+\mu^-$ pairs from $\gamma^*/Z$
  bosons produced in $p\bar{p}$ collisions at a center-of-momentum energy of
  1.96 TeV}},  {\em Phys. Rev.} {\bf D89} (2014), no.~7 072005,
  [\href{http://arxiv.org/abs/1402.2239}{{\tt arXiv:1402.2239}}].

\bibitem{Abazov:2014jti}
{\bf D0} Collaboration, V.~M. Abazov et~al., {\it {Measurement of the effective
  weak mixing angle in $p\bar{p}\rightarrow Z/\gamma^{*}\rightarrow e^{+}e^{-}$
  events}},  {\em Phys. Rev. Lett.} {\bf 115} (2015), no.~4 041801,
  [\href{http://arxiv.org/abs/1408.5016}{{\tt arXiv:1408.5016}}].

\bibitem{Aad:2015uau}
{\bf ATLAS} Collaboration, G.~Aad et~al., {\it {Measurement of the
  forward-backward asymmetry of electron and muon pair-production in $pp$
  collisions at $\sqrt{s}$ = 7 TeV with the ATLAS detector}},  {\em JHEP} {\bf
  09} (2015) 049, [\href{http://arxiv.org/abs/1503.03709}{{\tt
  arXiv:1503.03709}}].

\bibitem{Chatrchyan:2011ya}
{\bf CMS} Collaboration, S.~Chatrchyan et~al., {\it {Measurement of the weak
  mixing angle with the Drell-Yan process in proton-proton collisions at the
  LHC}},  {\em Phys. Rev.} {\bf D84} (2011) 112002,
  [\href{http://arxiv.org/abs/1110.2682}{{\tt arXiv:1110.2682}}].

\bibitem{Aaij:2015lka}
{\bf LHCb} Collaboration, R.~Aaij et~al., {\it {Measurement of the
  forward-backward asymmetry in $Z/\gamma^{\ast} \rightarrow \mu^{+}\mu^{-}$
  decays and determination of the effective weak mixing angle}},  {\em JHEP}
  {\bf 11} (2015) 190, [\href{http://arxiv.org/abs/1509.07645}{{\tt
  arXiv:1509.07645}}].

\bibitem{Freitas:2014hra}
A.~Freitas, {\it {Higher-order electroweak corrections to the partial widths
  and branching ratios of the Z boson}},  {\em JHEP} {\bf 04} (2014) 070,
  [\href{http://arxiv.org/abs/1401.2447}{{\tt arXiv:1401.2447}}].

\bibitem{Schroder:2005db}
Y.~Schroder and M.~Steinhauser, {\it {Four-loop singlet contribution to the rho
  parameter}},  {\em Phys. Lett.} {\bf B622} (2005) 124--130,
  [\href{http://arxiv.org/abs/hep-ph/0504055}{{\tt hep-ph/0504055}}].

\bibitem{Chetyrkin:2006bj}
K.~G. Chetyrkin, M.~Faisst, J.~H. Kuhn, P.~Maierhofer, and C.~Sturm, {\it
  {Four-Loop QCD Corrections to the Rho Parameter}},  {\em Phys. Rev. Lett.}
  {\bf 97} (2006) 102003, [\href{http://arxiv.org/abs/hep-ph/0605201}{{\tt
  hep-ph/0605201}}].

\bibitem{Boughezal:2006xk}
R.~Boughezal and M.~Czakon, {\it {Single scale tadpoles and O($G_F m_t^2
  \alpha_s^3$) corrections to the $\rho$ parameter}},  {\em Nucl. Phys.} {\bf
  B755} (2006) 221--238, [\href{http://arxiv.org/abs/hep-ph/0606232}{{\tt
  hep-ph/0606232}}].

\bibitem{Awramik:2003rn}
M.~Awramik, M.~Czakon, A.~Freitas, and G.~Weiglein, {\it {Precise prediction
  for the W boson mass in the standard model}},  {\em Phys. Rev.} {\bf D69}
  (2004) 053006, [\href{http://arxiv.org/abs/hep-ph/0311148}{{\tt
  hep-ph/0311148}}].

\bibitem{Chatrchyan:2013haa}
{\bf CMS} Collaboration, S.~Chatrchyan et~al., {\it {Determination of the
  top-quark pole mass and strong coupling constant from the t t-bar production
  cross section in pp collisions at $\sqrt{s}$ = 7 TeV}},  {\em Phys. Lett.}
  {\bf B728} (2014) 496--517, [\href{http://arxiv.org/abs/1307.1907}{{\tt
  arXiv:1307.1907}}]. [Erratum: Phys. Lett.B738,526(2014)].

\bibitem{Aoki:2016frl}
S.~Aoki et~al., {\it {Review of lattice results concerning low-energy particle
  physics}},  \href{http://arxiv.org/abs/1607.00299}{{\tt arXiv:1607.00299}}.

\bibitem{Peskin:1990zt}
M.~E. Peskin and T.~Takeuchi, {\it {A New constraint on a strongly interacting
  Higgs sector}},  {\em Phys. Rev. Lett.} {\bf 65} (1990) 964--967.

\bibitem{Peskin:1991sw}
M.~E. Peskin and T.~Takeuchi, {\it {Estimation of oblique electroweak
  corrections}},  {\em Phys. Rev.} {\bf D46} (1992) 381--409.

\bibitem{Altarelli:1990zd}
G.~Altarelli and R.~Barbieri, {\it {Vacuum polarization effects of new physics
  on electroweak processes}},  {\em Phys. Lett.} {\bf B253} (1991) 161--167.

\bibitem{Altarelli:1991fk}
G.~Altarelli, R.~Barbieri, and S.~Jadach, {\it {Toward a model independent
  analysis of electroweak data}},  {\em Nucl. Phys.} {\bf B369} (1992) 3--32.
  [Erratum: Nucl. Phys.B376,444(1992)].

\bibitem{Altarelli:1993sz}
G.~Altarelli, R.~Barbieri, and F.~Caravaglios, {\it {Nonstandard analysis of
  electroweak precision data}},  {\em Nucl. Phys.} {\bf B405} (1993) 3--23.

\bibitem{Barbieri:2004qk}
R.~Barbieri, A.~Pomarol, R.~Rattazzi, and A.~Strumia, {\it {Electroweak
  symmetry breaking after LEP-1 and LEP-2}},  {\em Nucl. Phys.} {\bf B703}
  (2004) 127--146, [\href{http://arxiv.org/abs/hep-ph/0405040}{{\tt
  hep-ph/0405040}}].

\bibitem{Choudhury:2001hs}
D.~Choudhury, T.~M.~P. Tait, and C.~E.~M. Wagner, {\it {Beautiful mirrors and
  precision electroweak data}},  {\em Phys. Rev.} {\bf D65} (2002) 053002,
  [\href{http://arxiv.org/abs/hep-ph/0109097}{{\tt hep-ph/0109097}}].

\bibitem{Grojean:2013qca}
C.~Grojean, O.~Matsedonskyi, and G.~Panico, {\it {Light top partners and
  precision physics}},  {\em JHEP} {\bf 10} (2013) 160,
  [\href{http://arxiv.org/abs/1306.4655}{{\tt arXiv:1306.4655}}].

\bibitem{Ghosh:2015wiz}
D.~Ghosh, M.~Salvarezza, and F.~Senia, {\it {Extending the Analysis of
  Electroweak Precision Constraints in Composite Higgs Models}},
  \href{http://arxiv.org/abs/1511.08235}{{\tt arXiv:1511.08235}}.

\bibitem{Aad:2014eha}
{\bf ATLAS} Collaboration, G.~Aad et~al., {\it {Measurement of Higgs boson
  production in the diphoton decay channel in pp collisions at center-of-mass
  energies of 7 and 8 TeV with the ATLAS detector}},  {\em Phys. Rev.} {\bf
  D90} (2014), no.~11 112015, [\href{http://arxiv.org/abs/1408.7084}{{\tt
  arXiv:1408.7084}}].

\bibitem{Khachatryan:2014ira}
{\bf CMS} Collaboration, V.~Khachatryan et~al., {\it {Observation of the
  diphoton decay of the Higgs boson and measurement of its properties}},  {\em
  Eur. Phys. J.} {\bf C74} (2014), no.~10 3076,
  [\href{http://arxiv.org/abs/1407.0558}{{\tt arXiv:1407.0558}}].

\bibitem{Aad:2015vsa}
{\bf ATLAS} Collaboration, G.~Aad et~al., {\it {Evidence for the Higgs-boson
  Yukawa coupling to tau leptons with the ATLAS detector}},  {\em JHEP} {\bf
  04} (2015) 117, [\href{http://arxiv.org/abs/1501.04943}{{\tt
  arXiv:1501.04943}}].

\bibitem{Chatrchyan:2014nva}
{\bf CMS} Collaboration, S.~Chatrchyan et~al., {\it {Evidence for the 125 GeV
  Higgs boson decaying to a pair of $\tau$ leptons}},  {\em JHEP} {\bf 05}
  (2014) 104, [\href{http://arxiv.org/abs/1401.5041}{{\tt arXiv:1401.5041}}].

\bibitem{Aad:2014eva}
{\bf ATLAS} Collaboration, G.~Aad et~al., {\it {Measurements of Higgs boson
  production and couplings in the four-lepton channel in pp collisions at
  center-of-mass energies of 7 and 8 TeV with the ATLAS detector}},  {\em Phys.
  Rev.} {\bf D91} (2015), no.~1 012006,
  [\href{http://arxiv.org/abs/1408.5191}{{\tt arXiv:1408.5191}}].

\bibitem{Chatrchyan:2013mxa}
{\bf CMS} Collaboration, S.~Chatrchyan et~al., {\it {Measurement of the
  properties of a Higgs boson in the four-lepton final state}},  {\em Phys.
  Rev.} {\bf D89} (2014), no.~9 092007,
  [\href{http://arxiv.org/abs/1312.5353}{{\tt arXiv:1312.5353}}].

\bibitem{Khachatryan:2014jba}
{\bf CMS} Collaboration, V.~Khachatryan et~al., {\it {Precise determination of
  the mass of the Higgs boson and tests of compatibility of its couplings with
  the standard model predictions using proton collisions at 7 and 8 TeV}},
  {\em Eur. Phys. J.} {\bf C75} (2015), no.~5 212,
  [\href{http://arxiv.org/abs/1412.8662}{{\tt arXiv:1412.8662}}].

\bibitem{ATLAS:2014aga}
{\bf ATLAS} Collaboration, G.~Aad et~al., {\it {Observation and measurement of
  Higgs boson decays to WW$^*$ with the ATLAS detector}},  {\em Phys. Rev.}
  {\bf D92} (2015), no.~1 012006, [\href{http://arxiv.org/abs/1412.2641}{{\tt
  arXiv:1412.2641}}].

\bibitem{Aad:2015ona}
{\bf ATLAS} Collaboration, G.~Aad et~al., {\it {Study of (W/Z)H production and
  Higgs boson couplings using $H \rightarrow WW^{\ast}$ decays with the ATLAS
  detector}},  {\em JHEP} {\bf 08} (2015) 137,
  [\href{http://arxiv.org/abs/1506.06641}{{\tt arXiv:1506.06641}}].

\bibitem{Chatrchyan:2013iaa}
{\bf CMS} Collaboration, S.~Chatrchyan et~al., {\it {Measurement of Higgs boson
  production and properties in the WW decay channel with leptonic final
  states}},  {\em JHEP} {\bf 01} (2014) 096,
  [\href{http://arxiv.org/abs/1312.1129}{{\tt arXiv:1312.1129}}].

\bibitem{Aad:2014xzb}
{\bf ATLAS} Collaboration, G.~Aad et~al., {\it {Search for the $b\bar{b}$ decay
  of the Standard Model Higgs boson in associated $(W/Z)H$ production with the
  ATLAS detector}},  {\em JHEP} {\bf 01} (2015) 069,
  [\href{http://arxiv.org/abs/1409.6212}{{\tt arXiv:1409.6212}}].

\bibitem{Aad:2015gra}
{\bf ATLAS} Collaboration, G.~Aad et~al., {\it {Search for the Standard Model
  Higgs boson produced in association with top quarks and decaying into
  $b\bar{b}$ in pp collisions at $\sqrt{s}$ = 8 TeV with the ATLAS detector}},
  {\em Eur. Phys. J.} {\bf C75} (2015), no.~7 349,
  [\href{http://arxiv.org/abs/1503.05066}{{\tt arXiv:1503.05066}}].

\bibitem{Chatrchyan:2013zna}
{\bf CMS} Collaboration, S.~Chatrchyan et~al., {\it {Search for the standard
  model Higgs boson produced in association with a W or a Z boson and decaying
  to bottom quarks}},  {\em Phys. Rev.} {\bf D89} (2014), no.~1 012003,
  [\href{http://arxiv.org/abs/1310.3687}{{\tt arXiv:1310.3687}}].

\bibitem{Khachatryan:2014qaa}
{\bf CMS} Collaboration, V.~Khachatryan et~al., {\it {Search for the associated
  production of the Higgs boson with a top-quark pair}},  {\em JHEP} {\bf 09}
  (2014) 087, [\href{http://arxiv.org/abs/1408.1682}{{\tt arXiv:1408.1682}}].
  [Erratum: JHEP10,106(2014)].

\bibitem{Aaltonen:2013ipa}
{\bf CDF} Collaboration, T.~Aaltonen et~al., {\it {Combination fo Searches for
  the Higgs Boson Using the Full CDF Data Set}},  {\em Phys. Rev.} {\bf D88}
  (2013), no.~5 052013, [\href{http://arxiv.org/abs/1301.6668}{{\tt
  arXiv:1301.6668}}].

\bibitem{Abazov:2013gmz}
{\bf D0} Collaboration, V.~M. Abazov et~al., {\it {Combined search for the
  Higgs boson with the D0 experiment}},  {\em Phys. Rev.} {\bf D88} (2013),
  no.~5 052011, [\href{http://arxiv.org/abs/1303.0823}{{\tt arXiv:1303.0823}}].

\bibitem{Heinemeyer:2013tqa}
{\bf LHC Higgs Cross Section Working Group} Collaboration, J.~R. Andersen
  et~al., {\it {Handbook of LHC Higgs Cross Sections: 3. Higgs Properties}},
  \href{http://arxiv.org/abs/1307.1347}{{\tt arXiv:1307.1347}}.

\bibitem{Contino:2014aaa}
R.~Contino, M.~Ghezzi, C.~Grojean, M.~M{\"u}hlleitner, and M.~Spira, {\it
  {eHDECAY: an Implementation of the Higgs Effective Lagrangian into HDECAY}},
  {\em Comput. Phys. Commun.} {\bf 185} (2014) 3412--3423,
  [\href{http://arxiv.org/abs/1403.3381}{{\tt arXiv:1403.3381}}].

\bibitem{Contino:2010mh}
R.~Contino, C.~Grojean, M.~Moretti, F.~Piccinini, and R.~Rattazzi, {\it {Strong
  Double Higgs Production at the LHC}},  {\em JHEP} {\bf 05} (2010) 089,
  [\href{http://arxiv.org/abs/1002.1011}{{\tt arXiv:1002.1011}}].

\bibitem{Giudice:2007fh}
G.~F. Giudice, C.~Grojean, A.~Pomarol, and R.~Rattazzi, {\it {The
  Strongly-Interacting Light Higgs}},  {\em JHEP} {\bf 06} (2007) 045,
  [\href{http://arxiv.org/abs/hep-ph/0703164}{{\tt hep-ph/0703164}}].

\bibitem{Azatov:2012bz}
A.~Azatov, R.~Contino, and J.~Galloway, {\it {Model-Independent Bounds on a
  Light Higgs}},  {\em JHEP} {\bf 04} (2012) 127,
  [\href{http://arxiv.org/abs/1202.3415}{{\tt arXiv:1202.3415}}]. [Erratum:
  JHEP04,140(2013)].

\bibitem{Contino:2013kra}
R.~Contino, M.~Ghezzi, C.~Grojean, M.~M{\"u}hlleitner, and M.~Spira, {\it
  {Effective Lagrangian for a light Higgs-like scalar}},  {\em JHEP} {\bf 07}
  (2013) 035, [\href{http://arxiv.org/abs/1303.3876}{{\tt arXiv:1303.3876}}].

\bibitem{Barbieri:2007bh}
R.~Barbieri, B.~Bellazzini, V.~S. Rychkov, and A.~Varagnolo, {\it {The Higgs
  boson from an extended symmetry}},  {\em Phys. Rev.} {\bf D76} (2007) 115008,
  [\href{http://arxiv.org/abs/0706.0432}{{\tt arXiv:0706.0432}}].

\bibitem{Grojean:2006nn}
C.~Grojean, W.~Skiba, and J.~Terning, {\it {Disguising the oblique
  parameters}},  {\em Phys. Rev.} {\bf D73} (2006) 075008,
  [\href{http://arxiv.org/abs/hep-ph/0602154}{{\tt hep-ph/0602154}}].

\bibitem{Azatov:2013ura}
A.~Azatov, R.~Contino, A.~Di~Iura, and J.~Galloway, {\it {New Prospects for
  Higgs Compositeness in $h \to Z\gamma$}},  {\em Phys. Rev.} {\bf D88} (2013),
  no.~7 075019, [\href{http://arxiv.org/abs/1308.2676}{{\tt arXiv:1308.2676}}].

\bibitem{Pich:2012dv}
A.~Pich, I.~Rosell, and J.~J. Sanz-Cillero, {\it {Viability of strongly-coupled
  scenarios with a light Higgs-like boson}},  {\em Phys. Rev. Lett.} {\bf 110}
  (2013) 181801, [\href{http://arxiv.org/abs/1212.6769}{{\tt
  arXiv:1212.6769}}].

\bibitem{Pich:2013fea}
A.~Pich, I.~Rosell, and J.~J. Sanz-Cillero, {\it {Oblique S and T Constraints
  on Electroweak Strongly-Coupled Models with a Light Higgs}},  {\em JHEP} {\bf
  01} (2014) 157, [\href{http://arxiv.org/abs/1310.3121}{{\tt
  arXiv:1310.3121}}].

\bibitem{Khachatryan:2016vau}
{\bf ATLAS, CMS} Collaboration, G.~Aad et~al., {\it {Measurements of the Higgs
  boson production and decay rates and constraints on its couplings from a
  combined ATLAS and CMS analysis of the LHC $pp$ collision data at $\sqrt{s}=$
  7 and 8 TeV}},  \href{http://arxiv.org/abs/1606.02266}{{\tt
  arXiv:1606.02266}}.

\bibitem{Falkowski:2013dza}
A.~Falkowski, F.~Riva, and A.~Urbano, {\it {Higgs at last}},  {\em JHEP} {\bf
  11} (2013) 111, [\href{http://arxiv.org/abs/1303.1812}{{\tt
  arXiv:1303.1812}}].

\bibitem{Ellis:2013lra}
J.~Ellis and T.~You, {\it {Updated Global Analysis of Higgs Couplings}},  {\em
  JHEP} {\bf 06} (2013) 103, [\href{http://arxiv.org/abs/1303.3879}{{\tt
  arXiv:1303.3879}}].

\bibitem{Djouadi:2013qya}
A.~Djouadi and G.~Moreau, {\it {The couplings of the Higgs boson and its CP
  properties from fits of the signal strengths and their ratios at the 7+8 TeV
  LHC}},  {\em Eur. Phys. J.} {\bf C73} (2013), no.~9 2512,
  [\href{http://arxiv.org/abs/1303.6591}{{\tt arXiv:1303.6591}}].

\bibitem{Belanger:2013xza}
G.~Belanger, B.~Dumont, U.~Ellwanger, J.~F. Gunion, and S.~Kraml, {\it {Global
  fit to Higgs signal strengths and couplings and implications for extended
  Higgs sectors}},  {\em Phys. Rev.} {\bf D88} (2013) 075008,
  [\href{http://arxiv.org/abs/1306.2941}{{\tt arXiv:1306.2941}}].

\bibitem{Chpoi:2013wga}
S.~Choi, S.~Jung, and P.~Ko, {\it {Implications of LHC data on 125 GeV
  Higgs-like boson for the Standard Model and its various extensions}},  {\em
  JHEP} {\bf 10} (2013) 225, [\href{http://arxiv.org/abs/1307.3948}{{\tt
  arXiv:1307.3948}}].

\bibitem{Bechtle:2014ewa}
P.~Bechtle, S.~Heinemeyer, O.~St{\aa}l, T.~Stefaniak, and G.~Weiglein, {\it
  {Probing the Standard Model with Higgs signal rates from the Tevatron, the
  LHC and a future ILC}},  {\em JHEP} {\bf 11} (2014) 039,
  [\href{http://arxiv.org/abs/1403.1582}{{\tt arXiv:1403.1582}}].

\bibitem{Bergstrom:2014vla}
J.~Bergstrom and S.~Riad, {\it {Bayesian Model comparison of Higgs couplings}},
   {\em Phys. Rev.} {\bf D91} (2015), no.~7 075008,
  [\href{http://arxiv.org/abs/1411.4876}{{\tt arXiv:1411.4876}}].

\bibitem{Corbett:2015ksa}
T.~Corbett, O.~J.~P. Eboli, D.~Goncalves, J.~Gonzalez-Fraile, T.~Plehn, and
  M.~Rauch, {\it {The Higgs Legacy of the LHC Run I}},  {\em JHEP} {\bf 08}
  (2015) 156, [\href{http://arxiv.org/abs/1505.05516}{{\tt arXiv:1505.05516}}].

\bibitem{Gomez-Ceballos:2013zzn}
{\bf TLEP Design Study Working Group} Collaboration, M.~Bicer et~al., {\it
  {First Look at the Physics Case of TLEP}},  {\em JHEP} {\bf 01} (2014) 164,
  [\href{http://arxiv.org/abs/1308.6176}{{\tt arXiv:1308.6176}}].

\bibitem{Barklow:2015tja}
T.~Barklow, J.~Brau, K.~Fujii, J.~Gao, J.~List, N.~Walker, and K.~Yokoya, {\it
  {ILC Operating Scenarios}},  \href{http://arxiv.org/abs/1506.07830}{{\tt
  arXiv:1506.07830}}.

\bibitem{Fujii:2015jha}
K.~Fujii et~al., {\it {Physics Case for the International Linear Collider}},
  \href{http://arxiv.org/abs/1506.05992}{{\tt arXiv:1506.05992}}.

\bibitem{CepCreport}
{\bf CEPC-SPPC Study Group} Collaboration, {\it {CEPC-SPPC Preliminary
  Conceptual Design Report}},  2015.

\bibitem{CMS:2013xfa}
{\bf CMS} Collaboration, {\it {Projected Performance of an Upgraded CMS
  Detector at the LHC and HL-LHC: Contribution to the Snowmass Process}},  in
  {\em {Community Summer Study 2013: Snowmass on the Mississippi (CSS2013)
  Minneapolis, MN, USA, July 29-August 6, 2013}}, 2013.
\newblock \href{http://arxiv.org/abs/1307.7135}{{\tt arXiv:1307.7135}}.

\bibitem{ATL-PHYS-PUB-2013-014}
{\it {Projections for measurements of Higgs boson cross sections, branching
  ratios and coupling parameters with the ATLAS detector at a HL-LHC}},  Tech.
  Rep. ATL-PHYS-PUB-2013-014, CERN, Geneva, Oct, 2013.

\bibitem{ATL-PHYS-PUB-2014-011}
{\it {Prospects for the study of the Higgs boson in the VH(bb) channel at
  HL-LHC}},  Tech. Rep. ATL-PHYS-PUB-2014-011, CERN, Geneva, Jul, 2014.

\bibitem{ATL-PHYS-PUB-2014-016}
{\it {Projections for measurements of Higgs boson signal strengths and coupling
  parameters with the ATLAS detector at a HL-LHC}},  Tech. Rep.
  ATL-PHYS-PUB-2014-016, CERN, Geneva, Oct, 2014.

\bibitem{Erler:2000jg}
J.~Erler, S.~Heinemeyer, W.~Hollik, G.~Weiglein, and P.~M. Zerwas, {\it
  {Physics impact of GigaZ}},  {\em Phys. Lett.} {\bf B486} (2000) 125--133,
  [\href{http://arxiv.org/abs/hep-ph/0005024}{{\tt hep-ph/0005024}}].
  [,1389(2000)].

\bibitem{Freitas:2013xga}
A.~Freitas, K.~Hagiwara, S.~Heinemeyer, P.~Langacker, K.~Moenig, M.~Tanabashi,
  and G.~W. Wilson, {\it {Exploring Quantum Physics at the ILC}},  in {\em
  {Community Summer Study 2013: Snowmass on the Mississippi (CSS2013)
  Minneapolis, MN, USA, July 29-August 6, 2013}}, 2013.
\newblock \href{http://arxiv.org/abs/1307.3962}{{\tt arXiv:1307.3962}}.

\bibitem{Fan:2014vta}
J.~Fan, M.~Reece, and L.-T. Wang, {\it {Possible Futures of Electroweak
  Precision: ILC, FCC-ee, and CEPC}},  {\em JHEP} {\bf 09} (2015) 196,
  [\href{http://arxiv.org/abs/1411.1054}{{\tt arXiv:1411.1054}}].

\bibitem{Ge:2016zro}
S.-F. Ge, H.-J. He, and R.-Q. Xiao, {\it {Probing New Physics Scales from Higgs
  and Electroweak Observables at $e^+ e^-$ Higgs Factory}},
  \href{http://arxiv.org/abs/1603.03385}{{\tt arXiv:1603.03385}}.

\bibitem{AzziFCCweek}
P.~Azzi, ``{Progress in FCC-ee experimental studies}.''
  \href{https://indico.cern.ch/event/438866/contributions/1085033/attachments/1255970/1854053/FCCee_Rome2016_Azzi.pdf}{Talk
  given at the 2016 FCC Week in Rome}.

\bibitem{Dawson:2013bba}
S.~Dawson et~al., {\it {Working Group Report: Higgs Boson}},  in {\em
  {Community Summer Study 2013: Snowmass on the Mississippi (CSS2013)
  Minneapolis, MN, USA, July 29-August 6, 2013}}, 2013.
\newblock \href{http://arxiv.org/abs/1310.8361}{{\tt arXiv:1310.8361}}.

\bibitem{Baak:2013fwa}
M.~Baak et~al., {\it {Working Group Report: Precision Study of Electroweak
  Interactions}},  in {\em {Proceedings, Community Summer Study 2013: Snowmass
  on the Mississippi (CSS2013): Minneapolis, MN, USA, July 29-August 6, 2013}},
  2013.
\newblock \href{http://arxiv.org/abs/1310.6708}{{\tt arXiv:1310.6708}}.

\bibitem{Freitas:2014owa}
A.~Freitas, {\it {Electroweak precision tests in the LHC era and Z-decay form
  factors at two-loop level}},  in {\em {Proceedings, 12th DESY Workshop on
  Elementary Particle Physics: Loops and Legs in Quantum Field Theory
  (LL2014)}}, 2014.
\newblock \href{http://arxiv.org/abs/1406.6980}{{\tt arXiv:1406.6980}}.

\bibitem{Freitas:2016sty}
A.~Freitas, {\it {Numerical multi-loop integrals and applications}},
  \href{http://arxiv.org/abs/1604.00406}{{\tt arXiv:1604.00406}}.

\bibitem{Asner:2008nq}
D.~M. Asner et~al., {\it {Physics at BES-III}},  {\em Int. J. Mod. Phys.} {\bf
  A24} (2009) S1--794, [\href{http://arxiv.org/abs/0809.1869}{{\tt
  arXiv:0809.1869}}].

\bibitem{LubiczFantasy}
V. Lubicz, private communication based on \cite{Andreazza:2015bja}.

\bibitem{Beneke:1998rk}
M.~Beneke, {\it {A Quark mass definition adequate for threshold problems}},
  {\em Phys. Lett.} {\bf B434} (1998) 115--125,
  [\href{http://arxiv.org/abs/hep-ph/9804241}{{\tt hep-ph/9804241}}].

\bibitem{Hoang:1999zc}
A.~H. Hoang and T.~Teubner, {\it {Top quark pair production close to threshold:
  Top mass, width and momentum distribution}},  {\em Phys. Rev.} {\bf D60}
  (1999) 114027, [\href{http://arxiv.org/abs/hep-ph/9904468}{{\tt
  hep-ph/9904468}}].

\bibitem{Beneke:2015kwa}
M.~Beneke, Y.~Kiyo, P.~Marquard, A.~Penin, J.~Piclum, and M.~Steinhauser, {\it
  {Next-to-Next-to-Next-to-Leading Order QCD Prediction for the Top Antitop
  $S$-Wave Pair Production Cross Section Near Threshold in $e^+e^-$
  Annihilation}},  {\em Phys. Rev. Lett.} {\bf 115} (2015), no.~19 192001,
  [\href{http://arxiv.org/abs/1506.06864}{{\tt arXiv:1506.06864}}].

\bibitem{Beneke:2015lwa}
M.~Beneke, A.~Maier, J.~Piclum, and T.~Rauh, {\it {Higgs effects in top
  anti-top production near threshold in $e^+e^-$ annihilation}},  {\em Nucl.
  Phys.} {\bf B899} (2015) 180--193,
  [\href{http://arxiv.org/abs/1506.06865}{{\tt arXiv:1506.06865}}].

\bibitem{Marquard:2015qpa}
P.~Marquard, A.~V. Smirnov, V.~A. Smirnov, and M.~Steinhauser, {\it {Quark Mass
  Relations to Four-Loop Order in Perturbative QCD}},  {\em Phys. Rev. Lett.}
  {\bf 114} (2015), no.~14 142002, [\href{http://arxiv.org/abs/1502.01030}{{\tt
  arXiv:1502.01030}}].

\bibitem{deBlas:2016nqo}
J.~de~Blas, M.~Ciuchini, E.~Franco, S.~Mishima, M.~Pierini, L.~Reina, and
  L.~Silvestrini, {\it {Electroweak precision constraints at present and future
  colliders}},  2016.
\newblock \href{http://arxiv.org/abs/1611.05354}{{\tt arXiv:1611.05354}}.

\bibitem{Andreazza:2015bja}
A.~Andreazza et~al., {\it {What Next: White Paper of the INFN-CSN1}},  {\em
  Frascati Phys. Ser.} {\bf 60} (2015) 1--302.

\end{thebibliography}\endgroup

\end{document}